\documentclass[letterpaper,twocolumn,table,10pt]{article}
\usepackage{usenix-2020-09}
\input{package}

\newcommand{\xl}[1]{{\color{blue}{\{xl: #1\}}}}
\newcommand{\zl}[1]{{\color{purple}{\{zl: #1\}}}}
\newcommand{\qz}[1]{{\color{brown}{\{qz: #1\}}}}
\newcommand{\xb}[1]{{\color{cyan}{\{xb: #1\}}}}
\newcommand{\system}{{\sc ResolverFuzz}\xspace}
\newcommand{\systemb}{{\sc \textbf{ResolverFuzz}}\xspace}
\newcommand{\ignore}[1]{}

\newcommand{\revise}[1]{{#1}}

\definecolor{clthu}{RGB}{85, 43, 111}
\definecolor{cluci}{RGB}{219, 109, 0}
\definecolor{clzgc}{RGB}{226, 32, 17}
\definecolor{LimeGreen}{RGB}{51, 205, 51}
\definecolor{clqc}{RGB}{75, 102, 131}
\newcommand{\mkthu}[0]{{\color{clthu}{$^*$}}}
\newcommand{\mkuci}[0]{{\color{cluci}{$^\dag$}}}
\newcommand{\mkzgc}[0]{{\color{clzgc}{$^\S$}}}
\newcommand{\mkqc}[0]{{\color{clqc}{$^\P$}}}
\newcommand{\mkletter}[0]{{\color{LimeGreen}{\Envelope}}}
\newcommand{\usenixhref}[3][black]{\href{#2}{\color{#1}{#3}}}%
\renewcommand\footnotemark{}

\begin{document}

\title{\Large \bf \systemb: Automated Discovery of DNS Resolver Vulnerabilities with Query-Response Fuzzing}

\author{
    \usenixhref{https://faculty.sites.uci.edu/zhouli/research/}{\rm Qifan Zhang}\mkuci ,
    \usenixhref{https://faculty.sites.uci.edu/zhouli/research/}{\rm Xuesong Bai}\mkuci ,
    \usenixhref{https://netsec.ccert.edu.cn/people/lx19}{\rm Xiang Li}\mkthu \mkletter \rm ,
    \usenixhref{https://netsec.ccert.edu.cn/people/duanhx/}{\rm Haixin Duan}\mkthu \mkzgc \mkqc \rm ,
    \usenixhref{https://netsec.ccert.edu.cn/people/qli/}{\rm Qi Li}\mkthu, and
    \usenixhref{https://faculty.sites.uci.edu/zhouli/}{\rm Zhou Li}\mkuci \mkletter \rm
    \thanks{\mkletter~Corresponding authors. Most of Xiang Li's work was done when visiting UCI as a project specialist.}
    \medskip
    \\
    \mkuci \usenixhref{https://uci.edu/}{University of California, Irvine}, 
    \mkthu \usenixhref{https://www.tsinghua.edu.cn/en/}{Tsinghua University}
    \\
    \mkzgc Zhongguancun Laboratory,
    \mkqc \usenixhref{https://www.qcl.edu.cn/}{Quan Cheng Laboratory}
}

\maketitle

\begin{abstract}
Domain Name System (DNS) is a critical component of the Internet. DNS resolvers, which act as the cache between DNS clients and DNS nameservers, are the central piece of the DNS infrastructure, essential to the scalability of DNS. However, finding the resolver vulnerabilities is non-trivial, and this problem is not well addressed by the existing tools. To list a few reasons, first, most of the known resolver vulnerabilities are non-crash bugs that cannot be directly detected by the existing oracles (or sanitizers). Second, there lacks rigorous specifications to be used as references to classify a test case as a resolver bug. Third, DNS resolvers are stateful, and stateful fuzzing is still challenging due to the large input space.

In this paper, we present a new fuzzing system termed \system to address the aforementioned challenges related to DNS resolvers, with a suite of new techniques being developed. First, \system performs constrained stateful fuzzing by focusing on the short query-response sequence, which has been demonstrated as the most effective way to find resolver bugs, based on our study of the published DNS CVEs. Second, to generate test cases that are more likely to trigger resolver bugs, we combine probabilistic context-free grammar (PCFG) based input generation with byte-level mutation for both queries and responses. Third, we leverage differential testing and clustering to identify non-crash bugs like cache poisoning bugs.
We evaluated \system against 6 mainstream DNS software under 4 resolver modes.
\revise{Overall, we identify 23 vulnerabilities that can result in cache poisoning, resource consumption, and crash attacks.
After responsible disclosure, 19 of them have been confirmed or fixed, and 15 CVE numbers have been assigned. 
}

\end{abstract}

\section{Introduction}
\label{sec:intro}


Domain Name System (DNS) is central to Internet activities, translating human-friendly domain names to machine-friendly IP addresses. When a client user issues a DNS query to an authoritative server with the answers, a DNS \textit{resolver} is often encountered on the resolution path, which provides caching service that is essential to the scalability of DNS infrastructure~\cite{jung2001dns}. Numerous resolvers, including public resolvers like Google DNS and local resolvers set up by ISPs, have been deployed~\cite{allman2018comments}, running various DNS software (e.g., BIND~\cite{bind} and Unbound~\cite{unbound}). 

Yet, despite decades of development of DNS infrastructure, resolver vulnerabilities are still continuously uncovered, with some causing severe damage if exploited by attackers. For example, by sending just one client query containing \texttt{RRSIG}, the attacker can crash a BIND resolver entirely (CVE-2022-3736~\cite{cve20223736stateless}). As another example, CVE-2022-2881~\cite{cve20222881infoleak} could be exploited like the infamous TLS heart-bleed vulnerability~\cite{heartbleed} to read sensitive memory data.
We believe new tools should be developed to effectively uncover resolver vulnerabilities, in order to secure the DNS infrastructure.

\ignore{Though resolver software has been extensively tested with the existing tools (e.g., BIND9 has joined the Google OSS-Fuzz project to be automatically fuzzed~\cite{bind-oss}), zero-day vulnerabilities were still exploited in the wild ~\cite{}.
}

\vspace{2pt} \noindent \textbf{Understanding resolver vulnerabilities.} 
As the first step, we try to understand the characteristics of resolver vulnerabilities by mining the published CVEs (Section~\ref{subsec:cves}). By carefully examining \textit{239} CVEs published from 1999 to 2023 about 6 mainstream resolvers, we found a lot of them are \textit{semantic} bugs \revise{that violate high-level rules or invariants\ignore{~\cite{zou2021tcp}}~\cite{tan2014bug}}, leading to cache poisoning~\cite{kaminsky2008black}, resource consumption, etc. 
\revise{Detecting such bugs is still challenging as they usually do not trigger software crash, which is the major target of the existing fuzzers like AFL~\cite{afl}. 
}
Though software fuzzing can be applied to test resolver software, generating meaningful DNS messages is non-trivial due to the complex structure (Section~\ref{sec:overview}). Moreover, DNS resolver runs a \textit{stateful} caching service, but stateful fuzzing has always been a major challenge in network fuzzing~\cite{ba2022stateful}, due to computational complexity in covering a large state space.

\vspace{2pt} \noindent \textbf{Query-response fuzzing for resolvers.}
To address the aforementioned challenges unique to DNS resolvers, we develop a new fuzzing system termed \system (Section~\ref{sec:design}). To accommodate resolver software that is built under different programming languages and models, we choose \textit{blackbox} fuzzing and monitor the status of a resolver with lightweight tools like cache dump and tcpdump~\cite{tcpdump}, without code recompilation or binary rewriting~\cite{afl}. The main insight guiding our design of \system\ is that a \textit{short} message sequence (e.g., one query from the client and/or one response from the nameserver) is sufficient to trigger \revise{a large number} of resolver bugs, as revealed from our CVE study, so \system\ performs \textit{constrained} stateful fuzzing by only mutating a pair of query and response. To generate test cases that are likely to be accepted by the resolvers, we perform grammar-based fuzzing, by generating test cases with \textit{probabilistic context-free grammar (PCFG)}~\cite{jelinek1992basic}, and augmenting the test cases with byte-level mutations. Given that the existing oracles are unsuited to detect the non-crash bugs specific to resolvers, we develop new oracles to detect cache poisoning and resource consumption. Detecting cache poisoning is particularly challenging, due to the lack of rigorous DNS RFCs to strictly define the canonical behaviors of caching. We address this issue by performing \revise{\textit{differential testing}\ignore{~\cite{zou2021tcp} \zl{different citation}}~\cite{mckeeman1998differential}} and use the cache inconsistency to find potential vulnerabilities. Still, inconsistent behaviors are pervasive among resolvers~\cite{dual-stack}, and many of them are not related to vulnerabilities. Hence, we develop a new bug triaging method based on \textit{bisecting K-means} to cluster the inconsistent test cases, so the manual investigation efforts are greatly reduced. Finally, to increase the fuzzing throughput and avoid affecting the remote DNS nameservers, we build a new test infrastructure that \textit{localizes} the nameserver hierarchy and enables concurrent resolver testing.

\vspace{2pt} \noindent \textbf{Evaluation.}
We have generated over 700K test cases towards 4 resolver modes (recursive-only, forward-only, CDNS with fallback, and CDNS without fallback) (Section~\ref{sec:evaluation}). The evaluation results show the test generator has good coverage of valid DNS messages and the oracles are effective in pinpointing resolver bugs. \revise{We have discovered \textit{23} vulnerabilities with \system (Section~\ref{sec:cases}). With responsible disclosure, 19 of them have been confirmed or fixed, and 15 CVEs were assigned.} We even discovered a very powerful bug ($CP1$ in Section~\ref{subsec:cachebug}) that can entirely bypass the bailiwick checking rule, and poison any domain in a TLD zone (e.g., any \texttt{.com} domain can be compromised after the bug is exploited).

\vspace{2pt} \noindent \textbf{Contributions.} Our contributions are summarized below.

\begin{itemize}\setlength{\itemsep}{0pt}
    \item We conduct a comprehensive study of DNS CVEs.
    \item We develop a new blackbox fuzzing system \system, based on our insights from the CVE study. It performs constrained query-response fuzzing for efficient bug discovery on resolvers.
    \item We develop and/or adjust a set of techniques, including DNS localization, PCFG-based test generation, differential testing, etc., for \system.
    \item \revise{We evaluate \system\ against 6 mainstream resolvers and uncover 23 bugs. We discuss our findings with software vendors and received acknowledgment.}
    \item \system\ is open-sourced~\cite{resolverfuzz-github}. 
\end{itemize}


\section{Background}
\label{sec:background}

\subsection{DNS and Resolvers}
\label{subsec:dns}


\begin{figure}[t]	
    \centering
    \includegraphics[width=\columnwidth]{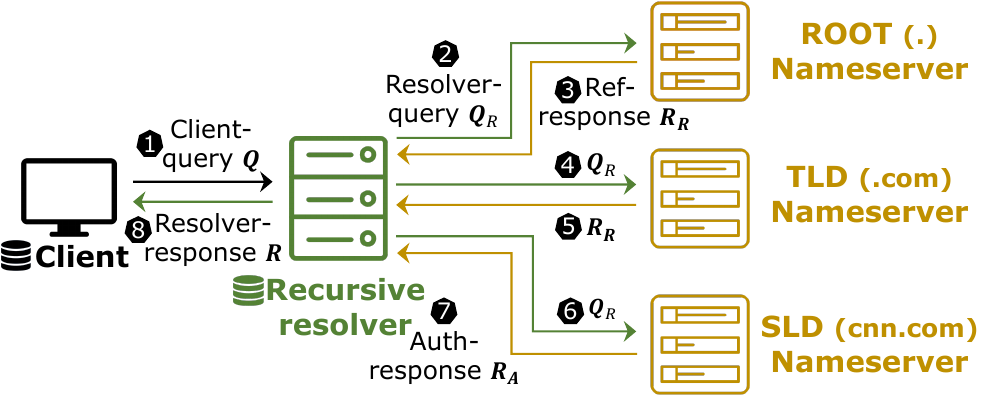}
    \caption{Example of DNS resolution process. ``ref-response'' and ``auth-response'' are both considered as ``ns-response''.
    }
     \vspace{-2mm}
    \label{fig:dns}
\end{figure}

DNS translates a user-friendly domain name to a numerical IP address. A domain name is written as a sequence of labels separated by ``\texttt{.}'', e.g., \texttt{cnn.com}. To resolve a domain name, a DNS client (or \textit{stub resolver}) usually issues a DNS query to a public DNS (e.g., Google Public DNS~\cite{google_dns}) or a local recursive resolver (e.g., Comcast ISP resolver), and lets the resolver contact nameservers iteratively.
Figure~\ref{fig:dns} illustrates the process for resolving \texttt{www.cnn.com}, during which the nameservers of root (denoted by ``\texttt{.}''), Top-Level Domain (TLD) \texttt{.com}, and Second-Level Domain (SLD) \texttt{cnn.com} are contacted. They answer the queries with \textit{resource record sets (RRSets)} in their \textit{zone} configurations. 

The format of the DNS query and response follows RFC 1034~\cite{mockapetris1987domain1034}.
In essence, the query and response share the same set of \textit{sections}, including ``Flags'', ``Question'', ``Answer'', ``Authority'', and ``Additional''. ``Flags'' signals the kind of message (e.g., \texttt{QR} represents response and \texttt{AA} represents authoritative answer). ``Question'' encodes the domain name to be resolved and the type of RRSet to be retrieved (e.g., \texttt{A} represents IPv4 address, \texttt{AAAA} represents IPv6 address, and \texttt{NS} represents nameserver domain). When the contacted nameserver has the authoritative answer, the ``Answer'' section of the response message encodes the requested RRSet. Otherwise, within the response, ``Authority'' fills \textit{referral} name that is ``closer'' (e.g., \texttt{.com} nameserver is closer to root server in answering queries about \texttt{example.com}) to give the authoritative answer, and ``Additional'' fills \textit{glue records} about the server addresses. Figure~\ref{fig:response} shows examples of DNS messages.


In this paper, we name the query from DNS client to resolver as \textit{client-query}, the query from resolver to nameserver as \textit{resolver-query}, response from nameserver to resolver as \textit{ns-response}, response from resolver to client as \textit{resolver-response}. For ns-response, it is classified into \textit{ref-response} which has the referral record, and \textit{auth-response} which has the authoritative answer. 

\vspace{2pt} \noindent \textbf{Resolver modes.} 
The resolver is central to the resolution procedure, which also caches the responses to reduce query latency. The standard mode of resolver runs recursive resolution and contacts nameservers. Alternatively, the resolver can run as \textit{forwarder}, which passes the query to other servers (e.g., upstream recursive resolvers). A resolver can also run the recursive and forwarder mode \textit{concurrently}, with each mode handling a subset of DNS namespace, and such resolver is called \textit{conditional DNS} (CDNS) ~\cite{cond_forwarder1, cond_forwarder2}. For some resolvers, a \textit{fallback} option~\cite{bind_fallback, unbound_fallback} can be enabled, to reissue the query in the recursive mode when forwarding the query fails.

\begin{table*}[t]
  \centering
  \small
  \caption{
  \revise{Study results of DNS CVEs for mainstream DNS software.} 
  }
  \begin{threeparttable}
    \begin{tabular}{c|cccc|ccc|c}
    \toprule
    \multirow{3}[3]{*}{\textbf{Software}\tnote{*}} & \multicolumn{8}{c}{\textbf{\# CVE}} \\
    \cmidrule{2-9}          & \multicolumn{4}{c|}{\textbf{Non-crash}} & \multicolumn{3}{c|}{\textbf{Crash}} & \multirow{2}[2]{*}{\textbf{Total}} \\
    \cmidrule{2-8}          & \multicolumn{1}{c|}{\textbf{Cache Poisoning}} & \multicolumn{1}{c|}{\textbf{Resource Consum.}\tnote{1}} & \multicolumn{1}{c|}{\textbf{Others}\tnote{2}} & \textbf{Total} & \multicolumn{1}{c|}{\textbf{Non-memory}\ignore{\tnote{3}}} & \multicolumn{1}{c|}{\textbf{Memory\ignore{\tnote{4}}}} & \textbf{Total} &  \\
    \midrule
    \textbf{BIND} & 18    & 18    & 11    & 47    & 75    & 22    & 97    & 144 \\
    \textbf{Unbound} & 4     & 5     & 4     & 13    & 5     & 8     & 13    & 26 \\
    \textbf{Knot Resolver} & 6     & 4     & 0     & 10    & 2     & 0     & 2     & 12 \\
    \textbf{PowerDNS Recursor} & 13    & 8     & 9     & 30    & 7     & 6     & 13    & 43 \\
    \textbf{MaraDNS} & 2     & 3     & 0     & 5     & 4     & 7     & 11    & 16 \\
    \textbf{Technitium} & 3     & 1     & 0     & 4     & 0     & 0     & 0     & 4 \\
    \midrule
    \textbf{Total} & 46    & 39    & 24    & 109   & 93    & 43    & 136   & 245 \\
    \bottomrule
    \end{tabular}%
    \begin{tablenotes}
      \footnotesize
      \item \tnote{*}: Recursive or forwarding modes. \tnote{1}: Resource consumption.
      \item \tnote{2}: An example of other non-crash bugs: CVE-2018-5738 that improperly permits recursion to all clients.
      \ignore{\tnote{3}: Crash caused by issues unrelated to memory corruption, like triggering assertion failure. \tnote{4}: Crash caused by memory corruption, like buffer overflow.}
      \item \# CVE of the forwarding mode only (7 in total): BIND (5), Unbound (0), Knot (1), PowerDNS (0), MaraDNS (0), and Technitium (1).
      \item \# CVE of the authoritative mode only (46 in total): BIND (19), Unbound (4), Knot (2), PowerDNS (20), MaraDNS (1), and Technitium (0).
      \item \# CVE of other software (132 in total): Microsoft DNS (90), Simple DNS Plus (1), Dnsmasq (34), CoreDNS (1), NSD (4), Yadifa (1), and TrustDNS (1).
    \end{tablenotes}
  \end{threeparttable}
          \vspace{-5mm}
  \label{tab:cve}%
\end{table*}%

\subsection{Study of DNS CVEs}
\label{subsec:cves}






Here we perform a comprehensive study to understand the distribution and root causes of DNS-related vulnerabilities, which also guides the design of \system. We crawl Common Vulnerabilities Exposures (CVE) databases~\cite{cve, cvedetails, cvebeta} and mainly analyze the CVE reports of six mainstream, open-sourced DNS software, including BIND~\cite{cvebind}, Unbound~\cite{cveunbound}, Knot~\cite{cveknot}, PowerDNS~\cite{cvepowerdns}, MaraDNS~\cite{cvemaradns}, and Technitium~\cite{cvetechnitium}. They are also extensively analyzed by previous work~\cite{zheng2020poison, man2020dns, lee2020longitudinal, jeitner2021injection}. 
\revise{We elaborate how we analyze CVE reports in Appendix~\ref{app:cvedetails}}.


Table~\ref{tab:cve} lists our study results of DNS CVEs related to resolver modes, and the CVE dates range from 1999 to 2023. We summarize our key findings ($F1$ to $F5$) below. 

\begin{itemize}\setlength{\itemsep}{0pt}
    \item \textbf{F1: Most of the CVEs are about resolvers.} In total, we identified \textit{291} CVEs related to the 6 studied DNS software (\textit{132} CVEs are related to other DNS software). Among them, \textit{245 (84\%)} are about resolvers (e.g., CVE-2019-6477 and CVE-2020-8621 for the recursive and forwarder mode). 
    Only 46 CVEs are about nameservers (e.g., CVE-2020-8619 and CVE-2017-3143).
    \item \textbf{F2: Diversified CVEs among DNS software.} Though BIND dominates in the number of CVEs\footnote{The high number of CVEs of BIND does not necessarily indicate it is more vulnerable. In fact, BIND has the largest market share~\cite{bind-market-share} and has been extensively tested~\cite{bind-test}.
    }, a prominent number of CVEs have also been found in other software (except Technitium).  Moreover, we found \textit{only 13} CVEs among the 245 CVEs affect all software (e.g., NXNSAttack  under CVE-2020-12662), suggesting the diverse  implementations of DNS software.
   \item 
   \revise{
   \textbf{F3: A significant portion of CVEs are not related to crash.} Like the results from a prior work that studies CVEs in TCP stacks~\cite{zou2021tcp}, we found the bugs that do not trigger software crash constitute a prominent portion (\textit{109} out of 245 CVEs).   
    The main consequences\ignore{ of them (\textit{167/172})} include cache poisoning (\textit{46} CVEs, e.g., caching illegal records under CVE-2002-2213 and CVE-2006-0527\ignore{~\cite{cve20022213cp, cve20060527cp}}) and resource consumption (\textit{39} CVEs, e.g., spending excessive resources to handle DNS queries under CVE-2022-2795 and CVE-2021-25219\ignore{~\cite{cve20222795rc, cve202125219rc}}).
    For the 136 crash-related bugs, only 43 are caused by memory corruption such as buffer overflow under CVE-2020-8625 and CVE-2021-25216. Others are mainly triggered by assertion failures (e.g., CVE-2022-0635 and CVE-2022-3080).
    
    \ignore{
    uninitiated memory~\cite{cve20173145unmem, cve20196470unmem}, and memory leaks~\cite{cve202238177memleak, cve202238178memleak}.}
    \ignore{In case of the remaining CVEs, we list them in the column ``Others'' that might lead to information leak~\cite{cve20222881infoleak}, restriction bypassing~\cite{cve20185738bypass}, privilege escalation~\cite{cve20173141privilege}, and more.
    }
    }
    \item \textbf{F4: Nearly every field of a DNS message has related CVEs.} Examples include \texttt{query name} (CVE-2020-8617), \texttt{query type} (CVE-2022-0667), \texttt{query flag} (CVE-2017-15105), \texttt{rcode} (CVE-2018-5734), \texttt{rdata} (CVE-2013-4854), \texttt{TTL} (CVE-2003-0914), etc.

    \ignore{e.g., \texttt{query}~\cite{cve20208623query}, \texttt{response}~\cite{cve20121033response}, \texttt{cache}~\cite{cve20103613cache}, \texttt{option}~\cite{cve20185744option}, \texttt{opcode}~\cite{cve20112464opcode}, \texttt{query name}~\cite{cve20208617qname}, \texttt{query type}~\cite{cve20220667qtype}, \texttt{query flag}~\cite{cve201715105flag}, \texttt{rcode}~\cite{cve20185734rcode}, \texttt{record}~\cite{cve20114313record}, \texttt{rdata}~\cite{cve20134854rdata}, \texttt{TTL}~\cite{cve20030914ttl}, \texttt{packet size}~\cite{cve20124244size}, \texttt{record count}~\cite{cve20064095count}, \texttt{packet number}~\cite{cve20064096number}, \texttt{TCP}~\cite{cve20220396tcp}, \texttt{fragmentation}~\cite{cve202125218frag}, \texttt{DNSSEC}~\cite{cve20196475dnssec}, \texttt{DoH}~\cite{cve20221183doh}, and others.
    }
    
    \item \textbf{F5: Most of the CVEs are triggered with a very short message sequence.} We found \textit{222/245} (91\%) CVEs could be triggered by sending \textit{just one client-query or ns-response}. One such example is CVE-2022-3736. For the other CVEs that require longer sequences (e.g., CVE-2022-3924), many client-queries are needed to trigger the bugs. We show their details in Appendix~\ref{app:cvedetails}.
    

    \ignore{For example, stateful bugs (CVE-2022-2881~\cite{cve20222881infoleak} and CVE-2012-3817~\cite{cve20123817stateful}) require sending many queries or reusing existing connections at a specific time.
}
\end{itemize}

With the above insights, we design \system\ and elaborate the design choices in Section~\ref{sec:overview}. \revise{We acknowledge that our CVE study could suffer from survivorship bias, and we discuss this issue in Section~\ref{sec:discussion}.}



\subsection{Prior Tools for DNS Bug Discovery}
\label{subsec:tools}

Here we survey the related tools that can automatically detect DNS bugs. In Section~\ref{sec:related}, we survey systems that can detect other network vulnerabilities.

First, we found fuzzing has been applied to test DNS resolvers. SnapFuzz aims to achieve high throughput in fuzzing network applications~\cite{andronidis2022snapfuzz}. It rewrites the tested program for greybox fuzzing and fast asynchronous communication. It was evaluated against a lightweight DNS software Dnsmasq~\cite{dnsmasq} that is usually deployed on routers, and 7 crashes were detected within 24 hours. However, it cannot directly detect non-crash bugs. 
DNS Fuzzer performs byte-level mutation by inserting new bytes into the seed DNS queries~\cite{dns-fuzzer}. Similar to SnapFuzz, it only detects crashes. As far as we know, the most related tool to \system is dns-fuzz-server~\cite{dns-fuzz-server}, which performs grammar-based and byte-level mutation on queries and responses. In Section~\ref{subsec:results}, we show the detailed comparison. 


In addition to resolvers, DNS \textit{nameservers} have also been found vulnerable when the zone files installed by the domain owners have mis-configurations.
A number of tools were developed to find such mis-configurations~\cite{romao1994tools, dnslint, dnssec_analyzer, mxtoolbox, dnsviz, pappas2004distributed}. Recently, formal methods have been applied by checking the DNS configurations with formal specifications. G-Root performs formal verification to prove the correctness of configurations and find counterexamples~\cite{kakarla2020groot}.  SCALE jointly generates zone files and corresponding queries that are specified by RFCs to discover implementation inconsistencies~\cite{kakarla2022scale}. Yet, for DNS resolvers, there lacks rigorous specifications to be used as references for vulnerability discovery~\cite{son2010hitchhiker, moon2021accurately}.

\section{Overview of \system}
\label{sec:overview}

\subsection{Problem Definition and Challenges}
\label{subsec:problem}

\noindent \textbf{General threat model and targeted vulnerabilities.} 
We consider a public or local recursive resolver to be targeted by the attacker. The attacker is able to control a downstream DNS client and/or an upstream nameserver on the resolution path of the resolver. Hence, DNS queries and responses in arbitrary format can be issued against the resolver. \revise{We consider 4 types of vulnerabilities as they are related to most CVEs (see Table~\ref{tab:cve}) and overview them below. In Section~\ref{sec:cases}, we elaborate the threat model for each type.

\begin{itemize}\setlength{\itemsep}{0pt}
\item \textbf{Cache poisoning.} The attacker tampers resolver's cache and directs victim clients to malicious servers.
\item \textbf{Resource consumption.} The attacker heavily consumes resolver's resources to impact its service quality.
\item \textbf{Non-memory crash.} The attacker terminates a resolver without memory corruption, e.g., by sending DNS messages to execute code with assertion failures. 
\item \textbf{Memory crash.} The attacker's DNS messages corrupt the resolver memory and terminate the resolver.
\end{itemize}

}
\ignore{The standard security mechanisms are all enabled on the resolver (e.g., bailiwick checks~\cite{} and DNSSEC validation~\cite{} \zl{DNSSEC is turned off}).
}
\ignore{
The attacker only needs to know the software and/or its version of the DNS resolver before the attack, and in Section~\ref{subsec:scanning} we show such information can be obtained through Internet-wide scanning. 
}


To notice, side-channel vulnerabilities are out of scope of this work, as these vulnerabilities often exist at the layers \textit{below} DNS.
 In Section~\ref{sec:related}, we review them under the theme of off-path cache poisoning attacks.

\vspace{2pt} \noindent \textbf{Design goals and challenges.} 
\system aims to uncover the vulnerabilities under the aforementioned threat model. We focus on four types of vulnerabilities including cache poisoning, resource consumption, service crash, and memory corruption, as our survey in Section~\ref{subsec:cves} suggests they are the major issues against DNS software. \system\ should be \textit{efficient} in testing resolvers at high throughput. Moreover, \system\ should be able to tell whether the test inputs could lead to vulnerability discovery at high \textit{accuracy}. We encounter a few key challenges towards meeting these goals:

\begin{itemize}\setlength{\itemsep}{0pt}
    \item \textbf{C1: Efficiency.} Notable latency is expected for a regular DNS resolution, as network communications are needed among the client, resolver, and multiple nameservers. Hence, achieving high throughput for resolver fuzzing is not trivial. 
    \item \textbf{C2: Mutation.} The widely used greybox fuzzers like AFL~\cite{afl} mutate the input with coverage-based metric. However, such metric does not provide sufficient guidance on which \textit{part} of the testing input should be mutated~\cite{peng2023gleefuzz}, but DNS messages contain many fields that are related to bugs (\textbf{F4} in Section~\ref{subsec:cves}).
    \item \textbf{C3: Stateful fuzzing.} Different from nameservers that run in a stateless mode, resolvers are \textit{stateful}~\cite{kakarla2022formal}, whose states depend on cache records, configurations, etc. Stateful services have been considered a major challenge for network fuzzing~\cite{ba2022stateful}, due to the large search space of input sequences.
    \item \textbf{C4: Oracle.} Non-crash bugs have a large share in resolver CVEs (\textbf{F3} in Section~\ref{subsec:cves}). However, there lacks an oracle to detect such bugs. Instead, crash bugs can be detected by oracles like AddressSanitizer~\cite{serebryany2012addresssanitizer}.  Differential testing has been used to uncover semantic bugs that are non-crash, but none of the prior works built the oracle for DNS. Moreover, our empirical analysis suggests inconsistencies among DNS resolvers are common, and many of them do not indicate vulnerabilities. In Section~\ref{subsec:results}, we show an example of normal inconsistencies. 
\end{itemize}



\subsection{Workflow of \systemb}
\label{subsec:flow}


\begin{figure*}[t]
    \centering
    \includegraphics[width=0.9\linewidth]{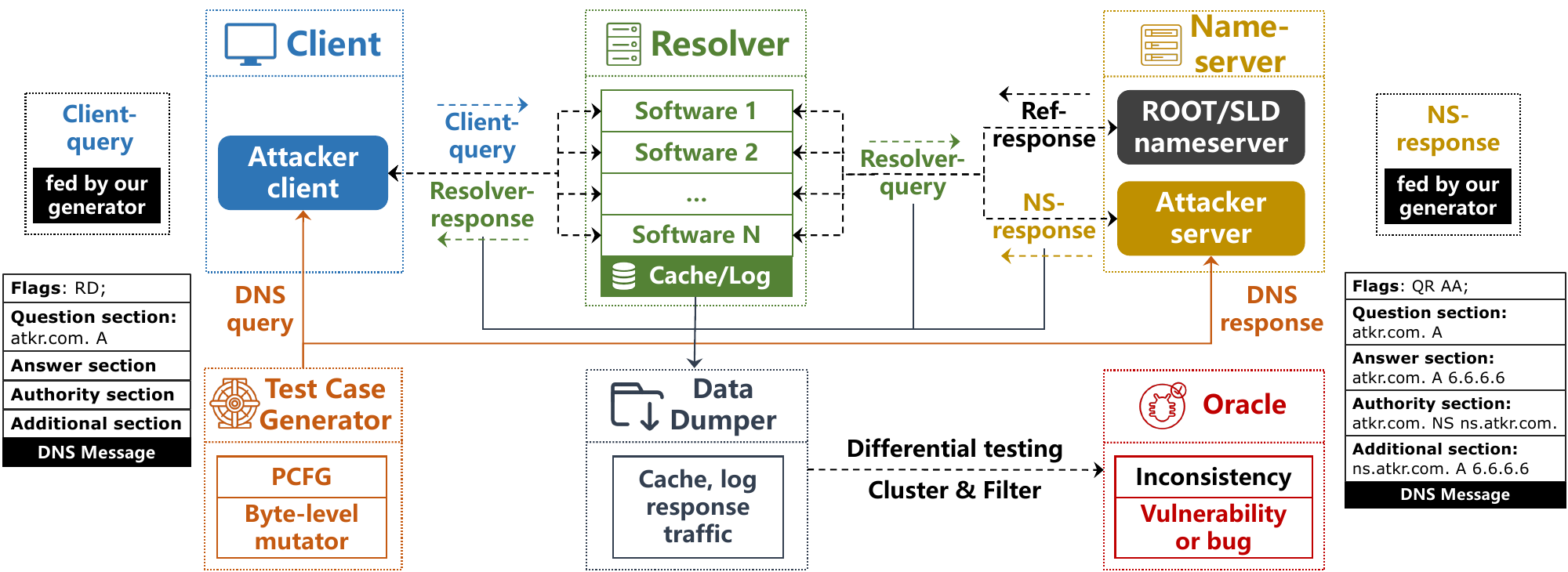}
    \caption{Workflow of \system.
    }
    \vspace{-5mm}
    \label{fig:work_flow}
\end{figure*}


\system addresses the aforementioned challenges with 3 newly designed components, which are elaborated in Section~\ref{sec:design}. The workflow of \system is also illustrated in Figure~\ref{fig:work_flow}.
In the first stage, the testing infrastructure (Section~\ref{subsec:env}) loads the resolvers of different implementations and configures the environment. A number of optimization techniques are applied to increase the throughput of DNS queries and responses, addressing \textbf{C1}. Then, the tests generator (Section~\ref{subsec:input}) performs grammar-based mutation on client-queries and ns-responses, addressing \textbf{C2}. 
Though resolver is stateful (\textbf{C3}), according to our study on CVEs (\textbf{F5} in Section~\ref{subsec:cves}), triggering a bug in most cases only requires a short sequence of one client-query and/or one ns-response. Hence, we simplify stateful fuzzing to simulate short sequences.
Finally, for data collected by our data dumper, including the cache, log, response, and traffic, we develop different oracles to detect different types of vulnerabilities mentioned in Section~\ref{subsec:problem}. We apply differential testing to identify inconsistencies among resolvers to capture cache poisoning bugs. To address \textbf{C4}, we develop a new method to cluster inconsistencies and alert on the abnormal ones, so the manual efforts in bug investigation will be significantly reduced. 

\ignore{
Therefore, in this study, we mainly focus on stateless vulnerabilities or bugs for both recursive and forwarding modes via fuzzing as many DNS fields as possible.
in contrast to stateful bugs.
\zl{cache poisoning has at least two rounds, right?}
\zl{we're indeed stateful~\cite{kakarla2022formal}, but we exploit a subset of possible states. only one round of query-response to find inconsistency. then, starting from the inconsistency, constructing the next set of query response to trigger the bugs. dns-fuzz-server is purely stateless, can we make a case that stateful bugs will be missed because of this?}
\zl{give an example of a stateful bug and why finding it can be done with our simple exploration}
}

\section{Design of \systemb}
\label{sec:design}

\subsection{Testing Infrastructure and Optimization}
\label{subsec:env}





We design a new infrastructure to simulate DNS queries and responses against resolvers and collect the traces to be later analyzed. The infrastructure mainly includes a DNS client and a nameserver that emit messages generated by our fuzzer (we call them \textit{attacker client} and \textit{attacker server}), a test scheduler, and a trace collector. Below we first describe these components and their network setup. Then, we overview the whole testing process. 

\vspace{2pt} \noindent \textbf{Attacker client and server.} 
Instead of using the off-the-shelf DNS software, we implement the attacker client and server with Python scripts. As such, we are able to send \textit{arbitrary} queries and responses that can even be incompliant with DNS RFCs~\cite{mockapetris1987domain1034} and reduce the processing latency with lightweight implementations. Specifically for the attacker server, though the standard implementations require a zone file to be hosted to answer the queries with the contained records, we choose to directly generate an ns-response given a client-query and skip the zone file (details are in Section~\ref{subsec:input}).
We notice that related systems like SCALE~\cite{kakarla2022scale} reuse the existing DNS software for nameservers and mutate the zone files for testing purposes. Our customized implementation is more flexible in response generation, and even allows low-level manipulation of DNS responses (e.g., answering DNS queries with UDP or TCP).


\vspace{2pt} \noindent \textbf{Test scheduler.} 
This central component initializes DNS components, including the attacker client, attacker server and resolvers. Between each round of test (i.e., one round-trip of DNS resolution), the scheduler resets these components. For efficient and complete resetting, we choose to host these components with lightweight Docker containers~\cite{merkel2014docker}. To increase the testing throughput, the scheduler will command the client to send queries to the resolvers \textit{in parallel} and replicate the resolver instances. Specifically, the scheduler groups an attacker client, an attacker server, and resolvers into a \textit{unit}, and runs multiple units concurrently (with different test cases). With container-based isolation, each resolver instance can be tested independently.  

\vspace{2pt} \noindent \textbf{Resolvers and analyzed data.} 
To detect the 4 types of vulnerabilities described in Section~\ref{subsec:problem}, we install a set of monitors inside each resolver container. First, we use a set of tools to export the resolver cache from memory to files (see Section~\ref{subsec:setup} for details of the tools) to detect cache-related bugs. Second, we use \texttt{tcpdump}~\cite{tcpdump} to collect the incoming and outgoing network traffic to detect bugs related to resource consumption. Third, we collect the log files generated by the DNS software. Finally, we also monitor the running status of the resolver process to detect service crashes and memory corruption with Linux command \texttt{ps}~\cite{ps}. 

\ignore{
\zl{to merge}
 To monitor the process status of a resolver, we use the command \texttt{ps -ef | grep -w [dns\_sw\_proc] | grep -v grep | wc -l}, where \texttt{[dns\_sw\_proc]} is replaced with the name of different software executable file (e.g., \emph{named} for BIND, \emph{unbound} for Unbound, etc.). This command will return either 1 or 0. We run this command after one payload is tested. If 1 is returned, it means the process of the resolver is normal, and we will continue for next round of testing. Otherwise, the container will be restarted, and then we start the resolver process again. 
}

\ignore{
Our implementation also includes traffic collection among the internal Docker network. tcpdump~\cite{tcpdump} is installed in the Docker containers of DNS software to collect network traffic during execution for further analysis. tcpdump is a tool that captures network packets on link layers of operating systems and does not depend on DNS software. Therefore, we can collect all the network traffic with DNS software Docker containers with the help of tcpdump regardless of DNS software implementations. On the host side, the scheduler integrates an input generator, which will be discussed in detail in Section~\ref{subsec:input}, 
}

In addition to testing the standard recursive mode (termed \textit{recursive-only}), we also test 3 alternative resolver modes: \textit{forward-only}, \textit{CDNS without fallback}, and \textit{CDNS with fallback} (explained in Section~\ref{subsec:dns}). 
We are motivated to test different modes due to that certain modes, e.g., forward, are more vulnerable as uncovered by previous work~\cite{zheng2020poison}. Changing a resolver from one mode to another can be easily done by loading a different configuration file into the resolver container. Figure~\ref{fig:mode_config} in Appendix~\ref{sec:config} shows configurations for all modes.
To notice, we did not test all resolver configurations: e.g., we disable DNSSEC~\cite{arends2005protocol} since we found it introduces a large number of inconsistencies among resolvers that are irrelevant to bugs. 
We discuss this limitation in Section~\ref{sec:discussion}.

\vspace{2pt} \noindent \textbf{Network configurations.} 
All the Docker containers are connected to a bridged Docker network assigned with /16 private IP addresses. 
Under a standard DNS resolution between the client and the nameserver of a registered domain, remote nameservers like root and TLD servers have to be contacted, so DNS round-trips could incur notable latency. Moreover, our tests could trigger bugs on those remote servers, raising ethical concerns. To address these issues, we choose to \textit{localize} the nameservers between the attacker client and attacker server in our lab network. We implement a server to simulate all these nameservers.
Each client-query asks about a domain name owned by us, and we use different subdomains to separate the test cases of different resolver modes.

\ignore{
Then, the authoritative server is set as the name server of three test domain names of the unit, i.e. \texttt{test-recursive.example.com}, \texttt{test-fwd-only.example.com} and \texttt{test-cdns.example\\.com}, with \texttt{NS} records.
}


\vspace{2pt} \noindent \textbf{Testing process.} 
1) The test scheduler initializes the Docker network and $n$ units of simulated DNS infrastructure. 2) The test scheduler obtains test cases from the generator and dispatches them to the attacker client and server. 3) After the start of each resolver container, the internal monitors collect the information about cache dump, network traffic, and process information. 4) When the attacker client receives the resolver-response, the resolver containers will be reset via cache flushing or software restarting, and the corresponding unit will go back to step 2 until all tests are completed.

\subsection{Test Case Generator}
\label{subsec:input}





We generate the test cases from \textit{two} dimensions, including client-queries for our attacker client and ns-responses for our attacker server.
Considering a DNS resolver as a stateful service, the ns-response is generated corresponding to a client-query. 
Though we only simulate a short message sequence (one client-query and one ns-response), we find it sufficient to uncover a variety of bugs, as suggested by \textbf{F5} in Section~\ref{subsec:cves}.

Though we can freely generate an ns-response disregarding the questions embedded in the client-query, our empirical analysis suggests such responses are often quickly dropped by the resolver. As a result, we restrict the ns-response to contain most of the fields (e.g., the ``Question'' Section and TxID) from the client-query and only add information to sections like ``Answer'', which significantly reduces the input space. Though under standard DNS resolution, the nameserver generates a response \textit{after} seeing the query, we found this process can be optimized by generating the pair of ns-response and client-query \textit{simultaneously} and dispatching them to the client and nameserver before the resolver is queried\footnote{A client-query can trigger a resolver to process multiple resolver-queries and ns-responses. In this case, we just use the same ns-response.}.

Given that DNS message has a complex structure, we apply \textit{probabilistic context-free grammar (PCFG)}~\cite{jelinek1992basic} \revise{to generate templates of client-queries and ns-responses first}.
A PCFG consists of a start symbol (denoted \texttt{<start>}), non-terminal symbols (symbols surrounded by \texttt{<>}), terminal symbols\ignore{ (symbols surrounded by \texttt{()})\xb{I didn't have () to mark terminal symbols, should I update the listings?}}, production rules, and rule probabilities. We assign high probabilities to certain fields \revise{after analyzing CVEs}, so the fuzzing process can be directed towards the code regions that are critical but error-prone~\cite{pcfg-fuzz}\footnote{\revise{The code regions of our interests are identified by analyzing CVE reports and reproducing CVE PoCs.}}. 
A DNS query is allowed to ask one or more questions (e.g., domain name) in the ``Question'' section\ignore{\xb{Only one allowed. DNS software will ignore the packet if contains more than 1 question.}\xl{not so right. some software accept more question. we needn't to emphasize here.}}, and a response can contain multiple answering records in the ``Answer'', ``Additional'', and ``Authority'' sections. However, letting the generator add an arbitrary number of questions or answering records will introduce a very large input space, which is also unlikely to trigger new bugs \ignore{\revise{(e.g., the same code logic is applied to handle 1,000 records and 100 records in ``Answer'' by BIND~\cite{})\zl{justify with code review}}}\footnote{\revise{For example, we reviewed the source code of BIND and found it rejects the query with multiple questions. In response, each answer is matched with the question but the number of answers does not impact the logic.}}
. Hence, we restrict the number of each section to range from 0 to 5 (for ``Question'', only 1 \texttt{QNAME} is inquired). We also force the counting field to contain the correct number of records (e.g., \texttt{ANCOUNT} contains the right number of records in ``Answer''). Field formats and additional rules are applied to fields like \texttt{QNAME} to ensure their validity.
In Appendix~\ref{app:cfg}, we list the complete PCFG.

Similar to previous work in fuzzing network services~\cite{jabiyev2022frameshifter, jabiyev2021t}, we perform byte-level mutation on \revise{the PCFG-generated message templates}. We are motivated to do so for DNS as recent work showed some DNS implementations fail to correctly decode strings with special characters embedded~\cite{jeitner2021injection}. We consider mutating each terminal symbol of PCFG with special bytes, such as \texttt{\textbackslash.}, \texttt{\textbackslash 000}, \texttt{@}, \texttt{/}, and \texttt{\textbackslash} (the first four characters are also exploited by ~\cite{jeitner2021injection}). The mutation operators include byte addition, deletion, and replacement, and we assign a probability to determine when the mutation should happen. We set the probability to conduct byte-mutation on a PCFG output as 0.1.
In Algorithm~\ref{alg:input_gen} of Appendix~\ref{app:input_gen} we summarize the whole process for test generation. 


\subsection{Data Dumper and Oracle}
\label{subsec:oracle}





\ignore{
In order to monitor the operation status of each software, we collect four types of output information: the resolver-response, the resolver's operation log, the resolver's cache, and the resolution traffic (both the client-resolver and resolver-nameserver side).
}


We first dump the traces collected from the resolver containers into the host and conduct data pre-processing. 
Then, we design 3 oracles to detect the 4 types of resolver bugs. 

\vspace{2pt} \noindent \textbf{Data dumper.}
Several previous works applied blackbox fuzzing on web services ~\cite{jabiyev2022frameshifter, jabiyev2021t} by only using information from the requests and responses. Though we also follow the direction of blackbox fuzzing without instrumenting and re-compiling the targeted resolvers, we collect information in addition to requests and responses, including cache dump, resolution traffic, software logs, and process status, to achieve better coverage of bugs. The formats of cache dumps and software logs vary for different resolver software, so we convert them to unified formats with the methods described below.

For cache dumps, the formats of our studied software are vastly different, as shown in Appendix \ref{app:cache_dump_format}. For instance, as shown in List~\ref{lis:cache_format_bind} and List~\ref{lis:cache_format_powerdns}, BIND caches basic DNS message information according to their sections, while PowerDNS is more verbose and keeps more results like zone and source. As the major goal of cache poisoning is to tamper the records associated with a domain name~\cite{kaminsky2008black}, we define a new cache structure that uses the cached domain names as the key and common record fields including \texttt{class}, \texttt{type}, \texttt{ttl}, and \texttt{rdata} as the values (shown in List~\ref{lis:cache_format_def}).
\revise{An alternative approach to cache dump is cache snooping~\cite{grangeia2004dns}, which infers the cache status of one record per query. We choose cache dump because 1) it provides more information like where the cached records come from and cache trust levels and 2) it is also more efficient.
}
For software logs, internal operations and their parameters are usually recorded,  which could be utilized to detect abnormal behaviors during resolution, e.g., excessive cache searching. We implement a pattern-matching method to search for log entries of our interests, and categorize them under keys such as \texttt{CACHE LOOKUP}, \texttt{QUERY}, and \texttt{SANITIZE RECORD}. For each test case, we assign the pre-processed cache dump, software logs, network traffic, and process status from all resolvers to it.


\vspace{2pt} \noindent \textbf{Cache poisoning oracle.}
We design this oracle based on the insights that the cache poisoning attack tampers the cache storage and usually \textit{inserts} forged records to hijack victim domains~\cite{kaminsky2008black}, so the cache records are likely to differ among the resolvers if some are vulnerable. 
We run \textit{differential testing} to find the cache anomalies. A number of previous work on differential testing rely on a \textit{``golden model''} to compare against. For example, DIFUZZRTL detects bugs in RISC-V CPU cores by comparing their RTL execution results with a golden model OpenRISC Or1ksim~\cite{hur2021difuzzrtl}. However, we cannot find a golden model for DNS resolvers: even the most widely used resolver BIND has more than 100 CVEs reported. 
Therefore, assuming the set of software studied by us is $\mathcal{S}$, for a software $s_i \in \mathcal{S}$, we consider $s_i$ is abnormal when its trace differs from any $s_j \in \mathcal{S} \setminus \{s_i\}$. 

Specifically for cache, for each test case, we check whether the cache records for all the resolver software are the same, by comparing the records' \texttt{NAME}, \texttt{TYPE}, and \texttt{RDATA} fields. After this stage, we found there are still many test cases with inconsistent cache records based on our evaluation, so we perform another round of bug triage by \textit{clustering} the test cases. 
We represent $s_i \in \mathcal{S}$ of each test case with 
the maximum number of different records of the software $i$ with other software. For instance, $<0, 0, 0, 5>$ means that software 1 to 3 has the same cache, but software 4 has 5 additional cache records. Then, we apply \textit{Bisecting K-Means}, which outperforms the basic K-Means in entropy measurement~\cite{kmeans}, on the cache vectors to generate clusters and investigate each cluster to look for vulnerabilities. Clustering is done for test cases under each resolver mode separately.

For the vulnerability analysis, we employ a semi-automatic method. By extracting information from the fields like \texttt{NAME}, \texttt{TYPE}, \texttt{RDATA}, \texttt{ZONE}, etc., we create matching rules and use them to separate test cases within a cluster into sub-clusters iteratively. 
\revise{For example, the sub-cluster for bug CP2 described in Section~\ref{subsec:resbug} is generated based on the existence of the NS record of the domain in the forwarding zone.
Then each sub-cluster is manually analyzed to confirm if it is related to vulnerabilities: we randomly sample 1 test case per sub-cluster and try to construct exploit.  
}
\ignore{For the vulnerability analysis, we randomly sample test cases, manually analyze the information related to \texttt{NAME}, \texttt{TYPE}, \texttt{RDATA}, \texttt{ZONE}, etc., and write filters to group test cases. This process is iteratively done until no more test cases are left.}
Though our vulnerability analysis can be done without clustering, we found the investigation overhead is significantly reduced after clustering.
\vspace{2pt} \noindent \textbf{Resource consumption oracle.}
Previous studies show that attackers have the incentive to disrupt the operation of resolvers or use the resolver to conduct DNS amplification attacks against other servers~\cite{dns_amplification_attacks}. We measure the resource consumption with 4 metrics, derived from the resolver's network traffic and software logs, including the number of resolver-queries, the sizes of responses (both ns-response and resolver-response), the resolution timeout, and the frequency of internal operations (e.g., cache search).


For a metric (say $m_j$), we compute its value distribution within the \textit{same} software (say $s_i$), and consider a test case abnormal if $s_i$'s value on $m_j$ falls out of the normal range. Specifically, we represent the value distribution as a Cumulative Distribution Function (CDF), and consider the normal range as $[0, \theta]$, where $\theta$ is the threshold and we set it to $0.9$. For instance, if the frequency of the cache search operations of one test case is higher than 90\% of all test cases in CDF, this test case is considered abnormal. Then, similar to cache oracle, we perform an iterative process of random sampling and manual investigation \revise{(the clustering stage is skipped as no differential testing is conducted here)}.

\vspace{2pt} \noindent \textbf{Crash oracle.}
To detect bugs related to memory and non-memory crash, the resolver container simply monitors the resolver process and considers anomaly happens when the process is not running (i.e., not included in the output of \texttt{ps} command). 
We acknowledge that this oracle is simple and false negatives can happen (e.g., dangling pointers might not trigger a crash). Though more complex oracles like AddressSanitizer~\cite{serebryany2012addresssanitizer} can detect more types of memory bugs, they often require re-compilation, which is incompatible with our blackbox fuzzing setting.

\ignore{
In this oracle, we implement a method to automatically detect whether the software is still alive.
In each round, after our client receives a response or a timeout is reached, we will invoke a system call to check whether the resolver software is still running and also responsive to new queries.

Once oracle detects a crash, all the operation info at that round will be saved for further analysis, the software will be restarted for testing in the next round.
}
\section{Evaluation and Results}
\label{sec:evaluation}

\subsection{Implementation Details}
\label{subsec:setup}

For the tested resolvers, we choose the ones listed in Table~\ref{tab:cve}, and their versions are all latest during our evaluation period (BIND: 9.18.0, Unbound: 1.16.0, Knot Resolver: 5.5.0, PowerDNS Recursor: 4.7.0, MaraDNS: 3.5.0022, and Technitium: 10.0.1). For the testing infrastructure, the attacker client and attacker server generate customized DNS messages using Python language.
When running the experiment, we force the reset of a resolver container if it does not respond before a 5-second timeout. 
The test scheduler manages Docker containers with Python Docker SDK~\cite{pythondocker}. Each software is compiled based on Ubuntu 22.04 Docker image~\cite{ubuntu-image}.
Within the resolver containers, we are able to dump the cache from 4 software with existing tools or supported APIs: rndc~\cite{bind-rndc} for BIND, unbound-control~\cite{unbound-unbound-control} for Unbound, rec\_control~\cite{powerdns-rec-control} for PowerDNS, and Technitium's HTTP API~\cite{technitium_web_config}.
We are unable to dump or decode the cache from MaraDNS and Knot, so we evaluate them by replaying the tests that are proven to impact other software on them. 
For the other nameservers in the resolution change, we write them with the Go language (for better performance) and configure their zone files to make our attacker server reachable. Figure~\ref{fig:auth_srv_zone_file} in Appendix~\ref{sec:config} shows the zone file. 

For the tests generator, we use different attacker domain names (all starting with \texttt{test-}) to test different resolver modes. 
For the oracles, we use a Python library scikit-learn~\cite{scikit-learn} to implement the clustering method.



We write in total of 3,649 lines of code (LoC) in Python for the scheduler, tests generator, attacker client, and server.
We also write 318 LoC in Go and 203 LoC in JSON file for the other nameservers. 
We use one workstation to run \system, which has an AMD 5950x CPU with 16 cores, 128 GB memory, and runs on Ubuntu 22.04. 


\ignore{
. Additionally, for each authoritative server, a domain will be assigned. 
demonstrates the configuration for the authoritative server in one unit. On the registrar side, \texttt{ns.example.com} will be pointed to the IP address of the authoritative server, i.e. \texttt{172.22.201.0} in Figure~\ref{fig:auth_srv_zone_file}, with an \texttt{A} record. 
Those test domains will be configured into DNS software to test different modes of DNS software during initialization. Zone files of those domains are set and maintained on Cloudflare~\cite{cloudflare}. Due to ethical considerations, a local authority server is deployed on the Docker network to respond to all the DNS queries from the Docker containers to contain all the testing payloads and network traffic within our Docker network. \qz{local authoritative server didn't do much help in efficiency, so I skip it.}
}

\subsection{Experiment Results}
\label{subsec:results}

In total, \system\ generates \textit{718.6K} test cases (each case consists of a pair of query and response) within 65.9 hours. The testing inputs are evenly distributed among the 4 resolver modes. The traces collected from the resolver containers occupy 1,892.9 GB of disk space. Below we first describe the analysis results and the runtime performance of \system. \revise{Then, we compare \system\ with the other DNS fuzzers and conduct an ablation study to assess the impact of several design choices.} Finally, we conduct a large-scale scanning to discover the open resolvers that are impacted by our discovered vulnerabilities and present the results in Appendix~\ref{sec:scanning}.

\vspace{2pt} \noindent \textbf{Results from the oracles.}
The oracles for resource consumption and crash \& corruption are relatively simple (shown in Table~\ref{tab:bug}). Here we focus on the cache-related oracle\footnote{The cache oracle is able to discover both cache poisoning and cache-related resource consumption bugs, like $RC2$ in Section~\ref{sec:cases}.}.
In Table~\ref{tab:oracle_cluster} of Appendix~\ref{app:clustering}, we show the results at different filtering stages. 
Among the 718.6K testing case inputs, our differential analysis filters out \textit{461.2K} (64.2\%) inputs that trigger identical behaviors on the tested resolvers. 
The large ratio of the remaining inputs indicates there is a great variety among the resolver implementations, and an inconsistency usually does not imply vulnerability.
For example, there are 69,967 testing cases with legitimate differences related to \texttt{NSEC3} records (``R2'' in Table~\ref{tab:oracle_cluster}).
\texttt{NSEC3} records\cite{arends2005protocol} are used to indicate a non-exist domain in a secured way, preventing malicious actors from sending fake negative responses to queries. We find that \texttt{NSEC3} records are cached aggressively by Unbound, even when the DNSSEC validation option is turned off. 

Among the inconsistencies, our clustering method generates 22 clusters, and we investigate each cluster to identify the situation(s) about the inconsistencies, which are listed in Table~\ref{tab:oracle_cluster} as well. 
For example, the \textit{180K} data points for the \textit{forward-only} mode are grouped into 7 clusters by bisecting K-means. 
We also identify that only BIND and Unbound implement a fallback mechanism for when the forwarder cannot receive a response with cluster \textit{C13} under column ``R2''. 
\ignore{Fallback mechanism serves as a ``backup'' plan when the forwarding process fails. It can prevent the client gets no answer, thus significantly improving the quality of service of the DNS resolver, which is a normal behavior existing in a few software.}
Overall, the results show that our testing oracles can significantly reduce the manual efforts in locating the vulnerabilities.

\begin{figure}[t]	
    \centering
    \includegraphics[width=\columnwidth]{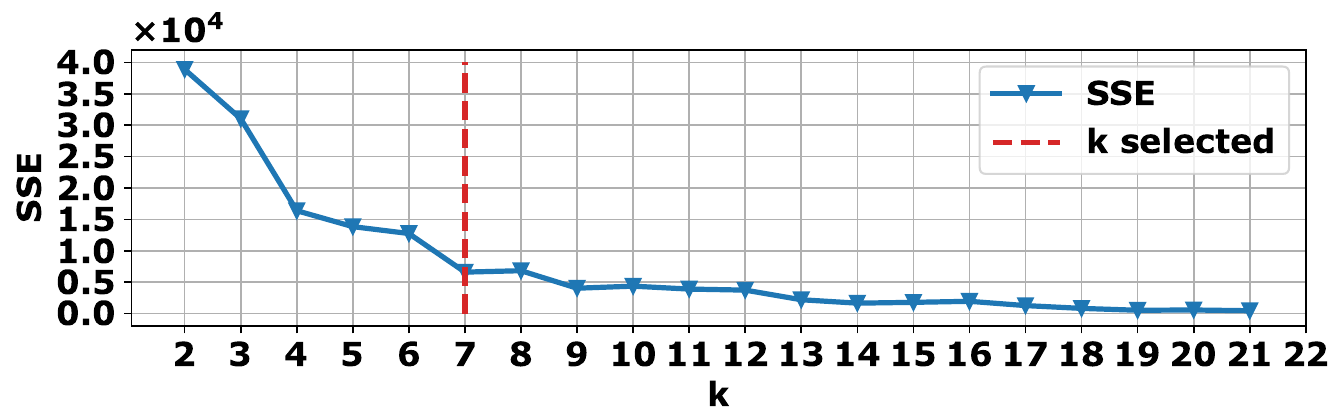}
    \caption{Sum of Squared Error (SSE) to different $k$ for bisecting K-means of \textit{forward-only} cache oracle. 
    }
    \vspace{-2mm}
    \label{fig:k_means}
\end{figure}

Finally, we justify how a key parameter, $k$ for K-means, is selected. For the \textit{forward-only} mode, $k=7$ because the SSE (sum of squared error) drops significantly at 7 and the slope is gradually flattened after that, as shown in Figure~\ref{fig:k_means}. \revise{According to elbow method~\cite{nainggolan2019improved}, $k$ should be set to such value to achieve the best performance in clustering.
}

\begin{figure}[t]
    \centering
    \subfigure[\footnotesize{Client-queries and NS-responses.}]{
		\begin{minipage}[t]{1\columnwidth}
		    \centering
			\includegraphics[width=\linewidth]{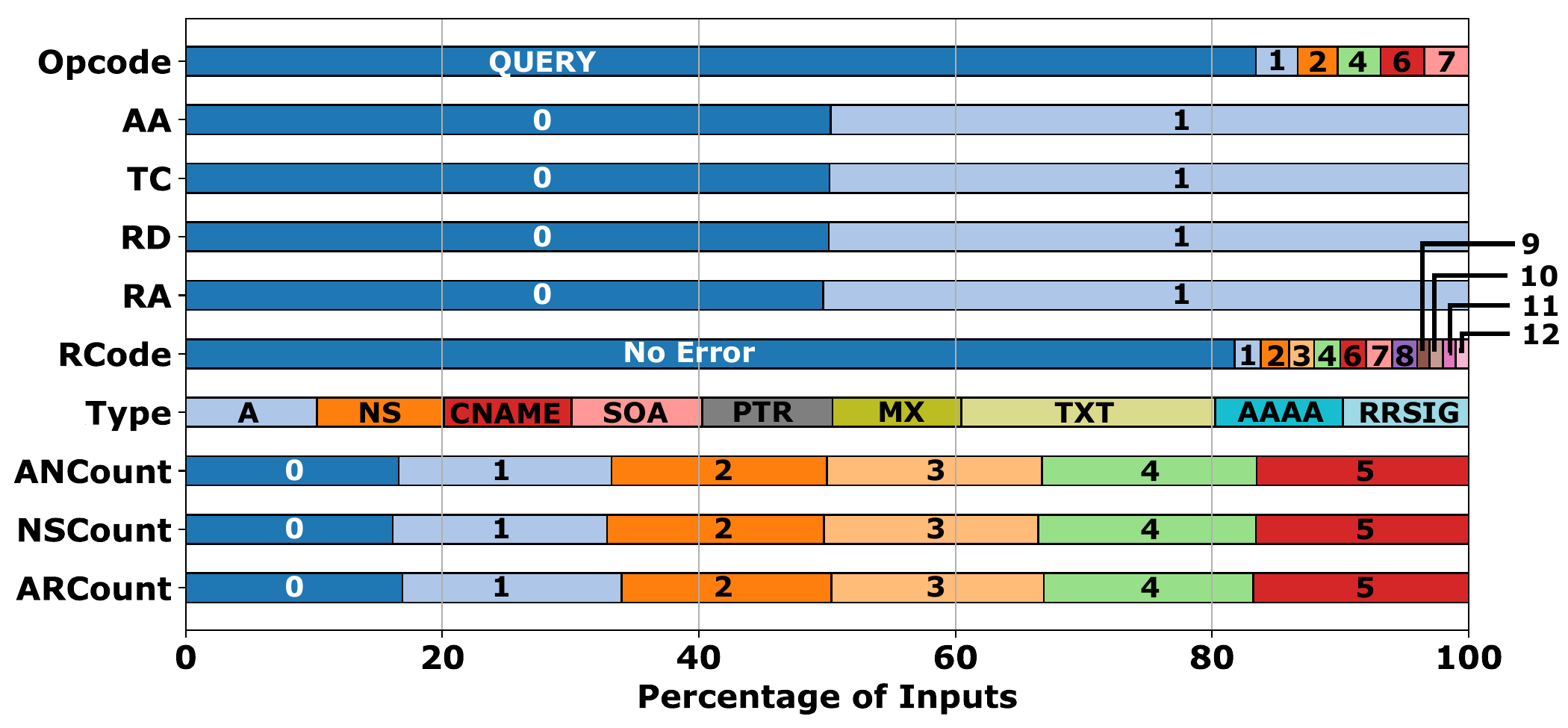}
		\end{minipage}
		\label{fig:auth_response_analysis}
	}
    \\
    \subfigure[\footnotesize{Resolver-responses. ``\textit{RCode \& T.o.}'' refers to ``RCODE and Timeouts''.}]{
		\begin{minipage}[t]{1\columnwidth}
		    \centering
			\includegraphics[width=\linewidth]{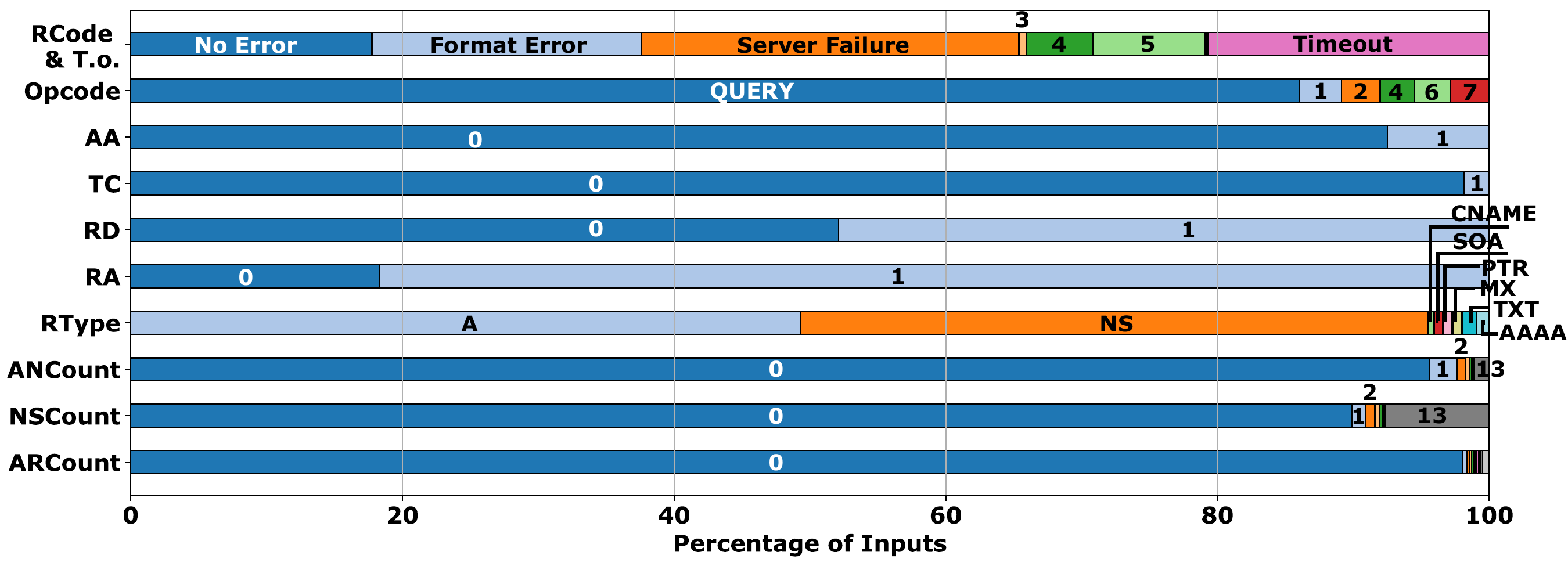}
		\end{minipage}
		\label{fig:resolver_response_analysis}
	}
	\caption{Input coverage analysis on: a) client-queries and ns-responses; b) resolver-responses. The client-query and ns-response have the similar distribution for fields from \texttt{OPCODE} to \texttt{TYPE}. \texttt{AN}/\texttt{NS}/\texttt{ARCOUNT} applies to ns-responses. The values marked on bars are standard DNS values from \cite{mockapetris1987domain1034}.
    }
    \vspace{-2mm}   
	\label{fig:input_analysis}
\end{figure}

\vspace{2pt} \noindent \textbf{Analysis of tests generation.}
We perform statistical analysis on test cases generated by \system\ to understand their main characteristics. We randomly sample 5K test cases for each mode, 20K in total, and parse the DNS messages of queries and responses. 
Figure~\ref{fig:auth_response_analysis} shows the distribution of key fields, including \texttt{TYPE}, \texttt{RCODE}, \texttt{OPCODE}, etc. Results show that our fuzzer achieves good coverage of different field values, and the rule probabilities of PCFG ensure certain code logic is tested more intensively. For example, about 80\% tests have \texttt{OPCODE} set to \texttt{QUERY}, the other DNS modes that are not related to resolution (like \texttt{NOTIFY}) have much fewer test cases.

\ignore{and the rest of \texttt{OPCODE} values are distributed equally, which satisfies our design shown in Listing~\ref{lis:cfg_query} and~\ref{lis:cfg_response}. Besides, other fields such as \texttt{QTYPE}, \texttt{RCODE}, \texttt{ANCOUNT}, etc., all follow the probabilities we set in the input generator. \zl{explain}
}

We also inspect the resolver-responses triggered by the client-queries and Figure~\ref{fig:resolver_response_analysis} shows the distribution. 
Only 17.8\% of the tests have \texttt{RCODE} that equals \texttt{NOERROR}, suggesting our test cases are prone to trigger errors, potentially bugs. 
We have also observed that 20.7\% of the test cases reach timeout without getting a reply.

\ignore{
Additionally, we also analyze the distribution of \textit{ID} and levels of subdomains in Figure~\ref{} and~\ref{}. \zl{we're running out space, you can talk about them in Appendix? or just skip them}
}





\begin{figure}[t]	
    \centering
    \includegraphics[width=\columnwidth]{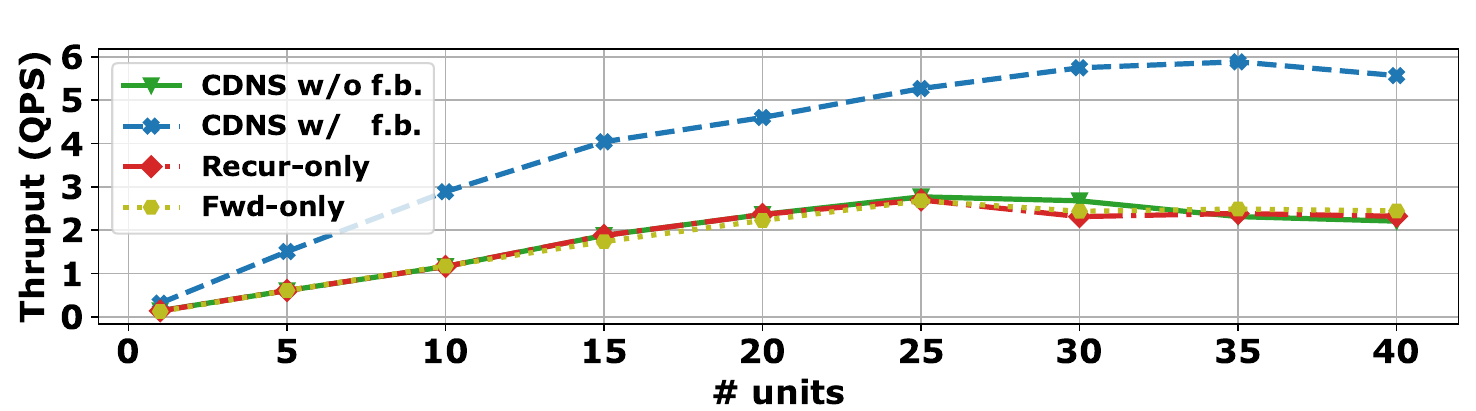}
    \caption{Throughput (``\textit{Thruput}'') of 4 modes with regard to the number of units. \textit{CDNS w/o f.b.}, \textit{CDNS w/ f.b.}, \textit{Recur-only} and \textit{Fwd-only} refers to \textit{CDNS without fallback}, \textit{CDNS with fallback}, \textit{Recursive-only}, and \textit{Forward-only}.
    }
    \vspace{-2mm}
    \label{fig:runtime_performance}
\end{figure}

\vspace{2pt} \noindent \textbf{Runtime performance.} 
After measuring the effectiveness of \system, we measure the efficiency of \system, focusing on its testing throughput. Here we employ \textit{Queries per second (QPS)} as the metric, which shows how many cases from the generator are tested per second. We measure the QPS for different resolver modes and different numbers of units, and Figure~\ref{fig:runtime_performance} shows the results. CDNS with fallback mode is only supported by BIND and Unbound. 
For the other 3 modes, all resolver software is tested. For the default setting (25 units), tests against CDNS with fallback have 2x throughput (5.9 QPS) than the other modes (2.7 - 2.8 QPS). The main reason is that MaraDNS and PowerDNS are slower in responding to client-queries, and \system\ resetting all resolvers synchronously for each test round. \revise{Besides, pre- and post-processing, such as nameserver initialization, slow down the tests and reduce QPS.}
Still, our result is comparable to other network fuzzers: e.g., T-Reqs achieves 7.9 QPS in HTTP fuzzing (``Request body'' of Table 2 in~\cite{jabiyev2021t}).


\ignore{
This performance boost is brought by excluding DNS software with longer response time, such as MaraDNS and PowerDNS, from the testing flow. In conclusion, we found that inefficient implementations of DNS software may be the bottleneck of the testing infrastructure. 
}

\ignore{
 lists the number of testing inputs and throughput for each mode. Apart from \textit{CDNS with fallback} mode, which is only supported by BIND and Unbound, we run all the six DNS software listed in Section~\ref{subsec:setup} on \textit{Recursive-only}, \textit{Forwarder-only} and \textit{CDNS without fallback} modes respectively. 25 units are simultaneously in execution for all four modes for the best performance. 
}

We also measure the impact of the unit number on the throughput, and Figure~\ref{fig:runtime_performance} shows the trend.
The throughput peaks for ``CDNS with fallback'' mode when 35 units are used. For the other modes, the throughput peaks at 25 units. Compared with single-unit execution, a 19-time throughput increase is observed. 
The peak throughput is limited by our workstation setup (only 32 threads can be run concurrently).



\begin{figure}[t]
    \centering
    \subfigure[\footnotesize{\revise{Recursive-only, forward-only and CDNS with/without fallback modes.}}]{
		\begin{minipage}[t]{1\columnwidth}
		    \centering
			\includegraphics[width=\linewidth]{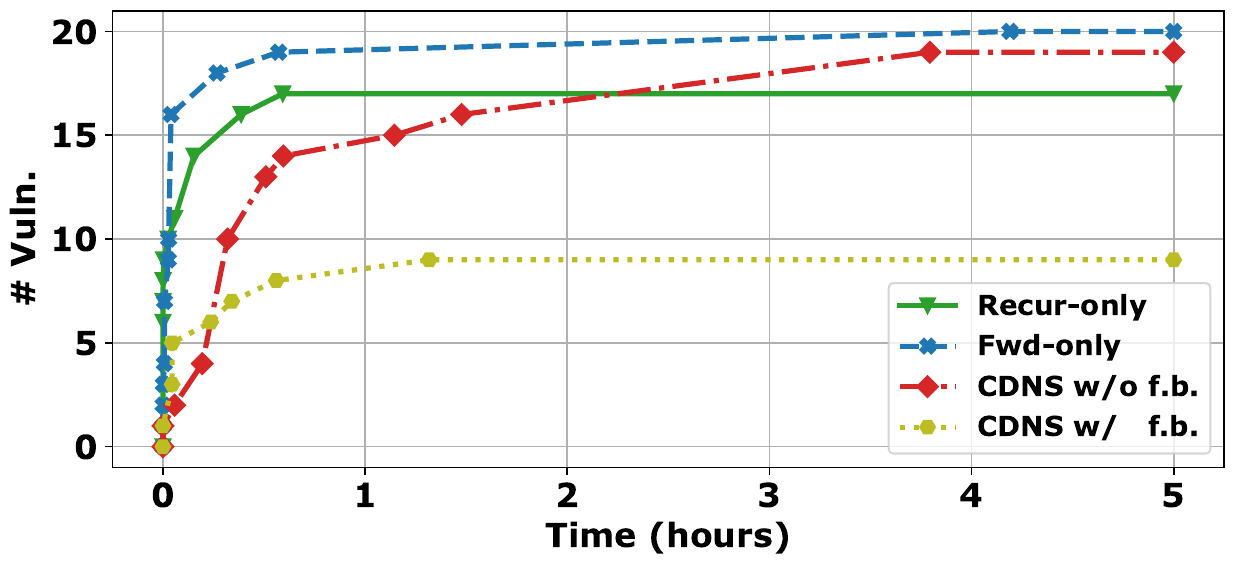}
		\end{minipage}
		\label{fig:perf_fig_1}
	}
    \\
    \subfigure[\footnotesize{\revise{CDNS with fallback under the default setting, long message sequence setting, and equal PCFG probabilities setting.}}]{
		\begin{minipage}[t]{1\columnwidth}
		    \centering
			\includegraphics[width=\linewidth]{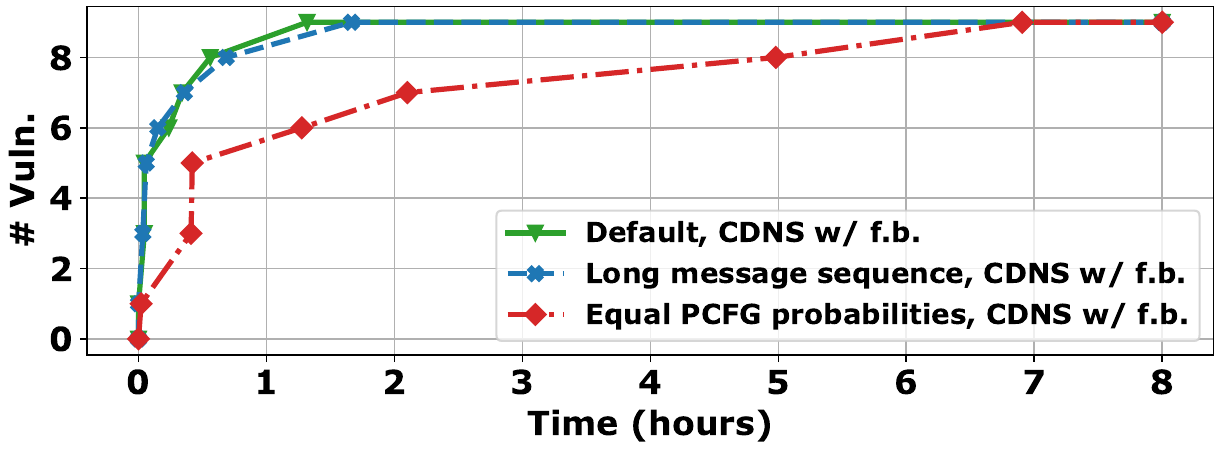}
		\end{minipage}
		\label{fig:perf_fig_2}
	}
	\caption{\revise{The number of vulnerabilities discovered along the time spent by \system.  
 }}
    \vspace{-2mm}
	\label{fig:runtime_perf}
\end{figure}

\revise{
\vspace{2pt} \noindent \textbf{Trend of vulnerability discovery.} 
We discovered 23 vulnerabilities in total and we elaborate them in Section~\ref{sec:cases}. Here we measure the number of vulnerabilities discovered from all the resolvers with regard to time spent. In Figure~\ref{fig:perf_fig_1}, we demonstrated our result in 4 resolver modes. As a vulnerability often relates to a number of test cases, we use the first test case to log the vulnerability discovery time. After 5 hours, all vulnerabilities were discovered, suggesting \system can discover vulnerabilities efficiently.
The number of vulnerabilities discovered differ by modes,  since 1) CDNS with fallback mode is only supported by BIND and Unbound, and 2) some vulnerabilities cannot be triggered in all modes.


}

\subsection{Comparison and Ablation Study}
\label{subsec:ablation}


\revise{
\noindent \textbf{Comparison with other fuzzers.}  
Section~\ref{subsec:tools} compares \system\ with other tools that discover DNS bugs briefly. Here, we provide detailed quantitative comparison with baseline DNS fuzzers including dns-fuzz-server, DNS Fuzzer and SnapFuzz. We run these systems and monitor crash, which is the default bug type considered by them. 
}

\revise{
We first compare with dns-fuzz-server~\cite{dns-fuzz-server}, which combines grammar-based and byte-level mutation, but tests a resolver in a \textit{stateless} way.
It consists of a fuzz-client, which sends client-queries, and a fuzz-server, which serves as an nameserver and sends ns-responses, however there is no coordination between these two components. 
\ignore{
To prove our speculation, we evaluated dns-fuzz-server under the same settings as \system. dns-fuzzer-server 
We set the testing interval on fuzz-client as 0.2 second to ensure no network congestion during testing. The default zone file provided was deployed on fuzz-server. fuzz-client and fuzz-server were deployed in separate containers during evaluation. 
}
We run dns-fuzzer-server to test BIND in the 4 resolver modes and set the interval between test cases to 0.2 seconds. Each mode was tested for 10 hours (180K test case generated for each mode, similar as \system), but no crash was triggered. 
\ignore{
We analyzed the traffic, and found that most of the client-queries are accepted by BIND. However, since the default zone file only includes a limited amount of \texttt{A}, \texttt{SOA}, \texttt{MX} and \texttt{NS} records, most of the resolver-queries will not get matched ns-responses from the nameserver of dns-fuzz-server. 
}
We found most of the ns-responses are simply refused by BIND because they do not have matched resolver-queries and client-queries.

Different from dns-fuzzer-server, DNS Fuzzer~\cite{dns-fuzzer} only implements fuzz-client and only performs byte-level mutation on seed DNS messages. We also test it against BIND and set the interval between test cases to 0.1 seconds, which yield 180K test cases for each resolver mode in 10 hours. Again, no crash was triggered.
The traffic analysis shows that most of the mutated client-queries fail to pass the resolution of BIND. 
The rest of client-queries \ignore{(we set the mutation probability to 50\%)} mostly ended up with timeout because the resolver does not receive ns-responses from a nameserver.
}


\revise{
We tried to run SnapFuzz on BIND but it turns out SnapFuzz does not support BIND\footnote{\revise{SnapFuzz hooks system calls to directly learn when the tested server is able to receive a new request, so there is no need to set a fixed interval between test cases. However, SnapFuzz does not support system calls invoked through \texttt{epoll}~\cite{snapfuzz_epoll}, which is extensively used by BIND.}}.  Meanwhile, we found SnapFuzz is built on top of AFLNet~\cite{pham2020aflnet} and speeds up AFLNet by 7x when fuzzing Dnsmasq~\cite{andronidis2022snapfuzz}. Hence, we take an alternative approach to run the AFLNet baseline inside the SnapFuzz repo for 7 days against BIND (also 4 modes), but no crash was detected. Though AFLNet conducts greybox fuzzing to improve the quality of test cases, it only simulates client-queries (implemented by the SnapFuzz repo), which is prone to generate DNS messages easily rejected by the resolver.


\ignore{The result is not surprising since BIND has provided original support for AFL-based fuzzing tools. Further on, BIND has joined OSS-Fuzz, a continuous fuzzing project to fuzz open source software using fuzzing engines including AFL-based ones. Several vulnerabilities have been identified via OSS-Fuzz~\cite{bind9_oss_fuzz_0, bind9_oss_fuzz_1, bind9_oss_fuzz_2}. Therefore, we think the vulnerabilities, which could be found by AFL-based tools, have already been detected and fixed during continuous fuzzing.
}
}

\revise{
To summarize, no crash has been identified by the baselines, suggesting finding crash bugs from the intensively tested resolvers like BIND is non-trivial. \system is able to trigger 1 crash (CC1 described in Section~\ref{subsec:servbug}).

\ignore{
The major advantage of \system over the baseline systems is the \textit{stateful} modeling of DNS resolvers. 
As an example, the triggering of the bug $RC6$ described in Section~\ref{subsec:resbug}
requires the nameserver to respond to 9 resolver-queries and wait up to 17 seconds. 
For stateless fuzzers like dns-fuzz-server, they cannot set the resolver into the long waiting time with the client-queries or ns-responses, which is the key condition for the bug. 
}

}

\revise{
\vspace{2pt} \noindent \textbf{Length of message sequence.}  
\system generates short message sequences based on the insights of our CVE study described in Section~\ref{subsec:cves}. Here we evaluate whether generating long message sequence can yield new vulnerabilities. We tested BIND under the CDNS with fallback mode and set the new sequence length to 5, such that 5 query-response pairs are sent to BIND in each round. Each message pair is independently generated. In the end, we did not discover any new vulnerability. Interestingly, \system triggers vulnerability faster than the short sequence, as shown in Figure~\ref{fig:perf_fig_2}. This is because all the message pairs in one round share the same cache, so the time spent on querying root and TLD servers can be saved. However, such a performance boost is rather small.
Admittedly, the way we generate the message pairs can be optimized by considering their dependencies, however, such change is non-trivial.

\vspace{2pt} \noindent \textbf{PCFG probabilities.}  
We assign different probabilities to different terminals when generating message templates under PCFG, as described in Section~\ref{subsec:input}. Here we assess the impact of this design choice, by evaluating a simpler setting that assigns the equal probability to each terminal shown in List~\ref{lis:cfg_query} and List~\ref{lis:cfg_response}. Again, no new vulnerabilities are discovered and we also found the pace of vulnerability discovery is significantly slowed down, as shown in Figure~\ref{fig:perf_fig_2}. The main reason is that messages are more likely to be rejected before reaching the deep code logic. For instance, the probability of \texttt{NOERROR} in \texttt{RCODE} is assigned to 0.8 under our default setting. 
However, if the probability is assigned equally, the probability of \texttt{NOERROR} is reduced to 0.091, and most of the responses will be trivially rejected.
}




\section{Discovered Vulnerabilities}
\label{sec:cases}



\begin{table*}[t]
  \centering
  \small
  \caption{
  Identified bugs and test cases of six mainstream DNS software.
  }
  \setlength{\tabcolsep}{5.25pt}
  \begin{threeparttable}
    \begin{tabular}{c|cccc|c|ccccccc|c|c|c}
    \toprule
    \multirow{3}[3]{*}{\textbf{Software}\tnote{*}} & \multicolumn{5}{c|}{\multirow{2}[1]{*}{\textbf{Cache poisoning}}} & \multicolumn{8}{c|}{\multirow{2}[1]{*}{\textbf{Resource consumption}}} & \textbf{Crash\&} & \multirow{3}[3]{*}{\textbf{Total}} \\
    & \multicolumn{5}{c|}{}                 & \multicolumn{8}{c|}{}                                         & \textbf{Corruption} &  \\
    \cmidrule{2-15}          & \multicolumn{1}{c|}{\textbf{CP1}} & \multicolumn{1}{c|}{\textbf{CP2}} & \multicolumn{1}{c|}{\textbf{CP3}} & \textbf{CP4}\tnote{1} & \textbf{Tot.}\tnote{2} & \multicolumn{1}{c|}{\textbf{RC1}} & \multicolumn{1}{c|}{\textbf{RC2}} & \multicolumn{1}{c|}{\textbf{RC3}} & \multicolumn{1}{c|}{\textbf{RC4}} & \multicolumn{1}{c|}{\textbf{RC5}} & \multicolumn{1}{c|}{\textbf{RC6}} & \textbf{RC7} & \textbf{Tot.} & \textbf{CC1} &  \\
    \midrule
    \textbf{BIND} & \cmark\tnote{\dag} & \greenxmark & \redcmark & \cmark & 3     & \greenxmark & \greenxmark & \greenxmark & \greenxmark & \greenxmark & \greenxmark & \greenxmark & 0     & \cmark & 4 \\
    \textbf{Unbound} & \greenxmark & \greenxmark & \redcmark & \cmark\tnote{\dag} & 2     & \greenxmark & \cmark & \cmark & \greenxmark & \cmark & \cmark & \greenxmark & 4     & -     & 6 \\
    \textbf{Knot} & \cmark\tnote{\dag} & \greenxmark & \cmark\tnote{\dag} & \cmark\tnote{\dag} & 3     & \greenxmark & \greenxmark & \greenxmark & \greenxmark & \greenxmark & \greenxmark & \cmark\tnote{\dag} & 1     & -     & 4 \\
    \textbf{PowerDNS} & \greenxmark & \cmark\tnote{\dag} & \greenxmark & \cmark\tnote{\dag} & 2     & \cmark\tnote{\dag} & \greenxmark & \cmark\tnote{\dag} & \greenxmark & \greenxmark & \greenxmark & \greenxmark & 2     & -     & 4 \\
    \textbf{MaraDNS} & \greenxmark & \greenxmark & -     & \cmark\tnote{\dag} & 1     & \greenxmark & \greenxmark & \greenxmark & \cmark\tnote{\dag} & \greenxmark & \greenxmark & \greenxmark & 1     & -     & 2 \\
    \textbf{Technitium} & \cmark\tnote{\dag} & \greenxmark & -     & \cmark\tnote{\dag} & 2     & \greenxmark & \greenxmark & \greenxmark & \cmark\tnote{\dag} & \greenxmark & \greenxmark & \greenxmark & 1     & -     & 3 \\
    \midrule
    \textbf{Total} & 3     & 1     & 3     & 6     & 13    & 1     & 2     & 1     & 2     & 1     & 1     & 1     & 9     & 1     & 23 \\
    \bottomrule
    \end{tabular}%
    \begin{tablenotes}
      \footnotesize
      \item \tnote{*}: Recursive or forwarding modes. \tnote{1}: They are triggered by different responses and their cache are inconsistent. \tnote{2}: Total. \cmark or \redcmark\color{black}: Vulnerable. 
      \item \redcmark\color{black}: \revise{Discussed but no immediate action.} \cmark: Confirmed and/or fixed by vendors. 
      \greenxmark\color{black}: Not vulnerable. \tnote{\dag}: CVEs assigned. `-': Not applicable.
      \item \# Amount of test cases: $CP1$ (19), $CP2$ (1,422), $CP3$ (111,328), $CP4$ (7,856), $RC1$ (539,745), $RC2$ (112,126), $RC3$ (88,935), $RC4$ (132), $RC5$ (272)
      \item \hspace{2.5cm} $RC6$ (6,264), $RC7$ (4,448), and $CC1$ (5).
    \end{tablenotes}
  \end{threeparttable}
      \vspace{-2mm}
  \label{tab:bug}%
\end{table*}%

We identify \textit{23} vulnerabilities with the help of \system: 
cache poisoning (13), resource consumption (9), and crash \& corruption (1).
\revise{After in-depth discussions with related vendors, 19 of these bugs have been confirmed or fixed, and \revise{15} CVEs have been assigned.} 
As shown in Table~\ref{tab:bug}, we categorize these bugs into 12 classes including $CP1$-$CP4$, $RC1$-$RC7$, and $CC1$, and list the number of vulnerable test cases in the note. Below, we first describe the concrete threat model and then elaborate on each bug group.

\subsection{Cache Poisoning Bugs}
\label{subsec:cachebug}


\revise{\noindent \textbf{Concrete threat model.} }
Different from the previous work that either considered recursive mode only~\cite{man2020dns} or forward mode only~\cite{zheng2020poison}, we consider the 4 resolver modes as described in Section~\ref{subsec:dns} and configure the cache accordingly. 
Specifically, the target resolver has two DNS zones.
Client-queries matching the forwarding zone $Z_{F}$ are directly forwarded to an upstream server (resolver-queries $Q_{F}$), and client-queries matching the recursive zone ($Z_{R}$) will trigger recursive resolution (resolver-queries $Q_{R}$).
Both forwarding and recursive mode \textit{share a global cache}.

\begin{figure}[t]	
    \centering
    \includegraphics[width=\columnwidth]{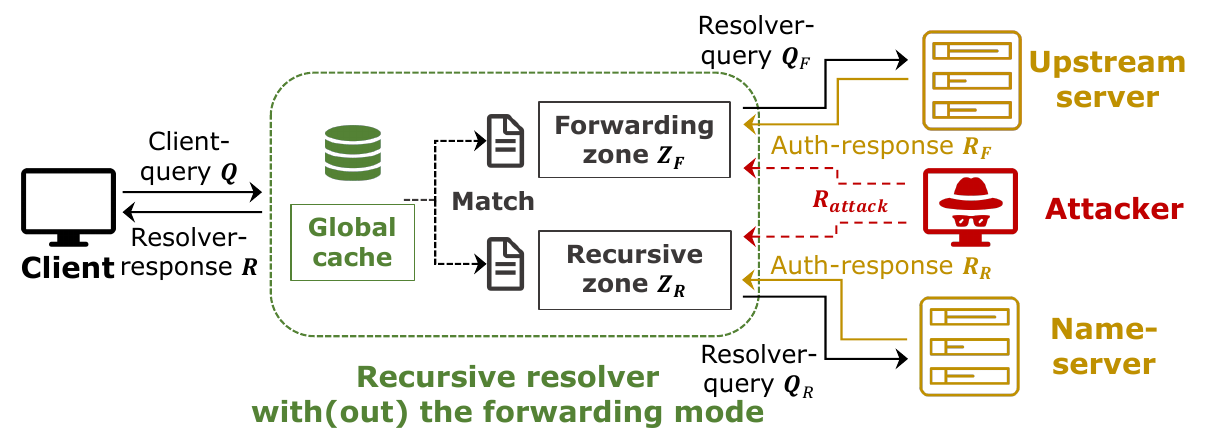}
    \caption{Threat model of cache poisoning bugs.
    }
    \vspace{-2mm}
    \label{fig:threat_model_cp}
\end{figure}

When launching an attack, the attacker sends a client-query ($Q$) to the target resolver, then provides malicious ns-responses ($R_{attack}$) to the resolver prior to the arrival of legal responses ($R_{F}$ or $R_{R}$). After accepting the malicious responses, the global cache and client would save the tampered answers. When the attacker is not on the resolution path between the client and the intended nameserver, the attacker can conduct IP spoofing and port guessing for the off-path cache poisoning~\cite{man2020dns, man2021dns, son2010hitchhiker}.
\revise{The concrete threat model is shown in Figure~\ref{fig:threat_model_cp}.}

\ignore{
Same to other cache poisoning attacks~\cite{kaminsky2008black, herzberg2013fragmentation, man2020dns, zheng2020poison, man2021dns}, we assume the attacker could send queries to the target resolver. As shown in Appendix~\ref{sec:scanning}, our found vulnerable targets are open resolvers that expose open access. 
In addition, under the off-path condition, attackers need to spoof the IP address of spoofed responses. During on-path attacks, the upstream server and authoritative server might return a malicious response to the target resolver.
}

\begin{figure}[t]
    \centering
    \subfigure[\footnotesize{Auth-response for $CP1$.}]{
		\begin{minipage}[t]{0.48\columnwidth}
		\centering
		\includegraphics[width=\columnwidth]{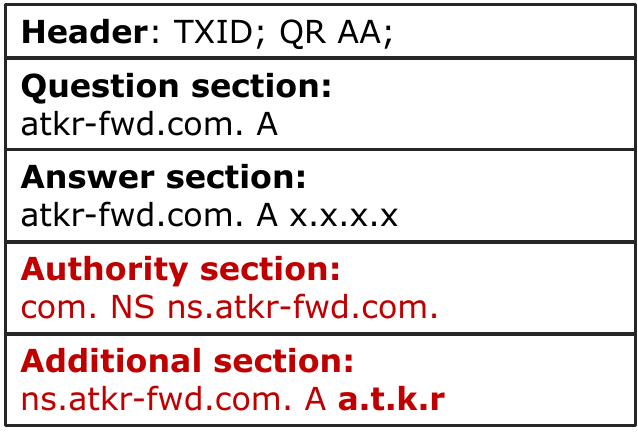} 
		\end{minipage}
		\label{fig:response_cp1}
	}
    \hspace{-4.25mm}
    \subfigure[\footnotesize{Auth-response for $CP2$.}]{
    	\begin{minipage}[t]{0.48\columnwidth}
    	\centering
   	 	\includegraphics[width=\columnwidth]{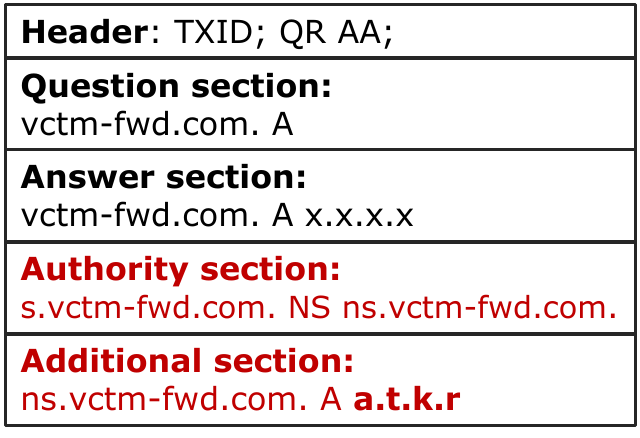}
    	\end{minipage}
		\label{fig:response_cp2}
    }
    \\
    \vspace{-1.25mm}
    \subfigure[\footnotesize{1st fragment for $CP3$.}]{
		\begin{minipage}[t]{0.48\columnwidth}
		\centering
		\includegraphics[width=\columnwidth]{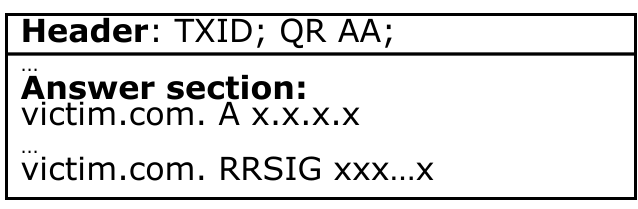} 
		\end{minipage}
		\label{fig:response_cp3_1}
	}
    \hspace{-4.25mm}
    \subfigure[\footnotesize{spoofed 2rd fragment for $CP3$.}]{
    	\begin{minipage}[t]{0.48\columnwidth}
    	\centering
   		\includegraphics[width=\columnwidth]{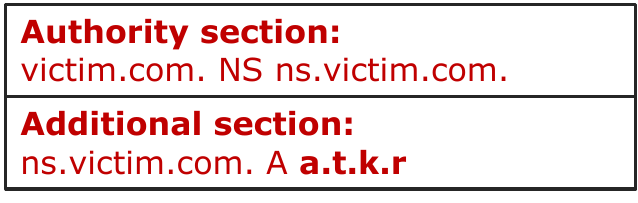}
    	\end{minipage}
		\label{fig:response_cp3_2}
    }
    \\
    \vspace{-1.25mm}
    \subfigure[\footnotesize{Auth-response for $CP4$.}]{
		\begin{minipage}[t]{0.48\columnwidth}
		\centering
		\includegraphics[width=\columnwidth]{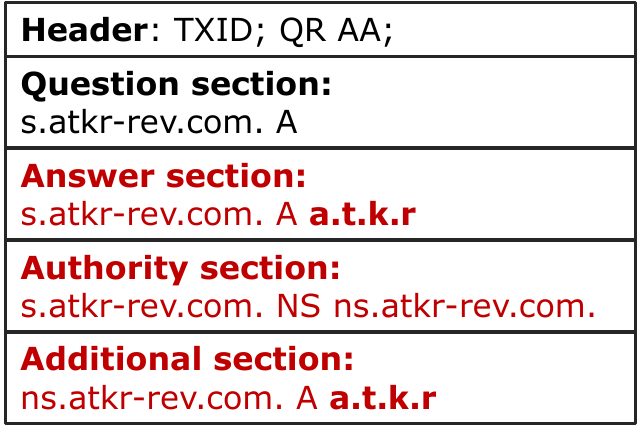} 
		\end{minipage}
		\label{fig:response_cp4_1}
	}
    \hspace{-4.25mm}
    \subfigure[\footnotesize{Ref-response for $CP4$.}]{
    	\begin{minipage}[t]{0.48\columnwidth}
    	\centering
   	\includegraphics[width=\columnwidth]{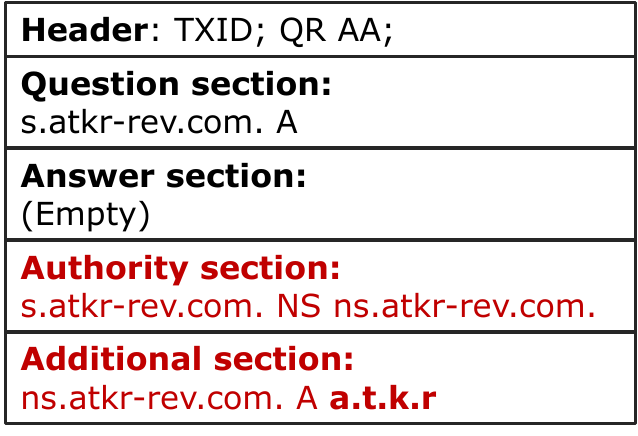}
    	\end{minipage}
		\label{fig:response_cp4_2}
    }
    \caption{DNS responses utilized for cache poisoning attacks. Red parts carry the attack payloads.
    }
    \label{fig:response}
\end{figure}

\vspace{2pt} \noindent
\textbf{CP1: Out-of-bailiwick cache poisoning.}
The bailiwick rule requires that authoritative servers should not return data outside of their controlled zones~\cite{elz1997clarifications}.
For example, responses from \texttt{.com} should not include data of other zones like \texttt{.net}.
Correspondingly, when receiving out-of-bailiwick data, resolvers should discard it before caching. However, when testing CDNSes with fallback, 
we found BIND accepts out-of-bailiwick data. Knot and Technitium are also identified to have this issue in further tests.

For a resolver-query $Q_F$ sent to our authoritative server, such as \texttt{atkr-fwd.com}, these three resolvers will \textit{cache every record} in the auth-responses generated by our mutator, even including the out-of-bailiwick records shown in Figure~\ref{fig:response_cp1}.
In this example, since the forged \texttt{NS} records of \texttt{.com} with a \texttt{AA} flag have a higher ranking~\cite{elz1997clarifications}, resolvers would opt to overwrite existing cached records and utilize them for future resolution.
Then, following the records of ``Authority'' and ``Additional'' Sections, the resolvers would request the attacker's nameserver \texttt{ns.atkr-fwd.com} for all queries under the \texttt{.com} zone, which allows the attacker to \textit{hijack the entire TLD zone}.
\ignore{Under the on-path conditions, attackers from the upstream server can directly return an out-of-bailiwick response to the vulnerable resolver for poisoning arbitrary zones.
While for off-path attacks, attackers still need to guess the UDP source port and DNS transaction ID (\texttt{TXID}) using existing methods~\cite{man2020dns, man2021dns, son2010hitchhiker}, which we will consider in our future work.
}
After discussion, all affected vendors confirmed this vulnerability and patched their software. 
We received 3 CVEs and the detailed study is presented in ~\cite{li2023maginot}.


\vspace{2pt} \noindent
\textbf{CP2: In-bailiwick cache poisoning.}
When the client-query matches a domain name in the forwarding zone, e.g., \texttt{vctm-fwd.com}, all software except PowerDNS simply forwards the query to the upstream server and waits for responses.
However, PowerDNS first searches its cache for nameservers. If there is a cache hit, it follows the nameserver records and finishes the resolution. Otherwise, it sends resolver-queries to the upstream server.
After receiving a response with additional nameserver information like Figure~\ref{fig:response_cp2}, PowerDNS will cache every record. Hence, this difference makes cache poisoning more powerful for PowerDNS: when the attacker conducts off-path cache poisoning, she just needs to tamper \textit{one} \texttt{NS} record (e.g., \texttt{s.vctm-fwd.com}) during forwarding, and all the follow-up queries under the zone of \texttt{s.vctm-fwd.com} will be tampered (e.g., redirected to the attacker server \texttt{a.k.t.r}). For the other software, the attacker has to tamper \textit{every} query.

\ignore{
Then any further queries under the \texttt{s.atkr-fwd.com} zone will be sent to the attacker's nameserver \texttt{ns.atkr-fwd.com}, though the \texttt{s.atkr-fwd.com} zone could belong to another party.

To exploit this bug, attackers also need to conduct an off-path attack similar to $CP1$. Nonetheless, for a forwarding zone, attackers only need to perform the attack once in order to hijack all queries to their controlled servers. For other software, attackers must launch a new attack each time they attempt to poison a forwarding zone, since they are always forwarded to upstream servers.
}

\vspace{2pt} \noindent
\textbf{CP3: Fragmentation-based cache poisoning.}
According to~\cite{brandt2018domain, zheng2020poison}, attackers could leverage IP fragmentation to initiate a DNS cache poisoning attack.
This attack exploits the fact that the second fragment of a fragmented DNS response packet contains neither UDP nor DNS headers, thus it is much easier to spoof this fragment as there is no need to guess the UDP source port or DNS \texttt{TXID}.
During an attack, attackers first send the spoofed second fragment (e.g., Figure~\ref{fig:response_cp3_2}) to the target resolver, then issue a query for the victim domain whose nameserver will return fragmented DNS packets. After receiving the first fragment shown in Figure~\ref{fig:response_cp3_1}, the target resolver will resemble it with the previously cached second fragment, resulting in a rogue DNS response to be accepted.


\ignore{
To force fragmentation, there are two approaches.
One strategy is to decrease the Path Maximum Transmission Unit (PMTU) between the resolver and nameserver~\cite{brandt2018domain}.
However, recent work~\cite{zheng2020poison} shows that only 0.7\% of the nameservers for Alexa Top 100K domains are willing to drop their PMTU to less than 528B.
This approach is less practical.
The second method is to increase the packet size of responses so that they exceed the general MTU limit (e.g., 1,500 bytes for Ethernet)~\cite{herzberg2013fragmentation}, which~\cite{zheng2020poison} uses to conduct defragmentation based cache poisoning attacks.
}

 \revise{Though \system does not directly generate fragmented packets, this bug was discovered because \system can generate large-size DNS messages (e.g.,  exceeding the general 1,500-byte MTU limit for Ethernet)~\cite{herzberg2013fragmentation}.
}
Through traffic analysis, \system found that BIND, Unbound, and Knot allow fragmented ns-responses to be larger than 1,232 bytes and even 4,096 bytes, whereas the other software only accepts ns-responses less than 1,232 bytes. Attackers could exploit nameservers that return large DNS responses or utilize the techniques in~\cite{zheng2020poison} to conduct fragmentation-based cache poisoning attacks.

\vspace{2pt} \noindent
\textbf{CP4: Iterative subdomain caching.}
During fuzzing, we discover that software including BIND, Unbound, Knot, and PowerDNS will accept unsolicited records from auth-responses (e.g., records in the Authority and Additional section Figure~\ref{fig:response_cp4_1}), while MaraDNS and Technitium will store the records in the ref-responses like Figure~\ref{fig:response_cp4_2}. 
After caching these records, resolvers will use them to serve future queries. For example, upon receiving a query for \texttt{s.atkr-rec.com}, resolvers will send queries straight to the nameserver \texttt{ns.atkr-rec.com} rather than iteratively querying the root and TLD servers.

Inspired by this behavior, we introduce a new attack in which the attacker could iteratively inject \texttt{NS} records of subdomains into the resolver's cache. Especially, when \texttt{NS} records of \texttt{s.atkr-rec.com} are about to expire, attackers return nameserver data of \texttt{s.s.atkr-rec.com}, enabling the target resolver to still be able to resolve domains under it, and for \texttt{s.s.s.atkr-rec.com} so on. In this manner, even if \texttt{atkr-rec.com} is revoked from the \texttt{.com} zone, the target resolver will continue to resolve a group of subdomains of \texttt{atkr-rec.com} for a very long time. This attack is an extension of the previous ``ghost domain attack''~\cite{jiang2012ghost} that defeats domain sinkholing~\cite{alowaisheq2019cracking}. 
All vendors of tested software have confirmed this vulnerability and some have fixed it. We obtained 5 CVE numbers and the detailed study was presented in ~\cite{li2023ghost}.

\ignore{
Previous work~\cite{jiang2012ghost} uncovered a new type of ``cache poisoning'' attack, named ghost domain, which could make a revoked domain resolvable for a long time by targeting recursive resolvers.
This attack exploits a flawed DNS cache update policy that overwrites cached records with a new \texttt{TTL} value.
By doing so, the cached \texttt{NS} of the attacker's domain never expires.
As a result, the target resolver will continue to use cached nameservers to resolve this domain even after it has been removed from the TLD zone.
}


\subsection{Resource Consumption Bugs}
\label{subsec:resbug}

\begin{figure}[t]	
    \centering
    \includegraphics[width=\columnwidth]{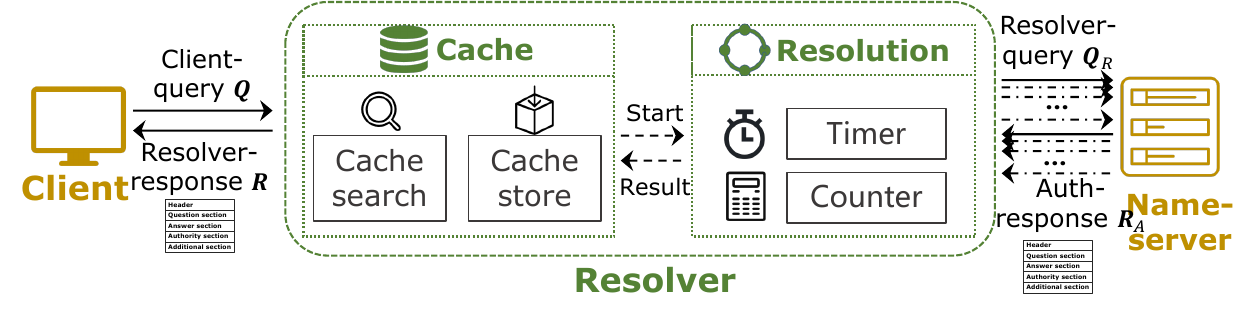}
    \caption{Threat model of resource consumption bugs.
    }
     \vspace{-2mm}
    \label{fig:threat_model_rc}
\end{figure}

\revise{\noindent \textbf{Concrete threat model.} }
Through a small number of client-queries and/or auth-responses, the attacker occupies a large portion of cache storage, triggers excessive resolver-queries, or consumes large computation overhead on the victim resolver.
\revise{Figure~\ref{fig:threat_model_rc} shows the threat model.}

\vspace{2pt} \noindent
\textbf{RC1: Excessive cache search operations.}
When running in forward-only mode, only PowerDNS looks up its local cache for trust anchors and \texttt{NS} records before sending it to a server, while the others just forward the client query to upstream servers.
For example, for a query of \texttt{s.atkr-fwd.com} under the forwarding zone \texttt{atkr-fwd.com}, PowerDNS searches its cache following the order of \texttt{s.atkr-fwd.com}, \texttt{atkr-fwd.com}, \texttt{.com}, and the ROOT server until finding an existing \texttt{NS} record, by removing one label each round. This process is repeated three times.
Therefore, attackers could construct a domain name containing 128 labels (the maximum number) to trick PowerDNS to perform 384 cache search operations, which is much more than other software (i.e., 1).

\vspace{2pt} \noindent
\textbf{RC2: Unlimited cache store operations.}
When receiving auth-responses, Unbound caches all in-bailiwick records from the ``Authority'' and ``Additional'' sections into the cache regardless of their validity.
According to DNS RFC~\cite{mockapetris1987domain1034, arends2005rfc}, only \texttt{NS}, \texttt{SOA}, and DNSSEC-related records are permitted in the ``Authority'' section, while glue records are allowed in the Additional section.
Unbound does not implement this rule and stores all types of records, allowing attackers to squander cache storage.
After discussion, Unbound's developers have included a sanitizer to filter out invalid records.

\vspace{2pt} \noindent
\textbf{RC3: Ignoring the RD flag.}
\revise{
DNS RFCs~\cite{mockapetris1987domain1034} require resolvers to issue queries to upstream servers only if they receive client queries with the Recursive-Desired flag set (\texttt{RD=1}), and refrain from follow-up queries when \texttt{RD=0}. However, Unbound and PowerDNS still forward client queries to upstream servers when \texttt{RD=0}, under the forwarding mode. The attacker can exploit this bug by directing the follow-up queries to her domain, and point the nameserver in her DNS zone back to the resolver, which results in a query loop and potential amplification attack.
The detailed study is presented in~\cite{xu2023tsuking}.
}


\vspace{2pt} \noindent
\textbf{RC4: Following a self-CNAME reference.}
The \texttt{CNAME} record is used to map an alias name to a canonical name (e.g., \texttt{www.cnn.com} to \texttt{cnn-tls.map.fastly.net})~\cite{mockapetris1987domain1034}.
Upon detecting a CNAME loop (i.e., responses with a CNAME record having the same alias name and canonical name), the current resolution process should terminate.
However, Unbound, MaraDNS, and technitium will chase the CNAME loop for 12, 113, and 289 times.
Moreover, technitium will provide the client with a response including 289 CNAME records.
Attackers could craft a self-CNAME record to consume resolvers' resources by triggering more outgoing queries and sending large responses for amplification attacks.
MaraDNS and Technitium have confirmed this vulnerability and patched their latest versions. 2 CVE numbers have been assigned.

\vspace{2pt} \noindent
\textbf{RC5: Large responses to clients.}
Large DNS responses have been used extensively in DNS reflection amplification attacks~\cite{moon2021accurately}.
When analyzing the packet size, our traffic oracle discovered that Unbound enables a maximum packet size of 4,096 bytes, whereas the other software restricts it to less than 1,232 bytes.
This suggests Unbound could provide a three-fold amplification ratio compared to other software.
After disclosing our findings, Unbound has changed its maximum UDP packet size to 1,232 bytes by default.

\vspace{2pt} \noindent
\textbf{RC6: Overlong waiting time over UDP.}
After receiving an ns-response \ignore{(\zl{can you show the example?}\xl{done})} modified by our byte-level mutator,
\ignore{, such as a malformed DNS payload that has an incorrect DNS format\zl{i mean the concrete payload, like Figure~\ref{fig:response}},}
Unbound continues to issue 9 resolver-queries and waits up to 17 seconds for a legal ns-response.
In contrast, other software just discards the invalid ns-responses and does not follow up with resolver-queries.
Hence, an attacker can send a large number of such malformatted ns-responses to Unbound, extend the resolver's waiting time and consume its resources.


\vspace{2pt} \noindent
\textbf{RC7: Excessive queries for resolution over TCP.}
Similar to $RC6$, upon receiving an invalid ns-response via TCP, Knot Resolver continues to issue 100 resolver-queries for a legitimate response, thus bypassing the restriction (5 resolver-queries at maximum) used to defend against the powerful NXNSDomain attacks~\cite{knot-nxns}.
After discussion, Knot acknowledged and fixed this vulnerability.
The detailed study is shown in~\cite{li2024tudoor}.

\subsection{Crash Bugs}
\label{subsec:servbug}

\ignore{
As with resource consumption-related bugs, in this part, we assume the attacker could send queries and responses directly to the target resolver.
The aim of this attack is to cause resolvers' normal service by exploiting logic flaws such as assertion failures.
With the query and response mutator, \system could provide massive legal and malformed DNS packets to the target resolver for detecting the service crash.}
\revise{\noindent \textbf{Concrete threat model.} }
The attacker sends client-queries or auth-responses to cause memory or non-memory crash. 

\vspace{2pt} \noindent
\textbf{CC1: Assertion failure when receiving queries.}
We found an assertion failure occurs when BIND receives the byte-mutated client-queries 
using the \texttt{udp\_recv} function, which crashes the resolver service.
After reading the source code, we identified this bug in the \texttt{udp\_recv} dispatching process, which returns a success code but cancels the query in the meantime, violating the assertion that ensures the current \texttt{udp\_recv} process receives a valid DNS packet. 
We found BIND fixed this issue in version 9.18.3~\cite{bind-crash-bug}, but their test case is different from ours. 
\section{Discussion}
\label{sec:discussion}


\noindent \textbf{Limitations and future work.} 
1) Our test cases are unable to cover all sorts of DNS messages (e.g., no \texttt{DNAME}), as it incurs extensive manual efforts in writing PCFG, and only a subset of DNS message types are actively used. 2) We did not test all DNS-related functionalities, like DNSSEC, due to their different logic from the normal resolver actions, i.e., caching records. 3) We test the stateful resolver with a pair of query and response. Admittedly, some complex bugs that are introduced by long sequences cannot be found\ignore{, but our study in Section~\ref{subsec:cves} suggests their ratio is small}. 4) We set a fixed timeout value to 5 seconds, but this is not optimal when the resolver is stuck before the timeout. SnapFuzz enables adaptive timeout on networking software by rewriting its code~\cite{andronidis2022snapfuzz} and our problem can be solved by this approach. 5) We conduct a blackbox fuzzing without relying on the code coverage as feedback. The main reason is that we have not found an ideal metric for the different types of semantic bugs. We plan to continue to explore the combination of coverage-based and grammar-based fuzzing, like~\cite{wang2019superion}. \revise{6) We follow other works (e.g., ~\cite{zou2021tcp}) to analyze CVEs and assess which type of bugs is more prevalent (e.g., short sequence triggers more bugs). Admittedly, such analysis could suffer from survivorship bias (e.g., bugs triggered by short sequence are easier to find, hence more CVEs). 
7) We use differential testing to discover cache poisoning bugs, assuming at least one implementation is correct. However, when the RFC is erroneous and all implementations follow the RFC, such bugs are unlikely to be discovered.
}

\vspace{2pt} \noindent \textbf{Ethical considerations.}
1) Fuzzing the resolver within the standard DNS infrastructure could affect the other remote nameservers, as described in Section~\ref{subsec:env}. Hence, we localize the root and TLD servers in our lab network, which improves the efficiency of \system.
\revise{
2) For the measurement study described in Appendix~\ref{sec:scanning}, we scanned the whole IPv4 network space. We follow the common practice in Internet-wide scanning, by setting the maximum probing rate to 10 kps and evenly distributing the traffic across the target. 
Like XMap which also scans IPv4 network space~\cite{li2021fast}, we created a website to receive the opt-out requests and a \texttt{PTR} record using our scanner's source IP to show our research intention. We received no opt-out requests during the period of experiments. We admit that such an approach cannot entirely mitigate the ethical issues, as getting informed consent for large-scale Internet measurement is very difficult~\cite{partridge2016ethical}. We tried our best to follow the best practices.
}


\section{Related Work}
\label{sec:related}

\noindent \textbf{Network protocol fuzzing.} Various fuzzing techniques have been applied to test network protocols in addition to DNS. 
AFLNet uses the response code from the server as the feedback to generate sequences of messages~\cite{pham2020aflnet}. SGFuzz uncovers the state space of a protocol (e.g., HTTP2) from the ``enum'' variables defined in source code~\cite{ba2022stateful}. For TCP, TCP-Fuzz improves the coverage of TCP states with a new branch transition metric~\cite{zou2021tcp}. 
For TLS, grammar-based fuzzing has been applied to generate test cases from specifications~\cite{somorovsky2016systematic, walz2017exploiting}.
Fiterau et al. extended~\cite{somorovsky2016systematic} to test DTLS~\cite{fiterau2020analysis}. For HTTP, T-Reqs detects HTTP smuggling vulnerabilities by finding the inconsistencies in how servers split an HTTP request~\cite{jabiyev2021t}. Frameshifter identifies HTTP/2-to-HTTP/1 protocol conversion anomalies by mutating the HTTP/2 frame sequence~\cite{jabiyev2022frameshifter}. For QUIC, DPIFuzz identifies new elusion methodologies for an attacker to evade QUIC-based Deep Packet Inspection (DPI)~\cite{reen2020dpifuzz}.

We found none of the prior works can be directly applied to DNS resolvers, due to the different protocol semantics and vulnerability types (e.g., cache poisoning). \system addresses these challenges with a new fuzzing framework and a set of techniques.

\vspace{2pt} \noindent \textbf{Differential testing.}
\system leverages the inconsistencies between DNS resolvers to identify semantic bugs, in particular cache bugs. Differential testing also exploits this insight to identify semantic bugs, which has been applied in various scenarios, including JIT compilers~\cite{bernhard2022jit}, transient execution~\cite{hur2022specdoctor}, browser rendering~\cite{song2022r2z2}, internal function models in symbolic execution~\cite{li2022sediff}, CPU bugs~\cite{hur2021difuzzrtl}, software time/space side-channels~\cite{nilizadeh2019diffuzz}, deep-learning stacks~\cite{guo2018dlfuzz},  hostname verification in SSL/TLS~\cite{sivakorn2017hvlearn}, packet parsing by front-/back-end servers~\cite{shen2022hdiff}, DPI~\cite{wang2020symtcp, wang2021themis}, JVM implementations~\cite{chen2019deep}, malware analysis~\cite{jana2012abusing}, file systems~\cite{min2015cross},  Ad blockers~\cite{zhu2018measuring}, etc. 
\revise{
Among these works, ~\cite{shen2022hdiff, wang2020symtcp, wang2021themis} are about network services, and we discuss them in details. 
HDiff generates HTTP messages to detect HTTP Request Smuggling attack, Host of Troubles attack and Cache-Poisoned Denial-of-Service Attack~\cite{shen2022hdiff}. It extracts syntax rules from RFC and use them to detect implementation discrepancies, however, there lacks precise specifications of resolver behaviors. SYMTCP automatically discovered insertion and evasion TCP packets against DPI middleboxes ~\cite{wang2020symtcp}. It uses symbolic execution to generate TCP state machines of endhosts and observe their discrepancies given TCP packets. Themis tackles a similar problem but detects the discrepancies by comparing the TCP state machines statically and finding the counterexample through a SAT solver~\cite{wang2021themis}. However, DNS resolvers do not have well-defined state machines.
}

In addition to the efforts of adjusting differential testing to specific scenarios, some works have investigated general strategies. Nezha proposed $\lambda$-diversity notation for fuzzing to increase the chances of finding inconsistencies~\cite{petsios2017nezha}.  HyDiff combines symbolic execution with greybox fuzzing to find semantic bugs~\cite{noller2020hydiff}. In our setting, since we conduct blackbox fuzzing, the metrics and methods proposed by those works do not directly apply.

\vspace{2pt} \noindent \textbf{DNS resolver vulnerabilities.}
Section~\ref{subsec:cves} reviews DNS vulnerabilities from the published CVEs. Here, we survey the related academic works. The major interests were centered around cache poisoning bugs, and many found that forwarders are more vulnerable~\cite{schuba1993addressing, stewart2003dns, son2010hitchhiker, herzberg2013vulnerable, schomp2014assessing, zheng2020poison}. Jeitner et al. identified semantic inconsistencies in DNS input validation and proposed new string injection attacks for cache poisoning~\cite{jeitner2021injection}. Recently, Jeitner et al. 
found special characters can be exploited for DNS cache poisoning attacks against routers~\cite{jeitner2022xdri}. So far, finding resolver bugs requires heavy manual analysis, and \system sheds light on how to automate this process with fuzzing.

When the attacker is not on the resolution path, a malicious response needs to be forged and raced against the legitimate response. In this case, the defense mechanisms based on randomization, like port randomization and TXIDs, have to be bypassed. Port brute-forcing~\cite{kaminsky2008black}, birthday attacks~\cite{stewart2003dns, son2010hitchhiker}, IP fragmentation~\cite{herzberg2013fragmentation, zheng2020poison}, ICMP-based side channels~\cite{man2020dns, man2021dns} and exploiting weak pseudo random number
generator (PRNG) in the Linux kernel~\cite{klein2021cross} have been proposed to achieve such goal. Dai et al. showed that following off-path cache poisoning, Internet resources like IP addresses, domains, certificates, and virtual platforms can be controlled by attackers~\cite{dai2021hijackers}.
The aforementioned vulnerabilities exploit side-channel information of DNS resolution. Though they cannot be directly discovered by \system, combined with the vulnerabilities discovered by \system, more powerful off-path cache poisoning attacks can be enabled (see CP2 of Section~\ref{subsec:cachebug}.

\section{Conclusion}

In this work, we develop a new blackbox fuzzing system \system\ that is tailored to find DNS resolver vulnerabilities. Based on our study of the published DNS CVEs, \system\ is designed with a set of novel techniques, including constrained stateful fuzzing, differential testing, and grammar-based fuzzing. Our evaluation results show that \system is effective in finding resolver bugs, \revise{with 23 vulnerabilities discovered and 15 CVEs assigned.} 

\vspace{2pt} \noindent \textbf{Lessons learnt.}
Despite that DNS resolvers were extensively tested (e.g., BIND has joined Google OSS-Fuzz project to be automatically fuzzed~\cite{bind-oss}), we can still discover many vulnerabilities in their latest versions. We believe the main reason is that bugs unique to DNS resolvers are still challenging to be discovered with the existing tools, and we hope this study can shed light on this understudied area. Besides, lacking rigorous specifications also contributes to the existence of resolver bugs ~\cite{son2010hitchhiker, moon2021accurately}, as reflected by the high number of inconsistencies observed during testing. Like prior work, we encourage the Internet community to work together and develop formal guidance about secured resolver implementations.

\section*{Acknowledgement}
\label{sec:acknowledgement}

We thank all the anonymous reviewers and our shepherd for their valuable comments. We thank Fish Wang for his suggestions in differential testing.
Authors from Tsinghua University were supported by the National Natural Science Foundation of China (U1836213, U19B2034, 62102218, and 62132011).
Authors from UCI were supported by NSF CNS-2047476.

\small
\bibliographystyle{abbrv}

\begin{thebibliography}{100}

\bibitem{allman2018comments}
M.~Allman.
\newblock Comments on dns robustness.
\newblock In {\em Proceedings of the Internet Measurement Conference 2018}, pages 84--90, 2018.

\bibitem{cve20223736stateless}
C.~Almond.
\newblock {CVE-2022-3736: named configured to answer from stale cache may terminate unexpectedly while processing RRSIG queries}.
\newblock \url{https://kb.isc.org/docs/cve-2022-3736}, 2022.

\bibitem{alowaisheq2019cracking}
E.~Alowaisheq, P.~Wang, S.~Alrwais, X.~Liao, X.~Wang, T.~Alowaisheq, X.~Mi, S.~Tang, and B.~Liu.
\newblock {Cracking the Wall of Confinement: Understanding and Analyzing Malicious Domain Take-downs}.
\newblock In {\em NDSS '19}.

\bibitem{cond_forwarder1}
A.~E. Alvarez.
\newblock {DNS Forwarding and Conditional Forwarding}.
\newblock \url{https://medium.com/tech-jobs-academy/dns-forwarding-and-conditional-forwarding-f3118bc93984}, 2016.

\bibitem{andronidis2022snapfuzz}
A.~Andronidis and C.~Cadar.
\newblock Snapfuzz: High-throughput fuzzing of network applications.
\newblock In {\em Proceedings of the 31st ACM SIGSOFT International Symposium on Software Testing and Analysis (ISSTA)}, 2022.

\bibitem{arends2005protocol}
R.~Arends, R.~Austein, M.~Larson, D.~Massey, and S.~Rose.
\newblock {RFC 4035: Protocol Modifications for the DNS Security Extensions}.
\newblock {\em RFC Proposed Standard}.

\bibitem{arends2005rfc}
R.~Arends, R.~Austein, M.~Larson, D.~Massey, and S.~Rose.
\newblock Rfc 4033: Dns security introduction and requirements, 2005.

\bibitem{ba2022stateful}
J.~Ba, M.~B{\"o}hme, Z.~Mirzamomen, and A.~Roychoudhury.
\newblock Stateful greybox fuzzing.
\newblock In {\em 31st USENIX Security Symposium (USENIX Security 22)}, pages 3255--3272, Boston, MA, Aug. 2022. USENIX Association.

\bibitem{bernhard2022jit}
L.~Bernhard, T.~Scharnowski, M.~Schloegel, T.~Blazytko, and T.~Holz.
\newblock Jit-picking: Differential fuzzing of javascript engines.
\newblock In {\em Proceedings of the 2022 ACM SIGSAC Conference on Computer and Communications Security (CCS)}, 2022.

\bibitem{bind-crash-bug}
{BIND}.
\newblock "xferquota" system test fails intermittently.
\newblock \url{https://gitlab.isc.org/isc-projects/bind9/-/issues/3300}.

\bibitem{dnsversion}
{BIND}.
\newblock {How do I change the version that BIND reports when queried for version.bind?}
\newblock \url{https://kb.isc.org/docs/aa-00359}, 2021.

\bibitem{bind}
BIND.
\newblock \url{https://www.isc.org/bind/}, 2022.

\bibitem{bind_fallback}
BIND.
\newblock {BIND Document: type forward}.
\newblock \url{https://bind9.readthedocs.io/en/latest/reference.html\#namedconf-statement-type\%20forward}, 2023.

\bibitem{brandt2018domain}
M.~Brandt, T.~Dai, A.~Klein, H.~Shulman, and M.~Waidner.
\newblock {Domain Validation++ For MitM-Resilient PKI}.
\newblock In {\em CCS '18}.

\bibitem{chen2019deep}
Y.~Chen, T.~Su, and Z.~Su.
\newblock Deep differential testing of jvm implementations.
\newblock In {\em 2019 IEEE/ACM 41st International Conference on Software Engineering (ICSE)}, pages 1257--1268. IEEE, 2019.

\bibitem{bind-rndc}
I.~S. Consortium.
\newblock {rndc.conf - rndc configuration file}.
\newblock \url{https://bind9.readthedocs.io/en/v9_16_5/manpages.html#rndc-conf-rndc-configuration-file}.

\bibitem{heartbleed}
{CVE Details}.
\newblock {CVE-2014-0160}.
\newblock \url{https://www.cvedetails.com/cve/CVE-2014-0160/}, 2014.

\bibitem{cve20222881infoleak}
{CVE Details}.
\newblock {CVE-2022-2881}.
\newblock \url{https://www.cvedetails.com/cve/CVE-2022-2881/}, 2022.

\bibitem{cve20223924stateful}
{CVE Details}.
\newblock {CVE-2022-3924}.
\newblock \url{https://www.cvedetails.com/cve/CVE-2022-3924/}, 2022.

\bibitem{cvebind}
{CVE Details}.
\newblock {BIND: Vulnerability Statistics}.
\newblock \url{https://www.cvedetails.com/product/144/ISC-Bind.html?vendor_id=64}, 2023.

\bibitem{cveknot}
{CVE Details}.
\newblock {Knot Resolver: Vulnerability Statistics}.
\newblock \url{https://www.cvedetails.com/product/63850/NIC-Knot-Resolver.html?vendor_id=20536}, 2023.

\bibitem{cvemaradns}
{CVE Details}.
\newblock {MaraDNS: Vulnerability Statistics}.
\newblock \url{https://www.cvedetails.com/vendor/1470/Maradns.html}, 2023.

\bibitem{cvepowerdns}
{CVE Details}.
\newblock {PowerDNS: Vulnerability Statistics}.
\newblock \url{https://www.cvedetails.com/vendor/2834/Powerdns.html}, 2023.

\bibitem{cvetechnitium}
{CVE Details}.
\newblock {Technitium: Vulnerability Statistics}.
\newblock \url{https://www.cvedetails.com/vendor/26782/Technitium.html}, 2023.

\bibitem{cveunbound}
{CVE Details}.
\newblock {Unbound: Vulnerability Statistics}.
\newblock \url{https://www.cvedetails.com/product/20882/Nlnetlabs-Unbound.html?vendor_id=9613}, 2023.

\bibitem{dai2021hijackers}
T.~Dai, P.~Jeitner, H.~Shulman, and M.~Waidner.
\newblock {The Hijackers Guide To The Galaxy: Off-Path Taking Over Internet Resources}.
\newblock In {\em USENIX Security '21}.

\bibitem{google_dns}
G.~P. DNS.
\newblock \url{https://developers.google.com/speed/public-dns}, 2022.

\bibitem{fpdns}
DNS-OARC.
\newblock {fpdns - DNS Fingerprinting Tool}.
\newblock \url{https://www.dns-oarc.net/tools/fpdns}, 2021.

\bibitem{dnsmasq}
Dnsmasq.
\newblock \url{https://thekelleys.org.uk/dnsmasq/doc.html}, 2022.

\bibitem{dnsviz}
DNSViz.
\newblock {A DNS visualization tool}.
\newblock \url{https://dnsviz.net/}, 2020.

\bibitem{elz1997clarifications}
R.~Elz and R.~Bush.
\newblock {RFC 2181: Clarifications to the DNS Specification}.
\newblock {\em RFC Proposed Standard}.

\bibitem{fiterau2020analysis}
P.~Fiterau-Brostean, B.~Jonsson, R.~Merget, J.~De~Ruiter, K.~Sagonas, and J.~Somorovsky.
\newblock Analysis of $\{$DTLS$\}$ implementations using protocol state fuzzing.
\newblock In {\em 29th USENIX Security Symposium (USENIX Security 20)}, pages 2523--2540, 2020.

\bibitem{afl}
{Google}.
\newblock {Fuzzing with afl-fuzz}.
\newblock \url{https://afl-1.readthedocs.io/en/latest/fuzzing.html}.

\bibitem{bind-oss}
{Google}.
\newblock {oss-fuzz/projects/bind9 at master · google/oss-fuzz}.
\newblock \url{https://github.com/google/oss-fuzz/tree/master/projects/bind9}, 2022.

\bibitem{grangeia2004dns}
L.~Grangeia.
\newblock Cache snooping or snooping the cache for fun and profit, 2004.

\bibitem{tcpdump}
T.~T. Group.
\newblock {Tcpdump and Libpcap}.
\newblock \url{https://www.tcpdump.org/}.

\bibitem{guo2018dlfuzz}
J.~Guo, Y.~Jiang, Y.~Zhao, Q.~Chen, and J.~Sun.
\newblock Dlfuzz: Differential fuzzing testing of deep learning systems.
\newblock In {\em Proceedings of the 2018 26th ACM Joint Meeting on European Software Engineering Conference and Symposium on the Foundations of Software Engineering}, pages 739--743, 2018.

\bibitem{dns-fuzzer}
{guyinatuxedo}.
\newblock \url{https://github.com/guyinatuxedo/dns-fuzzer}, 2019.

\bibitem{herzberg2013fragmentation}
A.~Herzberg and H.~Shulman.
\newblock {Fragmentation Considered Poisonous, or: One-domain-to-rule-them-all.org}.
\newblock In {\em CNS '13}.

\bibitem{herzberg2013vulnerable}
A.~Herzberg and H.~Shulman.
\newblock {Vulnerable Delegation of DNS Resolution}.
\newblock In {\em ESORICS '13}.

\bibitem{hur2022specdoctor}
J.~Hur, S.~Song, S.~Kim, and B.~Lee.
\newblock Specdoctor: Differential fuzz testing to find transient execution vulnerabilities.
\newblock In {\em Proceedings of the 2022 ACM SIGSAC Conference on Computer and Communications Security (CCS)}, pages 1473--1487, 2022.

\bibitem{hur2021difuzzrtl}
J.~Hur, S.~Song, D.~Kwon, E.~Baek, J.~Kim, and B.~Lee.
\newblock Difuzzrtl: Differential fuzz testing to find cpu bugs.
\newblock In {\em 2021 IEEE Symposium on Security and Privacy (SP)}, pages 1286--1303. IEEE, 2021.

\bibitem{pythondocker}
D.~Inc.
\newblock {Docker SDK for Python}.
\newblock \url{https://docker-py.readthedocs.io/}.

\bibitem{izhikevich2021lzr}
L.~Izhikevich, R.~Teixeira, and Z.~Durumeric.
\newblock {LZR: Identifying Unexpected Internet Services}.
\newblock In {\em Proceedings of the 30th USENIX Security Symposium (USENIX Security '21)}, 2021.

\bibitem{jabiyev2022frameshifter}
B.~Jabiyev, S.~Sprecher, A.~Gavazzi, T.~Innocenti, K.~Onarlioglu, and E.~Kirda.
\newblock $\{$FRAMESHIFTER$\}$: Security implications of $\{$HTTP/2-to-HTTP/1$\}$ conversion anomalies.
\newblock In {\em 31st USENIX Security Symposium (USENIX Security 22)}, pages 1061--1075, 2022.

\bibitem{jabiyev2021t}
B.~Jabiyev, S.~Sprecher, K.~Onarlioglu, and E.~Kirda.
\newblock T-reqs: Http request smuggling with differential fuzzing.
\newblock In {\em Proceedings of the 2021 ACM SIGSAC Conference on Computer and Communications Security (CCS)}, pages 1805--1820, 2021.

\bibitem{jana2012abusing}
S.~Jana and V.~Shmatikov.
\newblock Abusing file processing in malware detectors for fun and profit.
\newblock In {\em 2012 IEEE Symposium on Security and Privacy}, pages 80--94. IEEE, 2012.

\bibitem{jeitner2021injection}
P.~Jeitner and H.~Shulman.
\newblock {Injection Attacks Reloaded: Tunnelling Malicious Payloads over DNS}.
\newblock In {\em USENIX Security '21}.

\bibitem{jeitner2022xdri}
P.~Jeitner, H.~Shulman, L.~Teichmann, and M.~Waidner.
\newblock $\{$XDRI$\}$ attacks-and-how to enhance resilience of residential routers.
\newblock In {\em 31st USENIX Security Symposium (USENIX Security 22)}, pages 4473--4490, 2022.

\bibitem{jelinek1992basic}
F.~Jelinek, J.~D. Lafferty, and R.~L. Mercer.
\newblock {\em Basic methods of probabilistic context free grammars}.
\newblock Springer, 1992.

\bibitem{jiang2012ghost}
J.~Jiang, J.~Liang, K.~Li, J.~Li, H.-X. Duan, and J.~Wu.
\newblock {Ghost Domain Names: Revoked Yet Still Resolvable}.
\newblock In {\em NDSS '12}.

\bibitem{jung2001dns}
J.~Jung, E.~Sit, H.~Balakrishnan, and R.~Morris.
\newblock Dns performance and the effectiveness of caching.
\newblock In {\em Proceedings of the 1st ACM SIGCOMM Workshop on Internet Measurement}, pages 153--167, 2001.

\bibitem{kakarla2022formal}
S.~K.~R. Kakarla.
\newblock {\em Formal Methods for a Robust Domain Name System}.
\newblock University of California, Los Angeles, 2022.

\bibitem{kakarla2020groot}
S.~K.~R. Kakarla, R.~Beckett, B.~Arzani, T.~Millstein, and G.~Varghese.
\newblock Groot: Proactive verification of dns configurations.
\newblock In {\em Proceedings of the Annual conference of the ACM Special Interest Group on Data Communication on the applications, technologies, architectures, and protocols for computer communication (SIGCOMM)}, pages 310--328, 2020.

\bibitem{kakarla2022scale}
S.~K.~R. Kakarla, R.~Beckett, T.~Millstein, and G.~Varghese.
\newblock $\{$SCALE$\}$: Automatically finding $\{$RFC$\}$ compliance bugs in $\{$DNS$\}$ nameservers.
\newblock In {\em 19th USENIX Symposium on Networked Systems Design and Implementation (NSDI 22)}, pages 307--323, 2022.

\bibitem{kaminsky2008black}
D.~Kaminsky.
\newblock Black ops 2008: It’s the end of the cache as we know it.
\newblock {\em Black Hat USA}, 2, 2008.

\bibitem{bind-market-share}
S.~M. Kerner.
\newblock {BIND DNS Holds Lead}.
\newblock \url{https://www.serverwatch.com/server-news/bind-dns-holds-lead/}, 2020.

\bibitem{ps}
M.~Kerrisk.
\newblock {ps(1) — Linux manual page}.
\newblock \url{https://man7.org/linux/man-pages/man1/ps.1.html}.

\bibitem{klein2021cross}
A.~Klein.
\newblock {Cross Layer Attacks and How to Use Them (for DNS Cache Poisoning, Device Tracking and More)}.
\newblock In {\em S\&P '21}.

\bibitem{knot-nxns}
{Knot Resolver}.
\newblock Cz-nic/knot-resolver/lib/selection\_iter.c.
\newblock \url{https://github.com/CZ-NIC/knot-resolver/blob/v5.6.0/lib/selection_iter.c\#L281}.

\bibitem{bind-test}
J.~Knudsen.
\newblock {CyRC Case Study: Securing BIND 9}.
\newblock \url{https://www.synopsys.com/blogs/software-security/cyrc-case-study-securing-bind-9/}, 2022.

\bibitem{cve20223736patch}
M.~Kępień.
\newblock {Merge branch '3622-serve-stale-rrsig-fix-security' into 'security-main'}.
\newblock \url{https://gitlab.isc.org/isc-projects/bind9/-/commit/80ed02f935fbc2adcb8ba8632b0365375232b6cd}, 2022.

\bibitem{unbound-unbound-control}
S.~N. Labs.
\newblock {Unbound - unbound-control.8}.
\newblock \url{https://nlnetlabs.nl/documentation/unbound/unbound-control/}.

\bibitem{lawrence2020serving}
D.~C. Lawrence, W.~A. Kumari, and P.~Sood.
\newblock {RFC 8767: Serving Stale Data to Improve DNS Resiliency}.
\newblock {\em RFC Proposed Standard}, 2020.

\bibitem{lee2020longitudinal}
H.~Lee, A.~Gireesh, R.~v. Rijswijk-Deij, T.~T. Kwon, and T.~Chung.
\newblock {A Longitudinal and Comprehensive Study of the DANE Ecosystem in Email}.
\newblock In {\em USENIX Security '20}.

\bibitem{li2022sediff}
P.~Li, W.~Meng, and K.~Lu.
\newblock Sediff: scope-aware differential fuzzing to test internal function models in symbolic execution.
\newblock In {\em Proceedings of the 30th ACM Joint European Software Engineering Conference and Symposium on the Foundations of Software Engineering}, pages 57--69, 2022.

\bibitem{li2023ghost}
X.~Li, B.~Liu, X.~Bai, M.~Zhang, Q.~Zhang, Z.~Li, H.~Duan, and Q.~Li.
\newblock {Ghost Domain Reloaded: Vulnerable Links in Domain Name Delegation and Revocation}.
\newblock In {\em NDSS '23}.

\bibitem{li2021fast}
X.~Li, B.~Liu, X.~Zheng, H.~Duan, Q.~Li, and Y.~Huang.
\newblock {Fast IPv6 Network Periphery Discovery and Security Implications}.
\newblock In {\em DSN '21}.

\bibitem{li2023maginot}
X.~Li, C.~Lu, B.~Liu, Q.~Zhang, Z.~Li, H.~Duan, and Q.~Li.
\newblock {The Maginot Line: Attacking the Boundary of DNS Caching Protection}.
\newblock In {\em Proceedings of the 32nd USENIX Security Symposium (USENIX Security '23)}, 2023.

\bibitem{li2024tudoor}
X.~Li, W.~Xu, B.~Liu, M.~Zhang, Z.~Li, J.~Zhang, D.~Chang, X.~Zheng, C.~Wang, J.~Chen, H.~Duan, and Q.~Li.
\newblock {TuDoor Attack: Systematically Exploring and Exploiting Logic Vulnerabilities in DNS Response Pre-processing with Malformed Packets}.
\newblock In {\em Proceedings of 2024 IEEE Symposium on Security and Privacy}, Oakland S\&P '24, 2024.

\bibitem{ubuntu-image}
C.~Ltd.
\newblock ubuntu - official image | docker hub.
\newblock \url{https://hub.docker.com/_/ubuntu}.

\bibitem{man2020dns}
K.~Man, Z.~Qian, Z.~Wang, X.~Zheng, Y.~Huang, and H.~Duan.
\newblock {DNS Cache Poisoning Attack Reloaded: Revolutions with Side Channels}.
\newblock In {\em CCS '20}.

\bibitem{man2021dns}
K.~Man, X.~Zhou, and Z.~Qian.
\newblock {DNS Cache Poisoning Attack: Resurrections with Side Channels}.
\newblock In {\em CCS '21}.

\bibitem{mao2022assessing}
J.~Mao, M.~Rabinovich, and K.~Schomp.
\newblock {Assessing Support for DNS-over-TCP in the Wild}.
\newblock In {\em Proceedings of the 23rd International Conference on Passive and Active Measurement (PAM '22)}, 2022.

\bibitem{geolite2}
{MaxMind}.
\newblock {GeoLite2 Free Geolocation Data}.
\newblock \url{https://dev.maxmind.com/geoip/geolite2-free-geolocation-data}, 2023.

\bibitem{mckeeman1998differential}
W.~M. McKeeman.
\newblock Differential testing for software.
\newblock {\em Digital Technical Journal}, 10(1):100--107, 1998.

\bibitem{merkel2014docker}
D.~Merkel.
\newblock Docker: lightweight linux containers for consistent development and deployment.
\newblock {\em Linux journal}, 2014(239):2, 2014.

\bibitem{dnslint}
Microsoft.
\newblock {Description of the DNSLint utility}.
\newblock \url{https://support.microsoft.com/en-us/help/321045/description-of-the-dnslint-utility}, 2018.

\bibitem{min2015cross}
C.~Min, S.~Kashyap, B.~Lee, C.~Song, and T.~Kim.
\newblock Cross-checking semantic correctness: The case of finding file system bugs.
\newblock In {\em Proceedings of the 25th Symposium on Operating Systems Principles (SOSP)}, pages 361--377, 2015.

\bibitem{cve}
{MITRE}.
\newblock {CVE}.
\newblock \url{https://cve.mitre.org/}, 2023.

\bibitem{cvedetails}
{MITRE}.
\newblock {CVE Details}.
\newblock \url{https://www.cvedetails.com/}, 2023.

\bibitem{cvebeta}
{MITRE}.
\newblock {CVE List}.
\newblock \url{https://www.cve.org/}, 2023.

\bibitem{mockapetris1987domain1034}
P.~V. Mockapetris.
\newblock {RFC 1034: Domain Names - Concepts and Facilities}.
\newblock {\em RFC Standard}.

\bibitem{moon2021accurately}
S.-J. Moon, Y.~Yin, R.~A. Sharma, Y.~Yuan, J.~M. Spring, and V.~Sekar.
\newblock {Accurately Measuring Global Risk of Amplification Attacks using AmpMap}.
\newblock In {\em Proceedings of the 30th USENIX Security Symposium (USENIX Security '21)}, 2021.

\bibitem{mxtoolbox}
MxToolbox.
\newblock {Check your DNS MX Records online}.
\newblock \url{https://mxtoolbox.com/}, 2020.

\bibitem{nainggolan2019improved}
R.~Nainggolan, R.~Perangin-angin, E.~Simarmata, and A.~F. Tarigan.
\newblock Improved the performance of the k-means cluster using the sum of squared error (sse) optimized by using the elbow method.
\newblock In {\em Journal of Physics: Conference Series}, volume 1361, page 012015. IOP Publishing, 2019.

\bibitem{akhavan2021cache}
A.~A. Niaki, W.~R. Marczak, S.~Farhoodi, A.~McGregor, P.~Gill, and N.~Weaver.
\newblock {Cache Me Outside: A New Look at DNS Cache Probing}.
\newblock In {\em Proceedings of the 22nd International Conference on Passive and Active Measurement (PAM '21)}, 2021.

\bibitem{nilizadeh2019diffuzz}
S.~Nilizadeh, Y.~Noller, and C.~S. Pasareanu.
\newblock Diffuzz: differential fuzzing for side-channel analysis.
\newblock In {\em 2019 IEEE/ACM 41st International Conference on Software Engineering (ICSE)}, pages 176--187. IEEE, 2019.

\bibitem{noller2020hydiff}
Y.~Noller, C.~S. P{\u{a}}s{\u{a}}reanu, M.~B{\"o}hme, Y.~Sun, H.~L. Nguyen, and L.~Grunske.
\newblock Hydiff: Hybrid differential software analysis.
\newblock In {\em 2020 IEEE/ACM 42nd International Conference on Software Engineering (ICSE)}, pages 1273--1285. IEEE, 2020.

\bibitem{pappas2004distributed}
V.~Pappas, P.~F{\"a}ltstr{\"o}m, D.~Massey, and L.~Zhang.
\newblock {Distributed DNS troubleshooting}.
\newblock In {\em Proceedings of the ACM SIGCOMM workshop on Network troubleshooting: research, theory and operations practice meet malfunctioning reality}, pages 265--270, 2004.

\bibitem{partridge2016ethical}
C.~Partridge and M.~Allman.
\newblock Ethical considerations in network measurement papers.
\newblock {\em Communications of the ACM}, 59(10):58--64, 2016.

\bibitem{pearce2017global}
P.~Pearce, B.~Jones, F.~Li, R.~Ensafi, N.~Feamster, N.~Weaver, and V.~Paxson.
\newblock {Global Measurement of DNS Manipulation}.
\newblock In {\em Proceedings of the 26th USENIX Security Symposium (USENIX Security '17)}, 2017.

\bibitem{scikit-learn}
F.~Pedregosa, G.~Varoquaux, A.~Gramfort, V.~Michel, B.~Thirion, O.~Grisel, M.~Blondel, P.~Prettenhofer, R.~Weiss, V.~Dubourg, J.~Vanderplas, A.~Passos, D.~Cournapeau, M.~Brucher, M.~Perrot, and E.~Duchesnay.
\newblock Scikit-learn: Machine learning in {P}ython.
\newblock {\em Journal of Machine Learning Research}, 12:2825--2830, 2011.

\bibitem{peng2023gleefuzz}
H.~Peng, Z.~Yao, A.~A. Sani, D.~J. Tian, and M.~Payer.
\newblock Gleefuzz: Fuzzing webgl through error message guided mutation.
\newblock {\em USENIX Security'23}, 2023.

\bibitem{petsios2017nezha}
T.~Petsios, A.~Tang, S.~Stolfo, A.~D. Keromytis, and S.~Jana.
\newblock Nezha: Efficient domain-independent differential testing.
\newblock In {\em 2017 IEEE Symposium on security and privacy (SP)}, pages 615--632. IEEE, 2017.

\bibitem{pham2020aflnet}
V.-T. Pham, M.~B{\"o}hme, and A.~Roychoudhury.
\newblock Aflnet: a greybox fuzzer for network protocols.
\newblock In {\em 2020 IEEE 13th International Conference on Software Testing, Validation and Verification (ICST)}, pages 460--465. IEEE, 2020.

\bibitem{powerdns-rec-control}
PowerDNS.COM.
\newblock {rec\_control}.
\newblock \url{https://docs.powerdns.com/recursor/manpages/rec_control.1.html}.

\bibitem{reen2020dpifuzz}
G.~S. Reen and C.~Rossow.
\newblock Dpifuzz: a differential fuzzing framework to detect dpi elusion strategies for quic.
\newblock In {\em Annual Computer Security Applications Conference}, pages 332--344, 2020.

\bibitem{resolverfuzz-github}
{ResolverFuzz}.
\newblock \url{https://github.com/ResolverFuzz/ResolverFuzz}, 2023.

\bibitem{romao1994tools}
A.~Romao.
\newblock Tools for dns debugging.
\newblock Technical report, RFC 1713, FCCN, November, 1994.

\bibitem{snapfuzz_epoll}
{SaBRe}.
\newblock The binary rewriting plugin sabre used by snapfuzz.
\newblock \url{https://github.com/andronat/SaBRe/blob/4a41a5adaec89235e00adc5d339be308f5c8d57c/plugins/sbr-afl/main.c}, 2022.

\bibitem{schomp2014assessing}
K.~Schomp, T.~Callahan, M.~Rabinovich, and M.~Allman.
\newblock {Assessing DNS Vulnerability to Record Injection}.
\newblock In {\em PAM '14}.

\bibitem{schomp2013measuring}
K.~Schomp, T.~Callahan, M.~Rabinovich, and M.~Allman.
\newblock {On measuring the client-side DNS infrastructure}.
\newblock In {\em IMC '13}.

\bibitem{schuba1993addressing}
C.~Schuba and E.~H. Spafford.
\newblock {Addressing Weaknesses in the Domain Name System Protocol}.
\newblock {\em Master's thesis}.

\bibitem{serebryany2012addresssanitizer}
K.~Serebryany, D.~Bruening, A.~Potapenko, and D.~Vyukov.
\newblock Addresssanitizer: a fast address sanity checker.
\newblock In {\em Proceedings of the 2012 USENIX conference on Annual Technical Conference}, pages 28--28, 2012.

\bibitem{kmeans}
{sharadarao1999}.
\newblock {Bisecting K-Means Algorithm Introduction}.
\newblock \url{https://www.geeksforgeeks.org/bisecting-k-means-algorithm-introduction/}.

\bibitem{shen2022hdiff}
K.~Shen, J.~Lu, Y.~Yang, J.~Chen, M.~Zhang, H.~Duan, J.~Zhang, and X.~Zheng.
\newblock Hdiff: A semi-automatic framework for discovering semantic gap attack in http implementations.
\newblock In {\em 2022 52nd Annual IEEE/IFIP International Conference on Dependable Systems and Networks (DSN)}, pages 1--13. IEEE, 2022.

\bibitem{dual-stack}
A.~Singram and K.~Umashankar.
\newblock {Implementing dual stack recursive DNS at Microsoft: Challenges and Learning}.
\newblock \url{https://indico.dns-oarc.net/event/42/contributions/904/}, 2022.

\bibitem{dns-fuzz-server}
{sischkg}.
\newblock \url{https://github.com/sischkg/dns-fuzz-server}, 2019.

\bibitem{sivakorn2017hvlearn}
S.~Sivakorn, G.~Argyros, K.~Pei, A.~D. Keromytis, and S.~Jana.
\newblock Hvlearn: Automated black-box analysis of hostname verification in ssl/tls implementations.
\newblock In {\em 2017 IEEE Symposium on Security and Privacy (SP)}, pages 521--538. IEEE, 2017.

\bibitem{somorovsky2016systematic}
J.~Somorovsky.
\newblock Systematic fuzzing and testing of tls libraries.
\newblock In {\em Proceedings of the 2016 ACM SIGSAC Conference on Computer and Communications Security}, pages 1492--1504, 2016.

\bibitem{son2010hitchhiker}
S.~Son and V.~Shmatikov.
\newblock {The Hitchhiker's Guide to DNS Cache Poisoning}.
\newblock In {\em SecureComm '10}.

\bibitem{song2022r2z2}
S.~Song, J.~Hur, S.~Kim, P.~Rogers, and B.~Lee.
\newblock R2z2: Detecting rendering regressions in web browsers through differential fuzz testing.
\newblock In {\em Proceedings of the 2022 International Conference on Software Engineering (ICSE 2022)}, 2022.

\bibitem{stewart2003dns}
J.~Stewart.
\newblock {DNS Cache Poisoning – The Next Generation}.
\newblock {\em Secureworks}.

\bibitem{tan2014bug}
L.~Tan, C.~Liu, Z.~Li, X.~Wang, Y.~Zhou, and C.~Zhai.
\newblock Bug characteristics in open source software.
\newblock {\em Empirical software engineering}, 19:1665--1705, 2014.

\bibitem{technitium_web_config}
Technitium.com.
\newblock {Technitium DNS Server API Documentation}.
\newblock \url{https://github.com/TechnitiumSoftware/DnsServer/blob/master/APIDOCS.md}.

\bibitem{cond_forwarder2}
R.~Trader.
\newblock {Windows Server – How to configure a Conditional Forwarder in DNS}.
\newblock \url{https://www.interfacett.com/blogs/windows-server-how-to-configure-a-conditional-forwarder-in-dns/}, 2016.

\bibitem{unbound}
Unbound.
\newblock \url{https://nlnetlabs.nl/projects/unbound/about/}, 2022.

\bibitem{unbound_fallback}
Unbound.
\newblock {Unbound Document: forward-first}.
\newblock \url{https://unbound.docs.nlnetlabs.nl/en/latest/manpages/unbound.conf.html\#unbound-conf-forward-forward-first}, 2023.

\bibitem{dns_amplification_attacks}
US-CERT.
\newblock Alert (ta13-088a): Dns amplification attacks.
\newblock \url{https://www.cisa.gov/uscert/ncas/alerts/TA13-088A}, 2019.

\bibitem{dnssec_analyzer}
Verisign.
\newblock {DNSSEC Analyzer}.
\newblock \url{https://dnssec-analyzer.verisignlabs.com/}, 2020.

\bibitem{walz2017exploiting}
A.~Walz and A.~Sikora.
\newblock Exploiting dissent: towards fuzzing-based differential black-box testing of tls implementations.
\newblock {\em IEEE Transactions on Dependable and Secure Computing}, 17(2):278--291, 2017.

\bibitem{wang2019superion}
J.~Wang, B.~Chen, L.~Wei, and Y.~Liu.
\newblock Superion: Grammar-aware greybox fuzzing.
\newblock In {\em 2019 IEEE/ACM 41st International Conference on Software Engineering (ICSE)}, pages 724--735. IEEE, 2019.

\bibitem{wang2020symtcp}
Z.~Wang, S.~Zhu, Y.~Cao, Z.~Qian, C.~Song, S.~V. Krishnamurthy, K.~S. Chan, and T.~D. Braun.
\newblock Symtcp: Eluding stateful deep packet inspection with automated discrepancy discovery.
\newblock In {\em Network and Distributed System Security Symposium (NDSS)}, 2020.

\bibitem{wang2021themis}
Z.~Wang, S.~Zhu, K.~Man, P.~Zhu, Y.~Hao, Z.~Qian, S.~V. Krishnamurthy, T.~La~Porta, and M.~J. De~Lucia.
\newblock Themis: Ambiguity-aware network intrusion detection based on symbolic model comparison.
\newblock In {\em Proceedings of the 8th ACM Workshop on Moving Target Defense}, pages 31--32, 2021.

\bibitem{xu2023tsuking}
W.~Xu, X.~Li, C.~Lu, B.~Liu, J.~Zhang, J.~Chen, T.~Wan, and H.~Duan.
\newblock {TsuKing: Coordinating DNS Resolvers and Queries into Potent DoS Amplifiers}.
\newblock In {\em Proceedings of the 2023 ACM SIGSAC Conference on Computer and Communications Security}, CCS '23, 2023.

\bibitem{pcfg-fuzz}
A.~Zeller, R.~Gopinath, M.~Böhme, G.~Fraser, and C.~Holler.
\newblock Probabilistic grammar fuzzing.
\newblock \url{https://www.fuzzingbook.org/html/ProbabilisticGrammarFuzzer.html}.

\bibitem{zheng2020poison}
X.~Zheng, C.~Lu, J.~Peng, Q.~Yang, D.~Zhou, B.~Liu, K.~Man, S.~Hao, H.~Duan, and Z.~Qian.
\newblock {Poison Over Troubled Forwarders: A Cache Poisoning Attack Targeting DNS Forwarding Devices}.
\newblock In {\em USENIX Security '20}.

\bibitem{zhu2018measuring}
S.~Zhu, X.~Hu, Z.~Qian, Z.~Shafiq, and H.~Yin.
\newblock Measuring and disrupting anti-adblockers using differential execution analysis.
\newblock In {\em The Network and Distributed System Security Symposium (NDSS)}, 2018.

\bibitem{zou2021tcp}
Y.-H. Zou, J.-J. Bai, J.~Zhou, J.~Tan, C.~Qin, and S.-M. Hu.
\newblock $\{$TCP-Fuzz$\}$: Detecting memory and semantic bugs in $\{$TCP$\}$ stacks with fuzzing.
\newblock In {\em 2021 USENIX Annual Technical Conference (USENIX ATC 21)}, pages 489--502, 2021.

\end{thebibliography}

\appendix
\normalsize
\section{CVE Details}
\label{app:cvedetails}

\begin{figure}[t]	
    \centering
    \includegraphics[width=0.9\columnwidth]{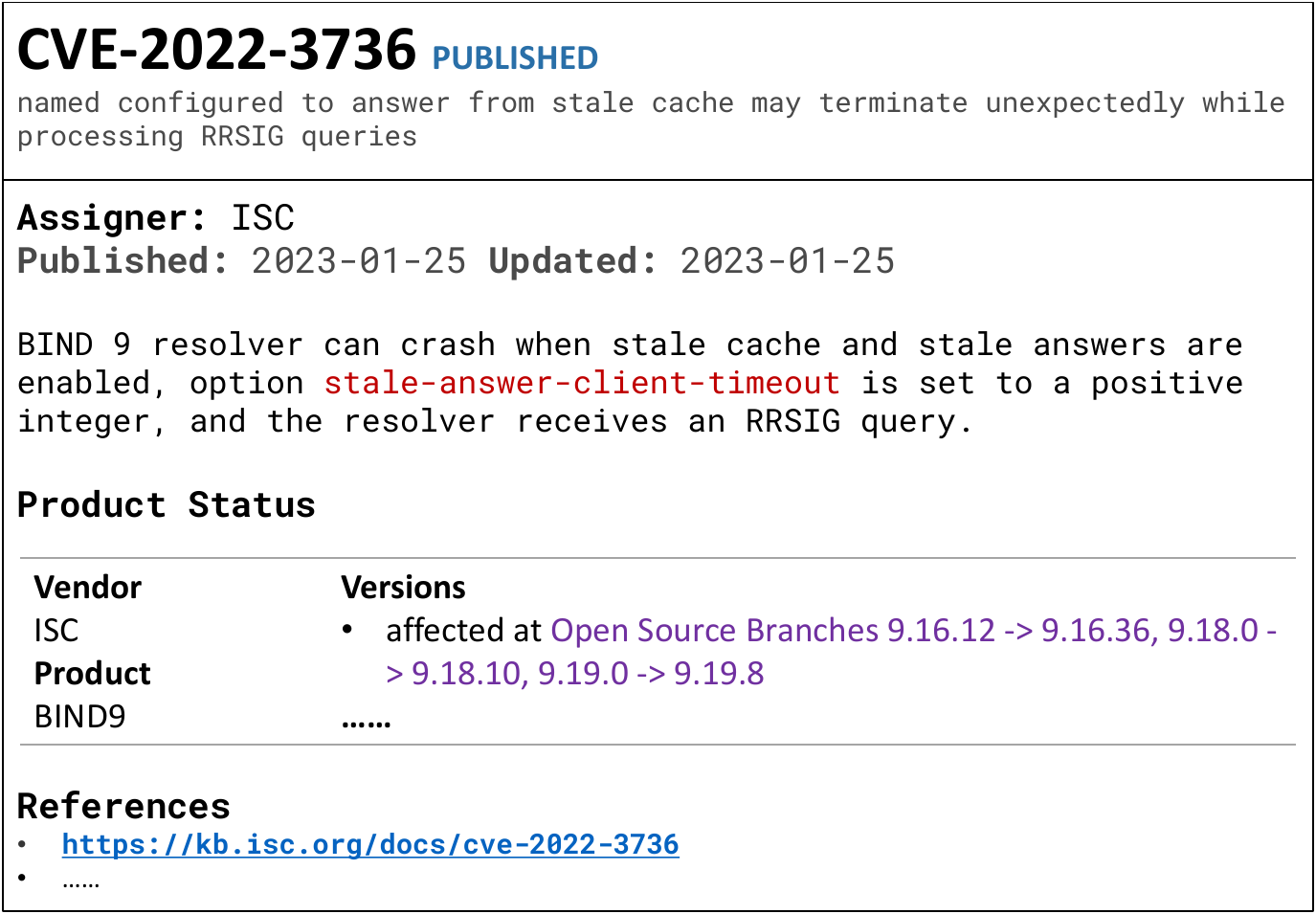}
    \caption{An example of a published CVE report.
    }
    \label{fig:cve}
\end{figure}

\begin{figure}[t]	
    \centering
    \includegraphics[width=1\columnwidth]{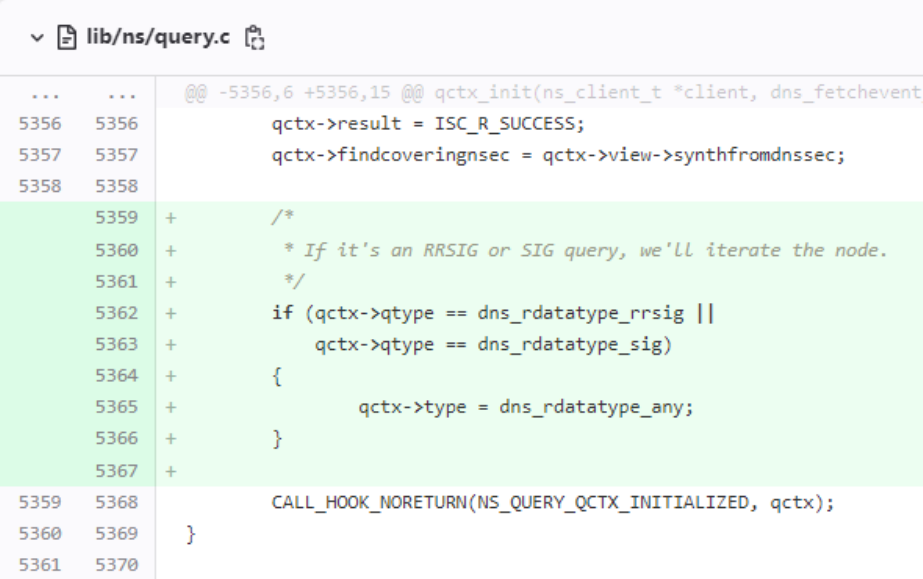}
    \caption{The Git commit related to CVE-2022-3736.}
    \label{fig:commit}
\end{figure}


\revise{
\noindent \textbf{Process of CVE analysis.} Our CVE study in Section~\ref{subsec:cves} characterizes each CVE by its impacted DNS modes (e.g., resolver or nameserver), impacted software, bug type, and the trigger condition (e.g., the related DNS field and the message sequence length). The first three categories can be readily learnt from the CVE reports. For the last one, we found some software vendors include detailed description about the trigger condition (e.g., the report of CVE-2022-3736 under BIND has the details about the trigger condition, as shown in Figure~\ref{fig:cve}). Besides, we can also synthesize a possible trigger condition from the patch described in the git commits (e.g., the commit to fix CVE-2022-3736, as shown in Figure~\ref{fig:commit}, indicates a \texttt{RRSIG} query is the trigger~\cite{cve20223736patch}).
}

\vspace{2pt} \noindent
\textbf{CVE-2022-3736~\cite{cve20223736stateless}.}
The attacker could trick a BIND resolver, which is configured to answer from a stale cache, to terminate unexpectedly by sending just one \texttt{RRSIG} query.
Under the stale cache configuration~\cite{lawrence2020serving}, BIND would attempt to return expired records in the cache to clients if authoritative servers do not reply.
However, BIND does not provide a complete and correct implementation of the stale cache mechanism for the \texttt{RRSIG} query.
Upon receiving a \texttt{RRSIG} query with the stale option set, BIND terminates due to the assertion failure \texttt{qtype != RRSIG}.

\vspace{2pt} \noindent
\textbf{CVE-2022-3924~\cite{cve20223924stateful}.}
By sending multiple queries to BIND, the attacker would drive BIND with the stale option enabled into a race condition and eventually crash.
After receiving a large number of queries for recursive resolution, BIND processes each one from the query queue separately and makes all clients wait for responses.
When a new client query is received, limited by the total number of clients, BIND will reply to the client who has the longest waiting time with the \texttt{SERVFAIL} response.
However, under the stale configuration, a race condition could occur between providing a stale answer and response timeout, which might cause BIND to crash.

\section{Resolver and Zone Configurations}
\label{sec:config}

\begin{figure*}[t]
	\centering
	\subfigcapskip=-1.5mm
	\subfigure[]{
		\begin{minipage}[t]{0.45\linewidth}
		    \centering
			\includegraphics[width=\linewidth]{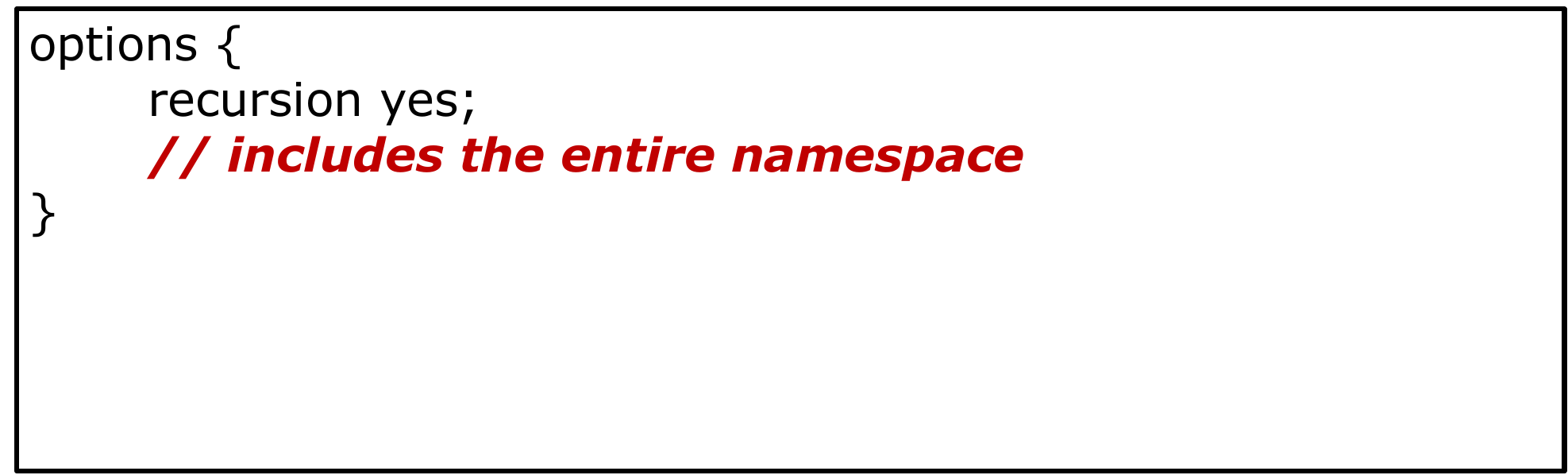}
		\end{minipage}
		\label{fig:mode_config_a}
	}
    \hspace{-3mm}
    \subfigure[]{
		\begin{minipage}[t]{0.45\linewidth}
		    \centering
			\includegraphics[width=\linewidth]{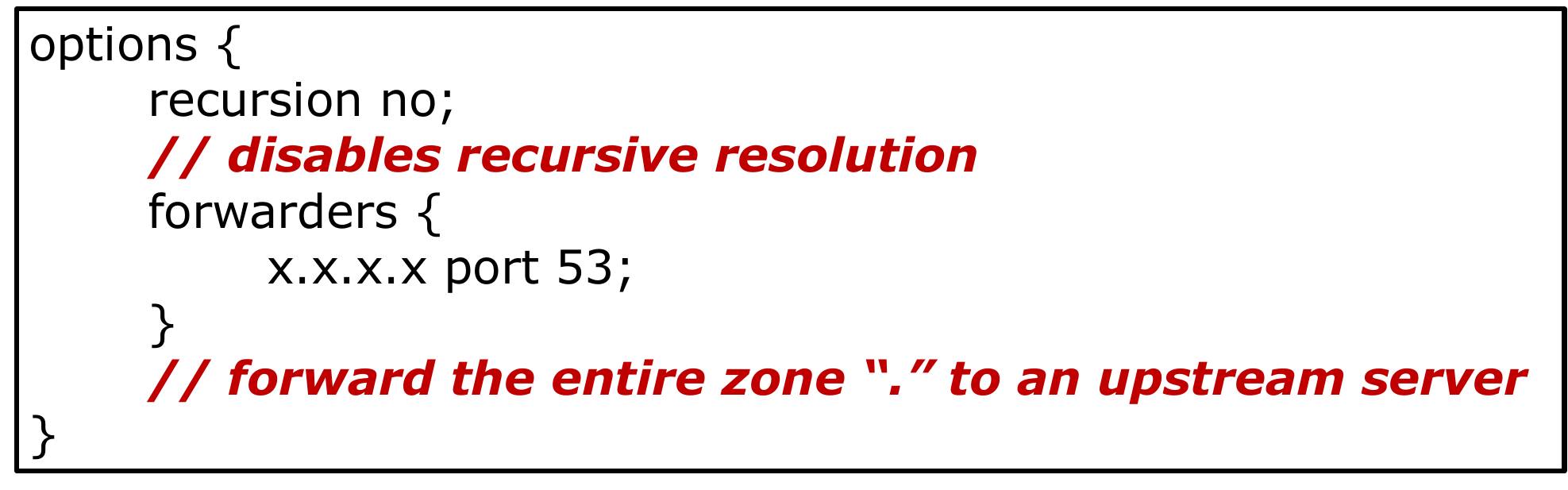}
		\end{minipage}
		\label{fig:mode_config_b}
	}
	\\
    \subfigure[]{
		\begin{minipage}[t]{0.45\linewidth}
		    \centering
			\includegraphics[width=\linewidth]{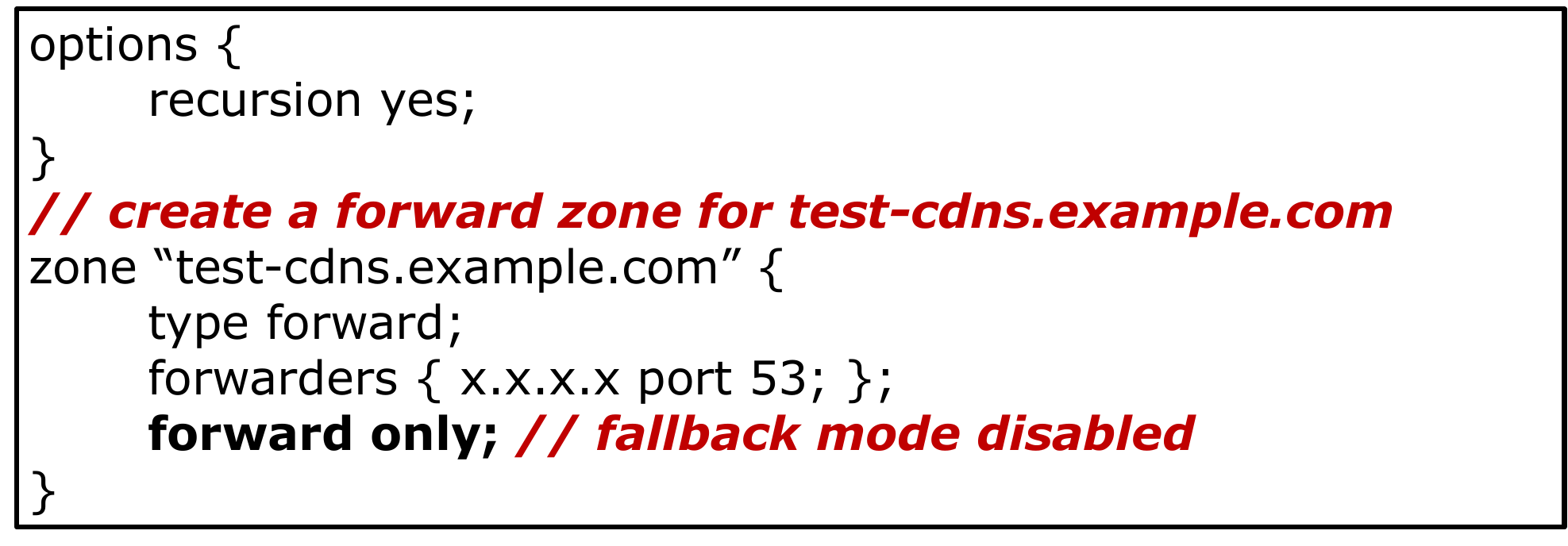}
		\end{minipage}
		\label{fig:mode_config_c}
	}
	\hspace{-3mm}
    \subfigure[]{
		\begin{minipage}[t]{0.45\linewidth}
		    \centering
			\includegraphics[width=\linewidth]{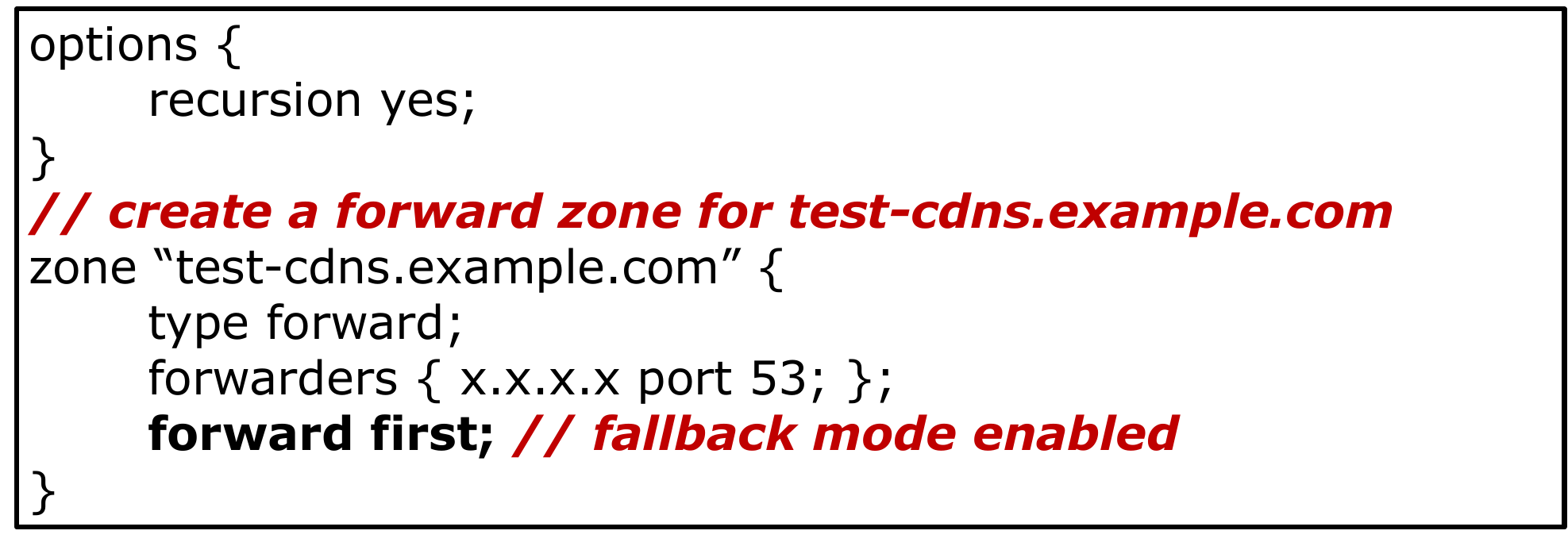}
		\end{minipage}
		\label{fig:mode_config_d}
	}
	\caption{Example BIND configs of a) recursive-only, b) forward-only, c) CDNS without fallback, and d) CDNS with fallback.
        }
	\vspace{-3mm}
	\label{fig:mode_config}
\end{figure*}

Figure~\ref{fig:mode_config} shows examples of the configuration files under the 4 tested modes. Figure~\ref{fig:mode_config_d} shows the zone file of our nameservers.

\begin{figure}[t]
    \centering
    \includegraphics[width=0.95\columnwidth]{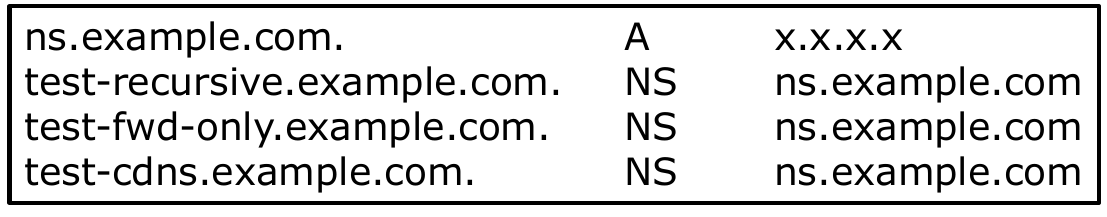}
    \caption{Zone file of \texttt{example.com}. \texttt{ns.example.com} is pointed to the nameserver with the \texttt{A} record. 
    Three \texttt{test-*} subdomains of \texttt{example.com} are used to receive the resolver-queries of different modes.
    Both \emph{CDNS without fallback} and \emph{CDNS with fallback} modes use \texttt{test-cdns.example.com}. 
    }
    \vspace{-2mm}
    \label{fig:auth_srv_zone_file}
\end{figure}

\section{PCFG Details}
\label{app:cfg}


List \ref{lis:cfg_query} and List \ref{lis:cfg_response} show the detailed PCFG for generating DNS query and response packet.

\lstdefinelanguage{BNF}{
    basicstyle=\ttfamily\footnotesize,
    keywordstyle=\bfseries\footnotesize,
    morekeywords={start, query, Header, Question, TransactionID, Flags, RRs, QR, OPCODE, AA, TC, RD, RA, Z, AD, CD, RCODE, QDCOUNT, ANCOUNT, NSCOUNT, ARCOUNT, QNAME, QTYPE, QCLASS},
    morecomment=[l]{//},
    morekeywords={::=},
    sensitive=true,
    breaklines=true,
        columns=flexible, 
    captionpos=b, 
    frame=tb,
    literate={->}{{$\rightarrow$}}1
          {<}{{\textbf{$\langle$}}}1
          {>}{{\textbf{$\rangle$}}}1
}

\begin{lstlisting}[language=BNF, caption={PCFG for DNS query.}, label=lis:cfg_query]
<start> ::= <query>
<query> ::= <Header><Question>
<Header> ::= <TransactionID><Flags><RRs>
<TransactionID> ::= (randomly generated 2-byte hex value)
<Flags> ::= <QR><OPCODE><AA><TC><RD><RA><Z><AD><CD><RCODE>
<QR> ::= 0
<OPCODE> ::= QUERY[.80] | IQUERY[.04] | STATUS[.04] | NOTIFY[.04] | UPDATE[.04] | DSO[.04]
<AA> ::= 0 | 1
<TC> ::= 0 | 1
<RD> ::= 0 | 1
<RA> ::= 0 | 1
<Z> ::= 0 | 1
<AD> ::= 0 | 1
<CD> ::= 0 | 1
<RCODE> ::= NOERROR[.80] | FORMERR[.01] | SERVFAIL[.01] | NXDOMAIN[.01] | NOTIMP[.01] | REFUSED[.01] | YXDOMAIN[.01] | YXRRSET[.01] | NXRRSET[.01] | NOTAUTH[.01] | NOTZONE[.01] | DSOTYPENI[.01] | BADVERS[.01] | BADKEY[.01] | BADTIME[.01] | BADMODE[.01] | BADNAME[.01] | BADALG[.01] | BADTRUNC[.01] | BADCOOKIE[.01]
<RRs> ::= <QDCOUNT><ANCOUNT><NSCOUNT><ARCOUNT>
<QDCOUNT> ::= 1
<ANCOUNT> ::= 0
<NSCOUNT> ::= 0
<ARCOUNT> ::= 0
<Question> ::= <QNAME><QTYPE><QCLASS>
<QNAME> ::= (base domain)[.40] | 
            (sub-domain)[.40] | 
            (2-9th sub-domain)[.10] | 
            (10-max sub-domain)[.10] |
<QTYPE> ::= A | NS | CNAME | SOA | PTR | MX | TXT | AAAA | RRSIG | SPF | ANY
<QCLASS> ::= IN
\end{lstlisting}

\lstdefinelanguage{BNF}{
    basicstyle=\ttfamily\footnotesize,
    keywordstyle=\bfseries\footnotesize,
    morekeywords={start, response, Header, Answer, Authority, Additional, TransactionID, Flags, RRs, QR, OPCODE, AA, TC, RD, RA, Z, AD, CD, RCODE, QDCOUNT, ANCOUNT, NSCOUNT, ARCOUNT, Record, NAME, TYPE, CLASS, TTL, RDLENGTH, RDATA},
    morecomment=[l]{//},
    morekeywords={::=},
    sensitive=true,
    breaklines=true,
    columns=flexible, 
    captionpos=b, 
    frame=tb,
    literate={->}{{$\rightarrow$}}1
          {<}{{\textbf{$\langle$}}}1
          {>}{{\textbf{$\rangle$}}}1
}

\begin{lstlisting}[language=BNF, caption={PCFG for DNS response.}, label=lis:cfg_response]
<start> ::= <response>
<response> ::= <Header><Answer><Authority><Additional>
<Header> ::= <Flags><RRs>
<Flags> ::= <QR><OPCODE><AA><TC><RD><RA><Z><AD><CD><RCODE>
<QR> ::= 1
<OPCODE> ::= QUERY[.80] | IQUERY[.04] | STATUS[.04] | NOTIFY[.04] | UPDATE[.04] | DSO[.04]
<AA> ::= 0 | 1
<TC> ::= 0 | 1
<RD> ::= 0 | 1
<RA> ::= 0 | 1
<Z> ::= 0 | 1
<AD> ::= 0 | 1
<CD> ::= 0 | 1
<RCODE> ::= NOERROR[.80] | FORMERR[.01] | SERVFAIL[.01] | NXDOMAIN[.01] | NOTIMP[.01] | REFUSED[.01] | YXDOMAIN[.01] | YXRRSET[.01] | NXRRSET[.01] | NOTAUTH[.01] | NOTZONE[.01] | DSOTYPENI[.01] | BADVERS[.01] | BADKEY[.01] | BADTIME[.01] | BADMODE[.01] | BADNAME[.01] | BADALG[.01] | BADTRUNC[.01] | BADCOOKIE[.01]
<RRs> ::= <ANCOUNT><NSCOUNT><ARCOUNT>
<ANCOUNT> ::= 0 | 1 | 2 | 3 | 4 | 5
<NSCOUNT> ::= 0 | 1 | 2 | 3 | 4 | 5
<ARCOUNT> ::= 0 | 1 | 2 | 3 | 4 | 5
<Answer> ::= "" | <Record> | <Record>*2 | <Record>*3 | <Record>*4 | <Record>*5
<Authority> ::= "" | <Record> | <Record>*2 | <Record>*3 | <Record>*4 | <Record>*5
<Additional> ::= "" | <Record> | <Record>*2 | <Record>*3 | <Record>*4 | <Record>*5
<Record> ::= <NAME><TYPE><CLASS><TTL><RDLENGTH><RDATA>
<NAME> ::= (domain queried)[.2] | 
            (sub-domain)[.2] | 
            (same-level domain)[.2] | 
            (parent domain)[.2] | 
            (unrelated domain)[.2]
<TYPE> ::= (TYPE queried)[.50] | A[.05] | CNAME[.05] | SOA[.05] | PTR[.05] | MX[.05] | TXT[.05] | AAAA[.05] | RRSIG[.05] | SPF[.05]
<CLASS> ::= IN
<TTL> ::= 60
<RDLENGTH> ::= (length of <RDATA>)[.90] |  (random value in [length, 2*length])[.05] |  (random value in [0, length])[.05] 
<RDATA> ::= (randomly generated data decided by <TYPE>)
\end{lstlisting}

\section{Pseudocode for Input Generation}
\label{app:input_gen}

Algorithm~\ref{alg:input_gen} shows the pseudo-code of \system's test generator.

\begin{algorithm}
    \footnotesize
    \SetKwInOut{Input}{Input}
    \SetKwInOut{Output}{Output}
    
    \underline{function input\_generator} $(base\_qname)$\;
    
    \Input{A domain as base qname for mutation $base\_qname$}
    \Output{A DNS query packet $query$ from client, \\
            A DNS $response$ from upstream, \\
            }
    $query \gets Base\_query$ \tcc{basic structure of DNS query.} 
    $response \gets Base\_response$ \tcc{basic structure of DNS response.}
    \For{$field$ $\in$ $[TXID, QR, OPcode, AA, TC, RD, RA, Z, AD, CD, RCODE$ $, QDCOUNT, ANCOUNT, NSCOUNT, ARCOUNT$]}{
        Set $field$ in $query$ with a generated value from PCFG grammar \\
        Set $field$ in $response$ with a generated value from PCFG grammar
    }

    \For{$field \in QD$}{
        Set $field$ in $query$ with generated records of the number of $QDCOUNT$.
    }
    
    \For{$field \in [AN, NS, AR]$}{
        Set $field$ in $response$ with generated records of the number of corresponding counts.
    }
    
    
    
    \If{$random(0, 1)> 0.9$}{
        Add, delete, or replace some characters in some parts of generated $query$ and $response$.
    }

    return {$query$, $response$}
    \caption{Logic of input generator.}
    \label{alg:input_gen}
\end{algorithm}

\section{Example Self-defined Cache Structure}
\label{app:cache_structure}

List~\ref{lis:cache_format_def} shows an example of a self-defined cache structure.

\begin{lstlisting}[basicstyle=\ttfamily\footnotesize, breaklines=true, captionpos=b, frame=tb, caption={Example of self-defined cache structure.}, label=lis:cache_format_def]
{
    ".": [
        {
            "name": ".",
            "class": "IN",
            "type": "NS",
            "ttl": "518400",
            "rdata": "a.root-servers.net."
        }
    ],
    "a.root-servers.net.": [
        {
            "name": "a.root-servers.net.",
            "class": "IN",
            "type": "A",
            "ttl": "518400",
            "rdata": "198.41.0.4"
        }
    ]
}
\end{lstlisting}

\section{Example Cache Dump Results}
\label{app:cache_dump_format}

List \ref{lis:cache_format_bind}, \ref{lis:cache_format_unbound}, \ref{lis:cache_format_powerdns}, \ref{lis:cache_format_technitium} present the example cache dump results of BIND, Unbound, PowerDNS, and Technitium. 

\begin{lstlisting}[basicstyle=\ttfamily\footnotesize, breaklines=true, captionpos=b, frame=tb, caption={Example of cache dump result of BIND.}, label=lis:cache_format_bind]
; Start view _default
; Cache dump of view '_default' (cache _default)
;
; using a 0 second stale ttl
$DATE 20220708100109
; authanswer
.   518399	IN NS	a.root-servers.net.
    518399	IN NS	b.root-servers.net.
; glue
app.    172799	NS	ns-tld1.charlestonroadregistry.com.
        172799	NS	ns-tld2.charlestonroadregistry.com.
; additional
    86399	DS	23684 8 2 (
    3A5CC8A31E02C94ABA6461912FABB7E9F5E3
    4957BB6114A55A864D96AEC31836 )
(truncated)
; Address database dump
;
; [edns success/timeout]
; [plain success/timeout]
;
; a.gtld-servers.net [v4 TTL 9] [v6 TTL 9] [v4 success] [v6 success]
;	192.5.6.30 [srtt 25] [flags 00000000] [edns 0/0] [plain 0/0] [ttl 1799]
;	2001:503:a83e::2:30 [srtt 39310] [flags 00000000] [edns 0/4] [plain 0/0] [ttl 1799]
\end{lstlisting}

\begin{lstlisting}[basicstyle=\ttfamily\footnotesize, breaklines=true, captionpos=b, frame=tb, caption={Example of cache dump result of Unbound.}, label=lis:cache_format_unbound]
START_RRSET_CACHE
;rrset 86398 13 1 8 0
.	86398	IN	NS   j.root-servers.net.
.	86398	IN	NS   e.root-servers.net.
.	86398	IN	NS   h.root-servers.net.
(truncated)
;rrset 86398 1 1 2 0
CK0POJMG874LJREF7EFN8430QVIT8BSM.com.	86398 IN NSEC3 1 1 0 - ck0q2d6ni4i7eqh8na30ns61o48ul8g5 NS SOA  RRSIG DNSKEY NSEC3PARAM ;{flags: optout}
CK0POJMG874LJREF7EFN8430QVIT8BSM.com.	86398 IN RRSIG NSEC3 8 2 86400 20220827042408 20220820031408 32298 com. DtbwR2L5wFUarqJkJIZuhJi03Kf+24qeQsxH0OZRKQED
    QMP9HAojgZWCx0UstF6MpmLu1ksizKkG35LexZQOqt3C
    2168e5tMVpNaXmcAfL7ZZMXB9M/pf2ngxyiVzRkMQ8cW
    31nYQYGrAqgiE0fYgfp99fIvxmlWghanFpGeCnPKZF15
    4TdIjMmlCdzc6cvodgw1iY4cYYKxWyo5+t81pw== 
    ;{id = 32298}
END_RRSET_CACHE
START_MSG_CACHE
msg . IN NS 32896 1 86398 0 1 0 26
. IN NS 0
m.root-servers.net. IN A 0
l.root-servers.net. IN A 0
k.root-servers.net. IN A 0
END_MSG_CACHE
EOF
\end{lstlisting}

\begin{lstlisting}[basicstyle=\ttfamily\footnotesize, breaklines=true, captionpos=b, frame=tb, caption={Example of cache dump result of PowerDNS.}, label=lis:cache_format_powerdns]
; main record cache dump follows
;
c.root-servers.net. 86400 86395 IN AAAA 2001:500:2::c ; (Indeterminate) auth=0 zone=. from=198.97.190.53  
c.root-servers.net. 86400 86395 IN A 192.33.4.12 ; (Indeterminate) auth=0 zone=. from=198.97.190.53  
com. 86400 86395 IN NS a.gtld-servers.net. ; (Indeterminate) auth=0 zone=. from=193.0.14.129  
net. 86400 86395 IN NS i.gtld-servers.net. ; (Indeterminate) auth=1 zone=net from=192.35.51.30  
m.gtld-servers.net. 86400 86395 IN A 192.55.83.30 ; (Indeterminate) auth=0 zone=. from=193.0.14.129  
; negcache dump follows
;
secpoll.powerdns.com. 3595 IN A VIA secpoll.powerdns.com. ; (Indeterminate) 
; main packet cache dump from thread follows
\end{lstlisting}

\begin{lstlisting}[basicstyle=\ttfamily\footnotesize, breaklines=true, captionpos=b, frame=tb, caption={Example of cache dump result of Technitium.}, label=lis:cache_format_technitium]
{
    "com": [
        {
            "name": "com",
            "type": "NS",
            "ttl": "172800 (2 days)",
            "rData": {
                "nameServer": "a.gtld-servers.net",
                "parentSideTtl": "86400 (1 day)"
            },
            "glueRecords": "192.5.6.30, 2001:503:a83e::2:30",
            "dnssecStatus": "Disabled",
            "lastUsedOn": "2023-01-23T17:51:38.320631Z"
        }
    ],
    "stephane.ns.cloudflare.com": [
        {
            "name": "stephane.ns.cloudflare.com",
            "type": "A",
            "ttl": "86353 (23 hours 59 mins 13 sec)",
            "rData": {
                "ipAddress": "108.162.194.112"
            },
            "dnssecStatus": "Disabled",
            "lastUsedOn": "2023-01-23T17:51:38.3435062Z"
        },
    ],
}
\end{lstlisting}


\section{Clustering Results from Cache Oracle}
\label{app:clustering}

Table~\ref{tab:oracle_cluster} shows the distribution of bugs and differences in the clusters for the cache oracle.

\begin{table}[t]
  \centering
  \small
  \caption{Top 10 regions and AS numbers of the discovered open resolvers during our scanning.}
  \setlength{\tabcolsep}{5pt}
  \begin{threeparttable}
    \begin{tabular}{cccccc}
    \toprule
    \multicolumn{1}{c|}{\textbf{Region}} & \multicolumn{1}{c|}{\textbf{ \# }} & \multicolumn{1}{c|}{\textbf{\%}} & \multicolumn{1}{c|}{\textbf{ASN}} & \multicolumn{1}{c|}{\textbf{ \# }} & \textbf{\%} \\
    \midrule
    China & \multicolumn{1}{r}{   665,328 } & \multicolumn{1}{r|}{36.7\%} & 4134  & \multicolumn{1}{r}{   245,061 } & \multicolumn{1}{r}{13.5\%} \\
    USA   & \multicolumn{1}{r}{   142,058 } & \multicolumn{1}{r|}{7.8\%} & 4837  & \multicolumn{1}{r}{   128,995 } & \multicolumn{1}{r}{7.1\%} \\
    India & \multicolumn{1}{r}{   109,436 } & \multicolumn{1}{r|}{6.0\%} & 4847  & \multicolumn{1}{r}{     58,465 } & \multicolumn{1}{r}{3.2\%} \\
    Russia & \multicolumn{1}{r}{     82,980 } & \multicolumn{1}{r|}{4.6\%} & 17488 & \multicolumn{1}{r}{     53,803 } & \multicolumn{1}{r}{3.0\%} \\
    South Korea & \multicolumn{1}{r}{     71,193 } & \multicolumn{1}{r|}{3.9\%} & 4538  & \multicolumn{1}{r}{     53,043 } & \multicolumn{1}{r}{2.9\%} \\
    Indonesia & \multicolumn{1}{r}{     66,809 } & \multicolumn{1}{r|}{3.7\%} & 4766  & \multicolumn{1}{r}{     40,297 } & \multicolumn{1}{r}{2.2\%} \\
    Brazil & \multicolumn{1}{r}{     51,904 } & \multicolumn{1}{r|}{2.9\%} & 4808  & \multicolumn{1}{r}{     37,304 } & \multicolumn{1}{r}{2.1\%} \\
    Bangladesh & \multicolumn{1}{r}{     43,059 } & \multicolumn{1}{r|}{2.4\%} & 24560 & \multicolumn{1}{r}{     29,830 } & \multicolumn{1}{r}{1.6\%} \\
    Iran  & \multicolumn{1}{r}{     41,578 } & \multicolumn{1}{r|}{2.3\%} & 58224 & \multicolumn{1}{r}{     29,712 } & \multicolumn{1}{r}{1.6\%} \\
    Taiwan & \multicolumn{1}{r}{     27,287 } & \multicolumn{1}{r|}{1.5\%} & 45090 & \multicolumn{1}{r}{     24,928 } & \multicolumn{1}{r}{1.4\%} \\
    \midrule
    \multicolumn{3}{c|}{\textbf{\# Total regions: 229}} & \multicolumn{3}{c}{\textbf{\# Total Ases: 25,342}} \\
    \midrule
    \multicolumn{6}{c}{\textbf{\# Discovered resolvers: 1,815,017}} \\
    \bottomrule
    \end{tabular}%
  \end{threeparttable}
  \label{tab:resolver_distribution}%
\end{table}%

\begin{table}[t]
  \centering
  \small
  \caption{Software used by the open resolvers.}
  \begin{threeparttable}
    \begin{tabular}{c|c|c|c}
    \toprule
    \multirow{3}[3]{*}{\textbf{Software}} & \multicolumn{3}{c}{\textbf{Identified resolvers}} \\
    \cmidrule{2-4}          & \textbf{\texttt{version.}} & \multirow{2}[1]{*}{\textbf{\texttt{fpdns}}} & \multirow{2}[1]{*}{\textbf{Total}} \\
          & \textbf{\texttt{bind}} &       &  \\
    \midrule
    BIND  & \multicolumn{1}{r}{35,987} & \multicolumn{1}{r}{12,420} & \multicolumn{1}{r}{48,407 (5.7\%)} \\
    Unbound & \multicolumn{1}{r}{9,447} & \multicolumn{1}{r}{2,265} & \multicolumn{1}{r}{11,712 (1.4\%)} \\
    Knot  & \multicolumn{1}{r}{43} & \multicolumn{1}{r}{0} & \multicolumn{1}{r}{43 (0.0\%)} \\
    PowerDNS & \multicolumn{1}{r}{10,787} & \multicolumn{1}{r}{749} & \multicolumn{1}{r}{11,536 (1.4\%)} \\
    Subtotal & \multicolumn{1}{r}{56,264} & \multicolumn{1}{r}{15,434} & \multicolumn{1}{r}{71,698 (8.4\%)} \\
    \midrule
    Others & \multicolumn{1}{r}{273,458} & \multicolumn{1}{r}{503,552} & \multicolumn{1}{r}{777,010 (91.6\%)} \\
    \midrule
    Total & \multicolumn{1}{r}{329,722} & \multicolumn{1}{r}{518,986} & \multicolumn{1}{r}{848,708 (100.0\%)} \\
    \bottomrule
    \end{tabular}%
    \begin{tablenotes}
      \footnotesize
      \item Others: Microsoft DNS, Dnsmasq, public DNS services, etc.
    \end{tablenotes}
  \end{threeparttable}
  \label{tab:resolver_version}%
\end{table}%

\begin{table*}[t]
  \centering
  \small
  \setlength{\tabcolsep}{4pt}
  \vspace{-1mm}
  \caption{\revise{Distribution of bugs and differences (non-bugs) for cache oracle.} 
  }
  \begin{threeparttable}
    \begin{tabular}{cc|rrrr|rrrrrr|r|r}
    \toprule
    \multicolumn{1}{c|}{\multirow{2}[2]{*}{\textbf{ Mode }}} & \multirow{2}[2]{*}{\textbf{ Cluster }} & \multicolumn{4}{c|}{\textbf{ Bug }} & \multicolumn{6}{c|}{\textbf{Non-bug Difference }} & \multicolumn{1}{c|}{\textbf{ Total }} & \multicolumn{1}{c}{\textbf{ \# Test }} \\
    \cmidrule{3-12}    \multicolumn{1}{c|}{} &   & \multicolumn{1}{c|}{\textbf{ CP1 }} & \multicolumn{1}{c|}{\textbf{ CP2 }} & \multicolumn{1}{c|}{\textbf{ CP4 }} & \multicolumn{1}{c|}{\textbf{ RC2 }} & \multicolumn{1}{c|}{\textbf{ B1 }} & \multicolumn{1}{c|}{\textbf{ B2 }} & \multicolumn{1}{c|}{\textbf{ B3 }} & \multicolumn{1}{c|}{\textbf{ B4 }} & \multicolumn{1}{c|}{\textbf{ B5 }} & \multicolumn{1}{c|}{\textbf{ B6 }} & \multicolumn{1}{c|}{\textbf{ Difference }} & \multicolumn{1}{c}{\textbf{ Cases }} \\
    \midrule
    \multicolumn{1}{c|}{} & \textbf{ C1 } & \multicolumn{1}{r}{0} & \multicolumn{1}{r}{100} & \multicolumn{1}{r}{2,663} & 11,765 & \multicolumn{1}{r}{11,685} & \multicolumn{1}{r}{11,765} & \multicolumn{1}{r}{11,765} & \multicolumn{1}{r}{11,765} & \multicolumn{1}{r}{11,476} & 11,624 & 11,765 & \multirow{5}[1]{*}{177,446} \\
    \multicolumn{1}{c|}{\textbf{ Recur- }} & \textbf{ C2 } & \multicolumn{1}{r}{0} & \multicolumn{1}{r}{236} & \multicolumn{1}{r}{625} & 22,593 & \multicolumn{1}{r}{9,213} & \multicolumn{1}{r}{22,593} & \multicolumn{1}{r}{22,593} & \multicolumn{1}{r}{22,593} & \multicolumn{1}{r}{4,227} & 9 & 22,593 &  \\
    \multicolumn{1}{c|}{\textbf{ sive-only }} & \textbf{ C3 } & \multicolumn{1}{r}{0} & \multicolumn{1}{r}{114} & \multicolumn{1}{r}{2,733} & 0 & \multicolumn{1}{r}{10,610} & \multicolumn{1}{r}{10,699} & \multicolumn{1}{r}{1} & \multicolumn{1}{r}{10,603} & \multicolumn{1}{r}{10,063} & 10,568 & 10,700 &  \\
    \multicolumn{1}{c|}{} & \textbf{ C4 } & \multicolumn{1}{r}{0} & \multicolumn{1}{r}{253} & \multicolumn{1}{r}{591} & 0 & \multicolumn{1}{r}{8,273} & \multicolumn{1}{r}{19,158} & \multicolumn{1}{r}{0} & \multicolumn{1}{r}{19,155} & \multicolumn{1}{r}{2,416} & 19 & 19,159 &  \\
    \multicolumn{1}{c|}{} & \textbf{ C5 } & \multicolumn{1}{r}{0} & \multicolumn{1}{r}{20} & \multicolumn{1}{r}{21} & 0 & \multicolumn{1}{r}{22,734} & \multicolumn{1}{r}{4,268} & \multicolumn{1}{r}{0} & \multicolumn{1}{r}{1,759} & \multicolumn{1}{r}{4,565} & 1,390 & 38,631 &  \\
    \midrule
    \multicolumn{1}{c|}{} & \textbf{ C6 } & \multicolumn{1}{r}{0} & \multicolumn{1}{r}{121} & \multicolumn{1}{r}{125} & 5,938 & \multicolumn{1}{r}{4,875} & \multicolumn{1}{r}{0} & \multicolumn{1}{r}{0} & \multicolumn{1}{r}{0} & \multicolumn{1}{r}{1,171} & 0 & 8,038 & \multirow{7}[1]{*}{180,000} \\
    \multicolumn{1}{c|}{} & \textbf{ C7 } & \multicolumn{1}{r}{0} & \multicolumn{1}{r}{96} & \multicolumn{1}{r}{97} & 475 & \multicolumn{1}{r}{469} & \multicolumn{1}{r}{0} & \multicolumn{1}{r}{0} & \multicolumn{1}{r}{0} & \multicolumn{1}{r}{181} & 0 & 704 &  \\
    \multicolumn{1}{c|}{} & \textbf{ C8 } & \multicolumn{1}{r}{0} & \multicolumn{1}{r}{34} & \multicolumn{1}{r}{34} & 0 & \multicolumn{1}{r}{157} & \multicolumn{1}{r}{0} & \multicolumn{1}{r}{0} & \multicolumn{1}{r}{0} & \multicolumn{1}{r}{45} & 0 & 850 &  \\
    \multicolumn{1}{c|}{\textbf{ forward }} & \textbf{ C9 } & \multicolumn{1}{r}{0} & \multicolumn{1}{r}{35} & \multicolumn{1}{r}{61} & 5,487 & \multicolumn{1}{r}{3,143} & \multicolumn{1}{r}{0} & \multicolumn{1}{r}{0} & \multicolumn{1}{r}{0} & \multicolumn{1}{r}{1,179} & 0 & 5,487 &  \\
    \multicolumn{1}{c|}{\textbf{ -only }} & \textbf{ C10 } & \multicolumn{1}{r}{0} & \multicolumn{1}{r}{96} & \multicolumn{1}{r}{111} & 8,666 & \multicolumn{1}{r}{5,362} & \multicolumn{1}{r}{0} & \multicolumn{1}{r}{0} & \multicolumn{1}{r}{0} & \multicolumn{1}{r}{1,841} & 0 & 9,259 &  \\
    \multicolumn{1}{c|}{} & \textbf{ C11 } & \multicolumn{1}{r}{0} & \multicolumn{1}{r}{69} & \multicolumn{1}{r}{85} & 7,308 & \multicolumn{1}{r}{4,163} & \multicolumn{1}{r}{0} & \multicolumn{1}{r}{0} & \multicolumn{1}{r}{0} & \multicolumn{1}{r}{1,571} & 0 & 7,308 &  \\
    \multicolumn{1}{c|}{} & \textbf{ C12 } & \multicolumn{1}{r}{0} & \multicolumn{1}{r}{199} & \multicolumn{1}{r}{199} & 1,447 & \multicolumn{1}{r}{1,004} & \multicolumn{1}{r}{0} & \multicolumn{1}{r}{0} & \multicolumn{1}{r}{0} & \multicolumn{1}{r}{481} & 0 & 1,520 &  \\
    \midrule
    \multicolumn{1}{c|}{} & \textbf{ C13 } & \multicolumn{1}{r}{1} & \multicolumn{1}{r}{0} & \multicolumn{1}{r}{14} & 4,132 & \multicolumn{1}{r}{5,842} & \multicolumn{1}{r}{1,389} & \multicolumn{1}{r}{6,961} & \multicolumn{1}{r}{4,299} & \multicolumn{1}{r}{4,122} & 1,402 & 6,962 & \multirow{5}[1]{*}{179,948} \\
    \multicolumn{1}{c|}{\textbf{ CDNS }} & \textbf{ C14 } & \multicolumn{1}{r}{0} & \multicolumn{1}{r}{4} & \multicolumn{1}{r}{159} & 2,014 & \multicolumn{1}{r}{8,915} & \multicolumn{1}{r}{0} & \multicolumn{1}{r}{340} & \multicolumn{1}{r}{0} & \multicolumn{1}{r}{4,203} & 3 & 11,983 &  \\
    \multicolumn{1}{c|}{\textbf{ with }} & \textbf{ C15 } & \multicolumn{1}{r}{9} & \multicolumn{1}{r}{40} & \multicolumn{1}{r}{109} & 8,999 & \multicolumn{1}{r}{24,923} & \multicolumn{1}{r}{0} & \multicolumn{1}{r}{27} & \multicolumn{1}{r}{38,050} & \multicolumn{1}{r}{13,811} & 44 & 39,426 &  \\
    \multicolumn{1}{c|}{\textbf{ fallback }} & \textbf{ C16 } & \multicolumn{1}{r}{0} & \multicolumn{1}{r}{1} & \multicolumn{1}{r}{1} & 770 & \multicolumn{1}{r}{1,971} & \multicolumn{1}{r}{0} & \multicolumn{1}{r}{0} & \multicolumn{1}{r}{4,161} & \multicolumn{1}{r}{714} & 0 & 4,318 &  \\
    \multicolumn{1}{c|}{} & \textbf{ C17 } & \multicolumn{1}{r}{4} & \multicolumn{1}{r}{4} & \multicolumn{1}{r}{6} & 4,224 & \multicolumn{1}{r}{5,893} & \multicolumn{1}{r}{1} & \multicolumn{1}{r}{8,877} & \multicolumn{1}{r}{9,521} & \multicolumn{1}{r}{3,163} & 14 & 9,560 &  \\
    \midrule
    \multicolumn{1}{c|}{} & \textbf{ C18 } & \multicolumn{1}{r}{2} & \multicolumn{1}{r}{0} & \multicolumn{1}{r}{127} & 17,598 & \multicolumn{1}{r}{2,964} & \multicolumn{1}{r}{63} & \multicolumn{1}{r}{95} & \multicolumn{1}{r}{63} & \multicolumn{1}{r}{3,092} & 0 & 17,598 & \multirow{5}[1]{*}{181,215} \\
    \multicolumn{1}{c|}{\textbf{ CDNS }} & \textbf{ C19 } & \multicolumn{1}{r}{0} & \multicolumn{1}{r}{0} & \multicolumn{1}{r}{95} & 9,373 & \multicolumn{1}{r}{6,153} & \multicolumn{1}{r}{0} & \multicolumn{1}{r}{0} & \multicolumn{1}{r}{0} & \multicolumn{1}{r}{5,101} & 0 & 22,954 &  \\
    \multicolumn{1}{c|}{\textbf{ without }} & \textbf{ C20 } & \multicolumn{1}{r}{0} & \multicolumn{1}{r}{0} & \multicolumn{1}{r}{0} & 1,115 & \multicolumn{1}{r}{478} & \multicolumn{1}{r}{0} & \multicolumn{1}{r}{0} & \multicolumn{1}{r}{0} & \multicolumn{1}{r}{255} & 0 & 6,280 &  \\
    \multicolumn{1}{c|}{\textbf{ fallback }} & \textbf{ C21 } & \multicolumn{1}{r}{3} & \multicolumn{1}{r}{0} & \multicolumn{1}{r}{0} & 95 & \multicolumn{1}{r}{95} & \multicolumn{1}{r}{31} & \multicolumn{1}{r}{0} & \multicolumn{1}{r}{31} & \multicolumn{1}{r}{63} & 255 & 1,498 &  \\
    \multicolumn{1}{c|}{} & \textbf{ C22 } & \multicolumn{1}{r}{0} & \multicolumn{1}{r}{0} & \multicolumn{1}{r}{0} & 127 & \multicolumn{1}{r}{31} & \multicolumn{1}{r}{0} & \multicolumn{1}{r}{0} & \multicolumn{1}{r}{0} & \multicolumn{1}{r}{0} & 0 & 797 &  \\
    \midrule
    \multicolumn{2}{c|}{\textbf{ Total }} & 19 & 1,422 & 7,856 & 112,126 & 138,953 & 69,967 & 50,659 & 122,000 & 73,740 & 25,328 & 257,390 & 718,609 \\
    \bottomrule
    \end{tabular}%
    \begin{tablenotes}
      \footnotesize
      \item $B1$: Resolver accepts records matched with queried domain and type, $B2$: Forward fallback mechanism is triggered, 
      \item $B3$: NSEC3 record is accepted, $B4$: NSEC record is accepted. $B5$: Resolver accepts records matched with the queried domain but different types.
      \item $B6$: Resolver validates glue records in the response. For $CP1$, $CP2$, $CP4$, and $RC1$, please check Section~\ref{sec:cases} for details.
    \end{tablenotes}
  \end{threeparttable}
  \vspace{-3mm}
  \label{tab:oracle_cluster}%
\end{table*}%

\section{Large-scale Resolver Scanning}
\label{sec:scanning}





The vulnerabilities identified by \system\ persist in all the studied DNS software (even in their latest versions), and we are interested in learning the impact if these vulnerabilities are exploited in the wild. To this end, we conduct a large-scale network scanning on open resolvers to discover the potentially vulnerable ones. We discuss the ethical issues and how they are addressed in Section~\ref{sec:discussion}.


\vspace{2pt} \noindent 
\textbf{The methodology of scanning.}
As revealed by ~\cite{schomp2013measuring}, the list of open resolvers that are actively operating is constantly changing. 
To obtain an up-to-date list of open resolvers, we conduct active network scanning using \texttt{XMap}~\cite{li2021fast}.
\texttt{XMap} is a fast Internet-wide scanner that allows customization of the probing requests, and we use \texttt{XMap} to send DNS queries from our lab machines to the whole IPv4 address space. Each query is a UDP packet, issued against port 53, and querying about our controlled domain names. We consider the IPs replying with valid responses to be candidates for open resolvers. To notice, scanning from a few vantage points cannot yield a complete resolver list. However, this is the common practice for the measurement studies about Internet services and DNS resolvers~\cite{izhikevich2021lzr, mao2022assessing, akhavan2021cache, pearce2017global}.


Due to ethical concerns, we cannot directly test if a resolver is vulnerable by sending packets generated by our fuzzer (otherwise, we might be considered as attackers by the resolver operators). Hence, we try to learn the software name of the open resolver and check if it matches our studied resolvers. We did not include the software version as we found the version information is usually not provided by the resolver, and all our discovered vulnerabilities are effective against the latest versions.

Specifically, we use both \texttt{version.bind}~\cite{dnsversion} query and a DNS fingerprinting tool \texttt{fpdns}~\cite{fpdns}.
\texttt{version.bind} is a special DNS query name configured with \texttt{TXT} type and \texttt{CHAOS} class to display the DNS software information.
If \texttt{version.bind} returns nothing, we utilize \texttt{fpdns} to identify the version.
\texttt{fpdns} is maintained by the DNS community and covers a wide range of DNS version fingerprints, including mainstream DNS software such as BIND and Unbound.

\vspace{2pt} \noindent
\textbf{Results of scanning.}
Our scanning ran for 5 days in 2022 and we discovered 1,815,017 open DNS resolvers, including both recursive resolvers and forwarders.
We first examined their geo-location and autonomous system (AS) using the GeoLite2 database~\cite{geolite2}.
Table~\ref{tab:resolver_distribution} show that the 1.8M resolvers are located in \textit{229} regions with the top three being China (36.7\%), USA (7.8\%), and India (6.0\%). \textit{25,342} AS numbers were associated with these resolvers, showing our list has broad coverage of resolvers.

Out of the 1.8M resolvers, we identified \textit{848,708} (46.8\%) returning software information to our probing methods of \texttt{version.bind} (329,722, 38.9\%) or \texttt{fpdns} (518,986, 61.1\%), as listed in Table~\ref{tab:resolver_version}.
Totally, \textit{71,698} (8.4\%) resolvers are potentially impacted by our discovered vulnerabilities, including BIND (48,407), Unbound (11,712), Knot (43), and PowerDNS (11,536).
We did not discover resolvers using MaraDNS and Technitium since they do not support the \texttt{version.bind} query or have fingerprints in the \texttt{fpdns} database. Noticeably, our scanning results give a conservative estimation of the vulnerable resolvers, due to the limitations of our probing methods.

\end{document}